\pdfoutput = 1

\documentclass{aastex62}

\received{May 27, 2020}
\revised{June 22, 2020}
\accepted{April 30, 2021}
\submitjournal{AJ}

\shorttitle{Variable Stars of Serpens Main}
\shortauthors{Yin et al.}

\begin{document}

\title{Census of Variable Stars toward Serpens Main}

\correspondingauthor{Jia Yin}
\email{jyin@nao.cas.cn}

\author[0000-0002-0786-7307]{Jia Yin}
\affil{ National Astronomical Observatories,
Chinese Academy of Sciences,
100101 Beijing, People's Republic of China}
\affiliation{ University of Chinese Academy of Sciences,
100049 Beijing, People's Republic of China}

\author{ZhiWei Chen}
\affil{ Purple Mountain Observatory,
Chinese Academy of Sciences,
210034 Nanjing, People's Republic of China}

\author{Rolf Chini}
\affil{ Astronomisches Institut, Ruhr–Universität Bochum,
Universitätsstrasse 150, 44801 Bochum, Germany}
\affiliation{ Instituto de Astronom\'{i}a, Universidad Cat\'{o}lica del Norte,
Avenida Angamos 0610, Casilla 1280, Antofagasta, Chile}

\author{Martin Haas}
\affil{ Astronomisches Institut, Ruhr–Universität Bochum,
Universitätsstrasse 150, D-44801 Bochum, Germany}

\author{Sadegh Noroozi}
\affil{ Centrum Astronomiczne im. Mikolaja Kopernika,
PAN, Bartycka 18, 00-716 Warsaw, Poland}

\author{YongQiang Yao}
\affil{ National Astronomical Observatories,
Chinese Academy of Sciences,
100101 Beijing, People's Republic of China}

\author{ZhiBo Jiang}
\affil{ Purple Mountain Observatory,
Chinese Academy of Sciences,
210034 Nanjing, People's Republic of China}

\author{Xuan Qian}
\affil{ National Astronomical Observatories,
Chinese Academy of Sciences,
100101 Beijing, People's Republic of China}

\author{LiYong Liu}
\affil{ National Astronomical Observatories,
Chinese Academy of Sciences,
100101 Beijing, People's Republic of China}

\author{Yao Li}
\affil{ National Astronomical Observatories,
Chinese Academy of Sciences,
100101 Beijing, People's Republic of China}
\affiliation{ University of Chinese Academy of Sciences,
100049 Beijing, People's Republic of China}



\begin{abstract}
We have monitored a 3 deg$^2$ area toward Serpens Main in the Pan-STARRS1 $r$, $i$, and $z$ bands from 2016 April to September. Light curves of more than 11,000 stars in each band were obtained, and 143 variables have been identified. Among those, 119 variables are new discoveries, while 24 variables were previously known. We present variability classes and periods of 99 stars. Of these, 81 are located in the upper giant branch, displaying long periods, while the remaining 18 variables are pre-main-sequence objects with short periods. We also identify eight eclipsing binary systems, including the known binary V0623\,Ser, and derive their physical parameters. According to a clustering analysis of Gaia DR2 stars in the observed field, there are 10 variable members in Serpens Main, where six members have been classified as young stellar objects in previous studies. Here we provide color--magnitude and color--color diagrams for these variables. The color variability of most variables in the color--magnitude diagrams produce the expected displacements, while the movements of cluster members point in different directions; this behavior may be associated with accretion spots or circumstellar disks.
\end{abstract}



\section{Introduction} \label{sec:intro}

Variability is a very common and complex characteristic of young stellar objects (YSOs). Due to thick envelopes and circumstellar accretion disks, YSOs appear reddened and have infrared (IR) excess. The variability of YSOs, which affects optical and IR wavelengths, occurs over different timescales \citep{2014AJ....148..122G} and originates from material exchange between the disks and the central stars driven by accretion and stellar winds \citep{2015AJ....150..145W, 2019A&A...627A.135B}. Various processes like obscuration by clumpy circumstellar material, unsteady accretion rates, and hot or cool photospheric spots may contribute simultaneously to the observed variability and thus complicate its interpretation. Moreover, variability is also influenced by disk structure \citep{2013ApJ...773..145W}, magnetic field intensity \citep{2012ApJ...756...68C}, stellar mass \citep{1999AJ....118.1043H}, and rotation rate \citep{2007A&A...461..183G}.

Optical monitoring is primarily sensitive to events near the stellar photosphere, where the photometric variability of the obtained light curves depends on the YSO's evolutionary stage. For example, weak-line T\,Tauri stars (WTTSs) display stable sinusoidal periodic light curves attributed to cold magnetic spots on their stellar surfaces \citep{1999AJ....117.2941S,2009A&A...508.1313F}, while classical T\,Tauri stars (CTTSs) exhibit stochastic or semiperiodic light curves due to more complex time-domain behavior \citep{1993AJ....106.1608V,2008MNRAS.391.1913R}. \citet{2014AJ....147...82C} presented a comprehensive analysis of young members in the cluster NGC 2264 with high-cadence time-series optical photometry obtained with CoRoT and in the mid-IR with Spitzer. They focused on 162 CTTSs and identified seven light-curve morphology classes, which they suggested represent different physical processes and geometric effects. For example, ``dippers'' and ``bursting'' were interpreted as changes in extinction and accretion events, which could explain the sudden downward dips and upward spikes in their light curves, respectively. Since contamination from nearby or line-of-sight objects ubiquitously exists in imaging studies of specific regions, light from evolved stars could be mixed with light from YSOs. Other evolved stars exhibit stable periodic light curves due to pulsations (e.g., RR Lyrae stars, classical Cepheids, and Mira stars) and due to the effects of rotation and eclipses in binary systems \citep{2008JPhCS.118a2010E}. In principle, the variabilities of different types of objects can be distinguished based on the shape of their optical and IR light curves.

\begin{figure}[!ht]
 \centering
          \includegraphics[width=.6\textwidth]{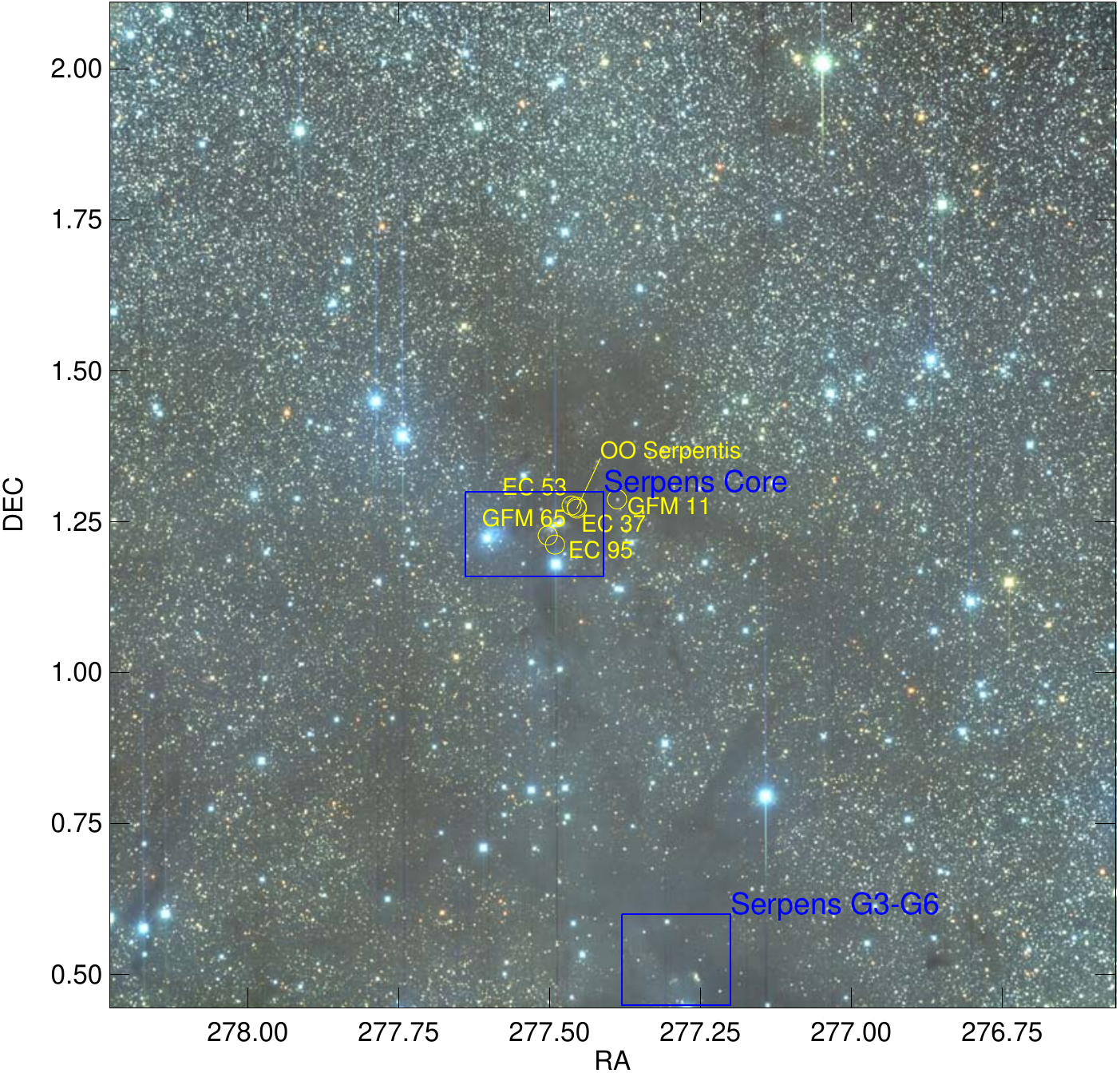}
      \caption{Color-composite image of Serpens Main (blue: $r$, green: $i$, and red: $z$). The clusters of Serpens Core and Serpens G3--G6 are shown as blue rectangles. Three radio sources, GFM\,11, EC\,95, and GFM\,65, which were used to measure the distance to Serpens by \citet{2017ApJ...834..143O} and the YSOVAR project (OO\,Serpentis, EC\,53, and EC\,37; \citealt{2012ApJ...744...56H}), are all marked with yellow circles.}\label{fig:comp}
\end{figure}

Time-domain imaging surveys of YSOs in star-forming regions (SFRs) are efficient in characterizing YSO variability. The near- and mid-IR wavelengths enable monitoring of the embedded nature of YSOs, where studies of sources in Orion \citep{2001AJ....121.3160C, 2015AJ....150..132R}, the Chamaeleon I molecular cloud \citep{2002AJ....124.1001C}, and Cyg OB7 \citep{2013ApJ...773..145W} have found that a very large fraction ($\geqslant90\%$) of YSOs are variables; and the other YSOs may be variables, since the small variability amplitudes could not be picked out from the noise of the light curves \citep{2014AJ....148..122G, 2015AJ....150..118P}. Consistent results have also been obtained for young clusters such as IRAS\,20050+2720 \citep{2015AJ....150..118P}, GGD\,12-15 \citep{2015AJ....150..145W}, and NGC\,1333 \citep{2015AJ....150..175R}. In these studies, $\sim$50--90\% of the cluster members were found to be variables, and the variable fraction for the most embedded objects (Class\,I and flat) with flat spectral energy distributions (SEDs) was higher than for less embedded objects (Class\,II and Class\,III). The most variable YSOs exhibit redder colors in fainter states associated with time-variable extinction. Some YSOs, however, become bluer when fainter due to changes in their accretion disk structure \citep{2013ApJ...773..145W, 2018AJ....155...99W}. For example, \citet{2019A&A...627A.135B} studied the photometric variability of pre-main-sequence (PMS) stars in the Pelican Nebula at optical wavelengths and found that the accreting CTTSs displayed significantly higher variability than the disk-free WTTSs. The combined variations of PMS stars in terms of luminosity and color were attributed to accretion, circumstellar extinction, or spot modulations.

The Serpens molecular cloud, which has attracted considerable research interest, contains objects in all stages of evolution prior to the main sequence (MS) stage \citep{2014ApJ...797...76L,2015AJ....149..103E}. Serpens Main (see Figure \ref{fig:comp}) is a small region in the Serpens cloud and contains three prominent subregions: the Serpens core (Cluster A), Serpens\,G3--G6 (Cluster B), and VV\,Serpentis \citep{2006ApJ...644..307H}. Numerous observations of Serpens Main, from X-ray to submillimeter, revealed a large population of YSOs from Class\,0 to Class\,III  \citep{2008hsf2.book..693E,2010ApJ...714..778O,2015ApJS..220...11D}. To study the physical mechanisms of accretion in the early phases of stellar evolution, IR-to-millimeter observations were carried out during the YSO VARiabilty (YSOVAR) project, including photometric mid-IR time series obtained with Spitzer \citep{2014AJ....148...92R}. Variabilities of deeply embedded protostars were studied by interferometric observations \citep{2019ApJ...871..149F}. The outflow activity of OO\,Serpentis and the spatial and temporal variations of the surrounding bipolar reflection nebula were observed by \citep{2012ApJ...744...56H}. The Class\,I YSO EC\,53 (V371\,Ser) shows similar light curves in the $K-$band and 850\,$\mu$m, both with periods of 543 days, which are attributed to disk accretion in a close binary system \citep{2012ApJ...744...56H,2017ApJ...849...69Y}. The deeply embedded outburst star (DEOS) in Serpens has been classified as an FU\,Orionis variable due to its irregular decline rates after the outburst. The object EC\,37 (V370\,Ser) shows a slow irregular variation, probably caused by mass accretion \citep{1996ApJ...468..861H,1999AJ....118.1338H}.

Time-domain observations are helpful for understanding the origins of the stellar variability of YSOs at different evolutionary stages. In this context, we made use of the robotic Berlin Exoplanet Search Telescope\,II (BEST\,II) to detect variable YSOs in Serpens Main. Section~\ref{sec:obs} presents details of the telescope and the observations. In Section~\ref{sec:data}, we present our procedures for data calibration, photometry, and data analysis. The results for the variable stars toward Serpens Main are presented in Section~\ref{sec:vs}. Section~\ref{sec:cluster} describes the clustering analysis of Serpens Main and the related variable stars. Section~\ref{sec:color} presents the color variability of the different types of variables using color--magnitude diagrams (CMDs) and color--color diagrams (C-CDs). A summary is then given in Section~\ref{sec:summary}.

\section{Data} \label{sec:obs}
\subsection{Telescope and Observations}

The monitoring observations of Serpens Main were conducted with the robotic BEST\,II system \citep{2004PASP..116...38R,2010AJ....139...53R} located at the Observatory Cerro Armazones (OCA) in Chile. The system consists of a 25\,cm Baker--Ritchey--Chr\`{e}tien telescope with a focal ratio of $f$/5.0, yielding a wide field of view (FOV) of $1.^{\circ}5\times1.^{\circ}5$. The telescope is equipped with a Peltier-cooled 4k$\times$4k Finger Lakes CCD imager (KAF-16801E2) with a pixel size of 9\,$\mu$m and a pixel scale of $1\farcs5$\,pixel$^{-1}$. Table~\label{tab:best} summarizes the main technical details of BEST\, II.

\begin{deluxetable*}{lc}
\tablecaption{Parameters of BEST\,II. \label{tab:best}}
\tablewidth{0pt}
\tablehead{\colhead{Parameter} & \colhead{Value}}
\startdata
Aperture & 250\,mm \\
Focal~ratio & $f$/5.0 \\
CCD~size & 4096 $\times$ 4096 \\
Pixel~size  & 9\,$\mu$m \\
Pixel~scale  & 1.5\,arcsec pixel$^{-1}$ \\
Field~of~view & $1\fdg5$ \\
Precision & $<1\%$ ($V<15$\,mag) \\
\enddata
\end{deluxetable*}

\begin{figure}[!ht]
 \centering
          \includegraphics[width=.6\textwidth]{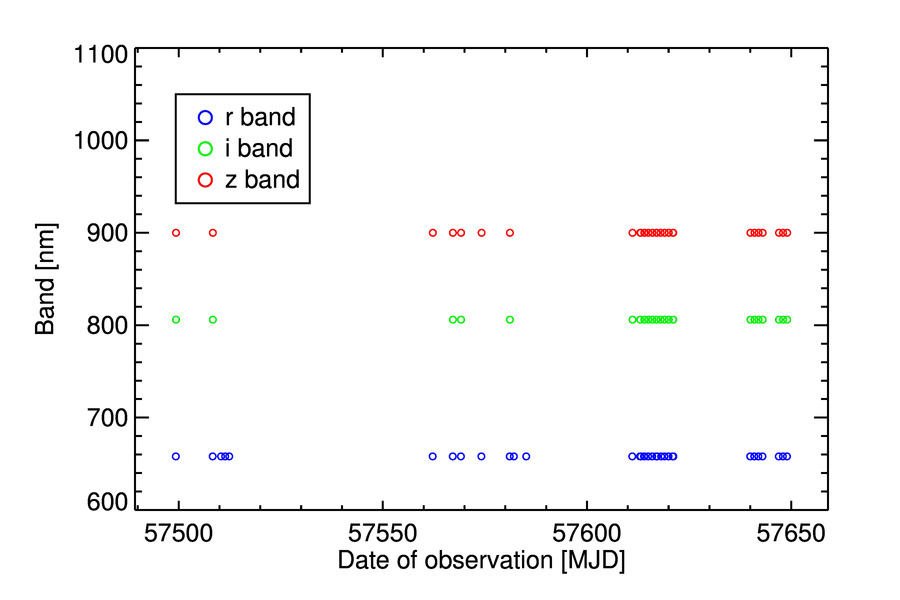}
      \caption{Observing epochs for Serpens Main in the $riz$ filters.}\label{fig:mjd}
\end{figure}

Serpens Main was observed in the Pan-STARRS1 $r$, $i$, and $z$ filters for 29, 27, and 24 nights, respectively, from 2016 April 20 to September 17 in (see Figure~\ref{fig:mjd}). The observed region was centered at R.A.(2000) = $18^{\rm h} 29^{\rm m} 35^{\rm s}$, decl.(2000) = $+01^{\circ} 16' 41''$, and the exposure time was 120\,s for each single frame. A pattern consisting of nine dithered positions was used in each sequence of $r$, $i$, and $z$ filters. To reduce background noise and improve the photometric quality, we coadded (averaged) images in each filter. The minimum time interval between observations was about 0.09\,days (2 hr) for each filter. Flat, bias, and dark images were taken before and after the observations.

\subsection{Auxiliary Data}
\label{sec:axldata}

In this study, we made use of the highly accurate positions, parallaxes, proper motions, astrophysical parameters, effective temperatures and variabilities from Gaia DR1 \citep{2016A&A...595A...2G} and DR2 \citep{2018A&A...616A...1G}. In the field of Serpens Main, the typical uncertainties of the celestial positions, parallaxes, and proper motions are about 0.14\,mas, 0.28\,mas, and 0.16\,mas yr$^{-1}$, respectively. The stellar effective temperatures were derived from three broadband filters ($G$, $G_{\rm BP}$, and $G_{\rm RP}$) and the parallaxes, where the typical uncertainty of a temperature is about 401\,K \citep{2018A&A...616A...1G}. \citet{2018A&A...618A..30H} provided variability classifications for the objects in Serpens Main, and the results of 23 objects have been used in this study as a reference. \citet{2018A&A...616A...2L} obtained the periods of the variables in Serpens Main from astrometric data covering 640 days and from which we selected the results of 10 objects as reference.

As one of the nearest SFRs, the YSOs of Serpens Main have mostly been identified according to their IR properties, analogous to identifying forming stars with disks/envelopes (\citealt{2007ApJ...663.1149H,2009ApJS..184...18G}). In previous studies of YSOs \citep{2007ApJ...669..493W,2009ApJS..181..321E,2013ApJ...762..128O,2015ApJS..220...11D}, YSO catalogs of Serpens Main were compiled based on the relevant observations; their evolutionary stages were assigned whenever their SEDs were well constructed.

\section{Data Reduction and the Variable Star Catalog} \label{sec:data}
\subsection{Data Reduction}\label{subsec:reduction}
The Bochum OCA VYSOS interactive pipeline was used to reduce and calibrate the images obtained with BEST\,II \citep{2012AN....333..706H}. Astrometric corrections were applied using SCAMP \citep{2006ASPC..351..112B}, SExtractor \citep{1996A&AS..117..393B} and SWARP \citep{2002ASPC..281..228B}; resampling was applied to the reduced images, yielding an image size of $\sim8000\times8000$\,pixels with a pixel scale of $0\farcs77$\,pixel$^{-1}$.

\begin{figure}[!h]
 \centering
          \includegraphics[width=.6\textwidth]{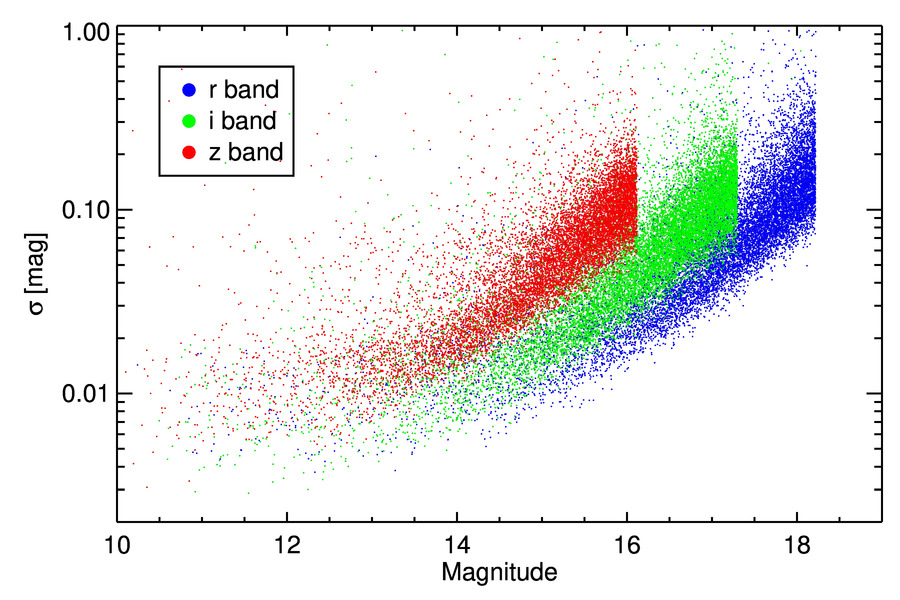}
      \caption{Photometric uncertainties of the detected sources in the $riz$ filters.}\label{fig:sigma}
\end{figure}

The SExtractor tool was employed to derive the magnitudes of the objects in the reduced $riz$ images of Serpens Main in each epoch. Then, relative light curves of the detected objects were constructed after correcting for atmospheric extinction and transmission. The absolute flux calibrations of the objects were achieved using standard-star photometry based on Pan-STARRS1 stars in the FOV. The completeness limits were 18.2, 17.3, and 16.1\,mag in the $riz$ filters, respectively. The numbers of detected objects whose median magnitudes were brighter than the completeness limits were 11,971 in $r$, 14,134 in $i$, and 11,179 in $z$. The photometric uncertainties and median magnitudes of these objects are shown in Figure~\ref{fig:sigma}. The photometric uncertainties of most of these objects were smaller than 0.1\,mag.

\subsection{Identification of Variable Objects}\label{subsec:detection}
As mentioned by \citet{2012AN....333..706H}, three methods can be used to find variable stars, i.e. the standard deviation ($S$) of the light curves (the $SD$ method), the amplitude ($A$) of the light curves (the $A$ method), and the Stetson's variability $J$ Index \citep[the $J$ method;][]{1996PASP..108..851S}. For the $SD$ method and $A$ method, the standard deviations and amplitudes increase as the brightness decreases. Threshold values of $5\sigma$ above the average level, as determined from their brightness distributions, are typically used. If the $S$ or $A$ of an object is larger than the threshold value, it is identified as a variable object. For the $J$ method, the relatively strict criterion ($J \geqslant 0.5$) is used.

\begin{figure}[!h]
 \centering
          \includegraphics[width=.3\textwidth]{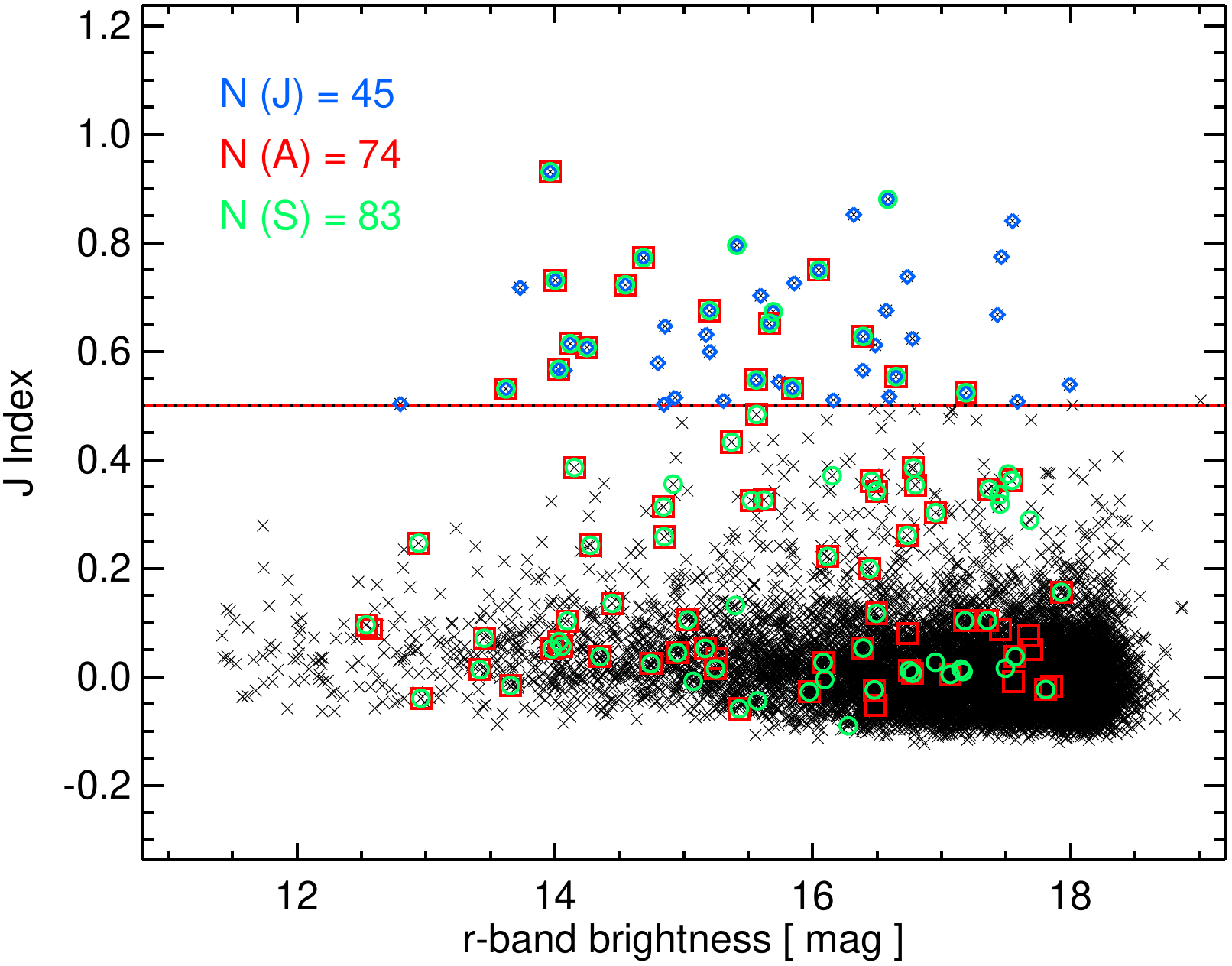}
          $\quad$
          \includegraphics[width=.3\textwidth]{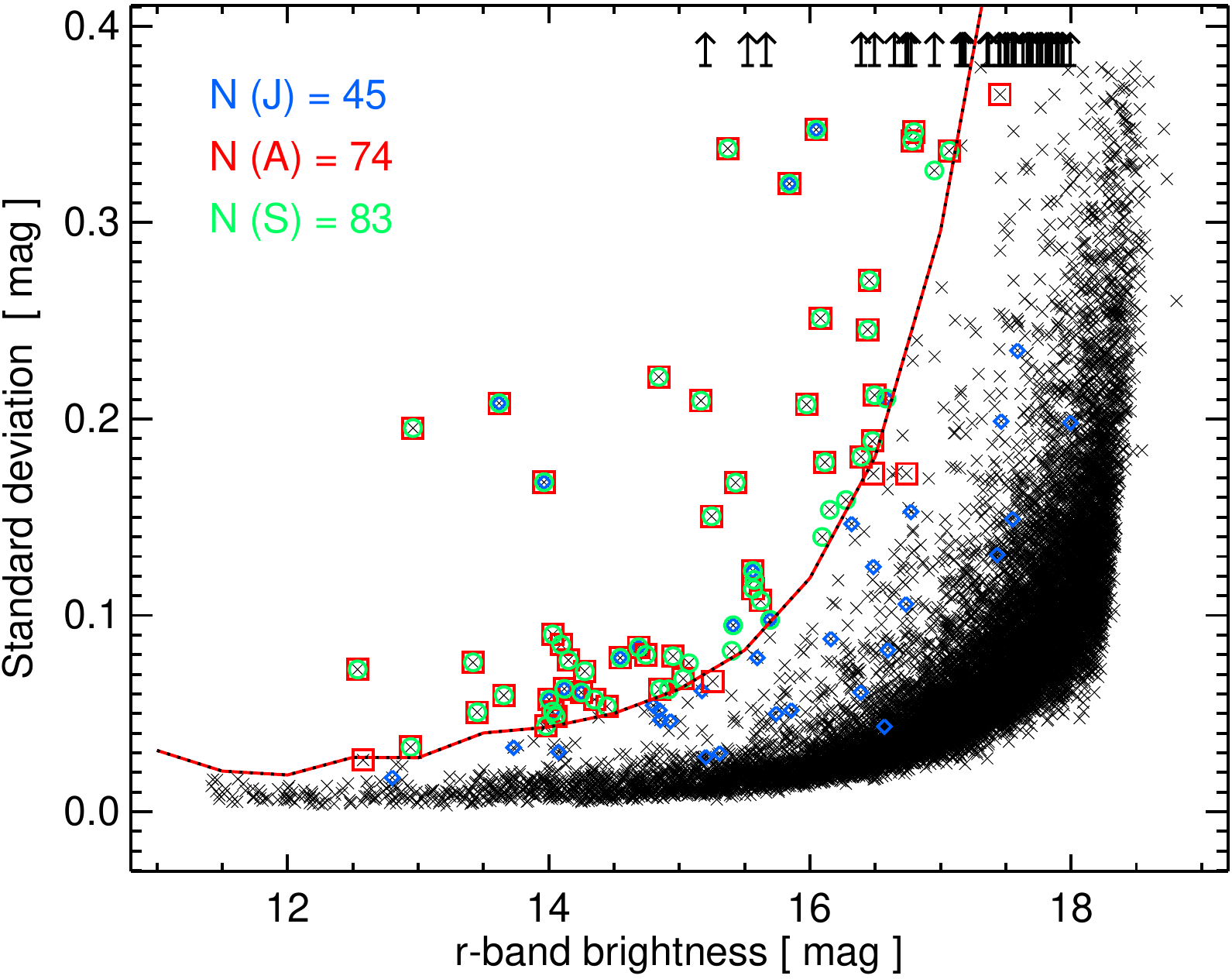}
          $\quad$
          \includegraphics[width=.3\textwidth]{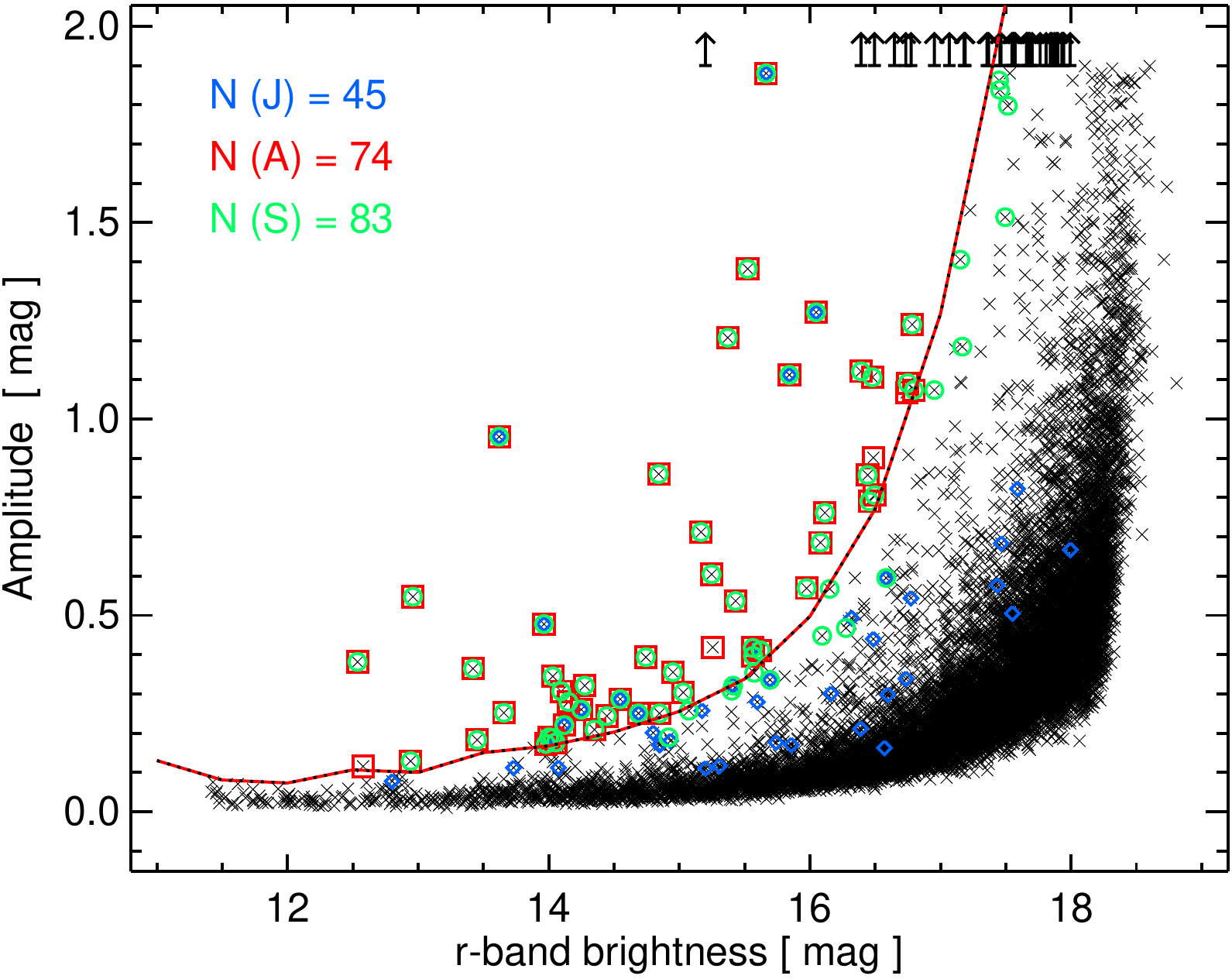}
          \\ $\quad$ \\
          \includegraphics[width=.3\textwidth]{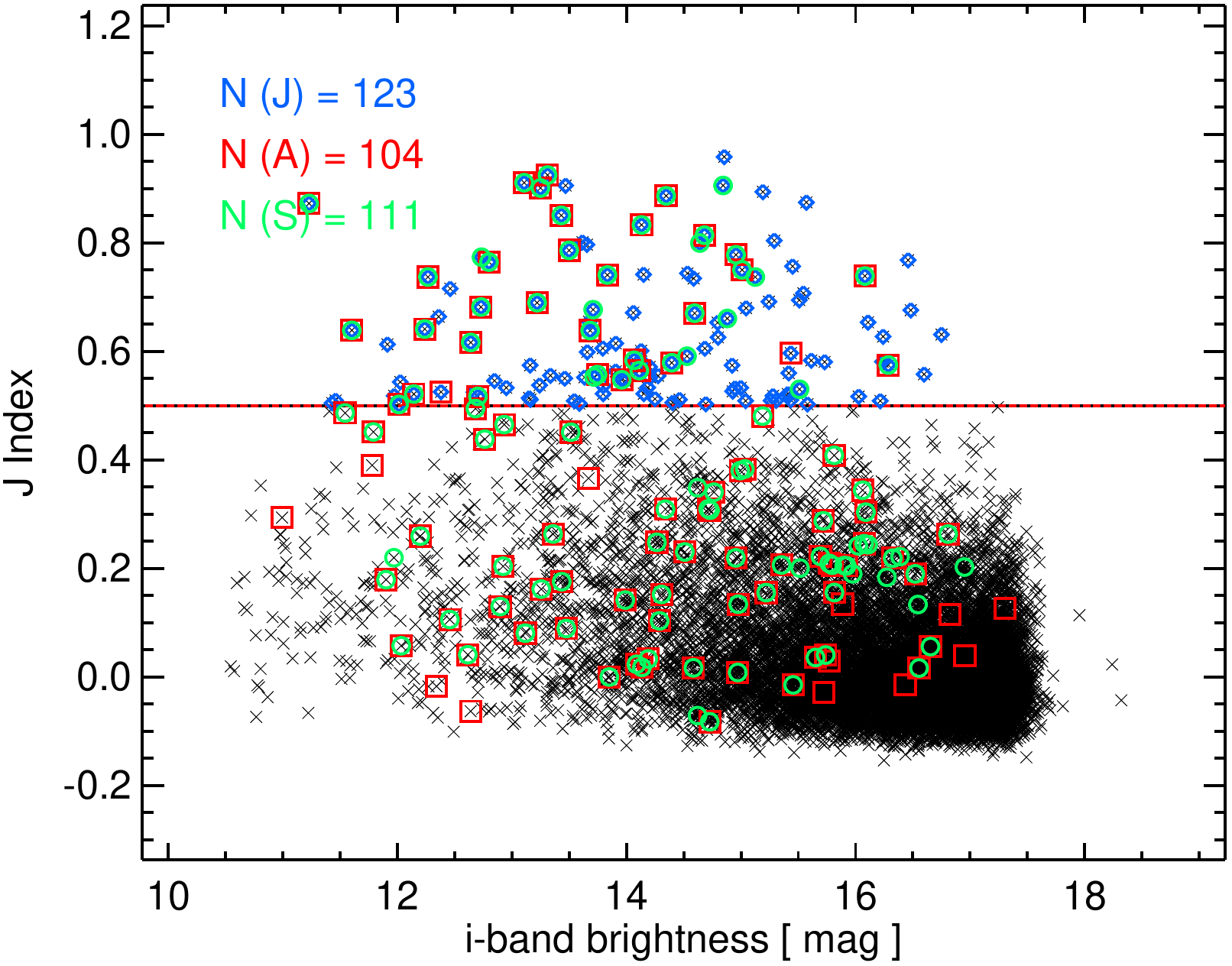}
          $\quad$
          \includegraphics[width=.3\textwidth]{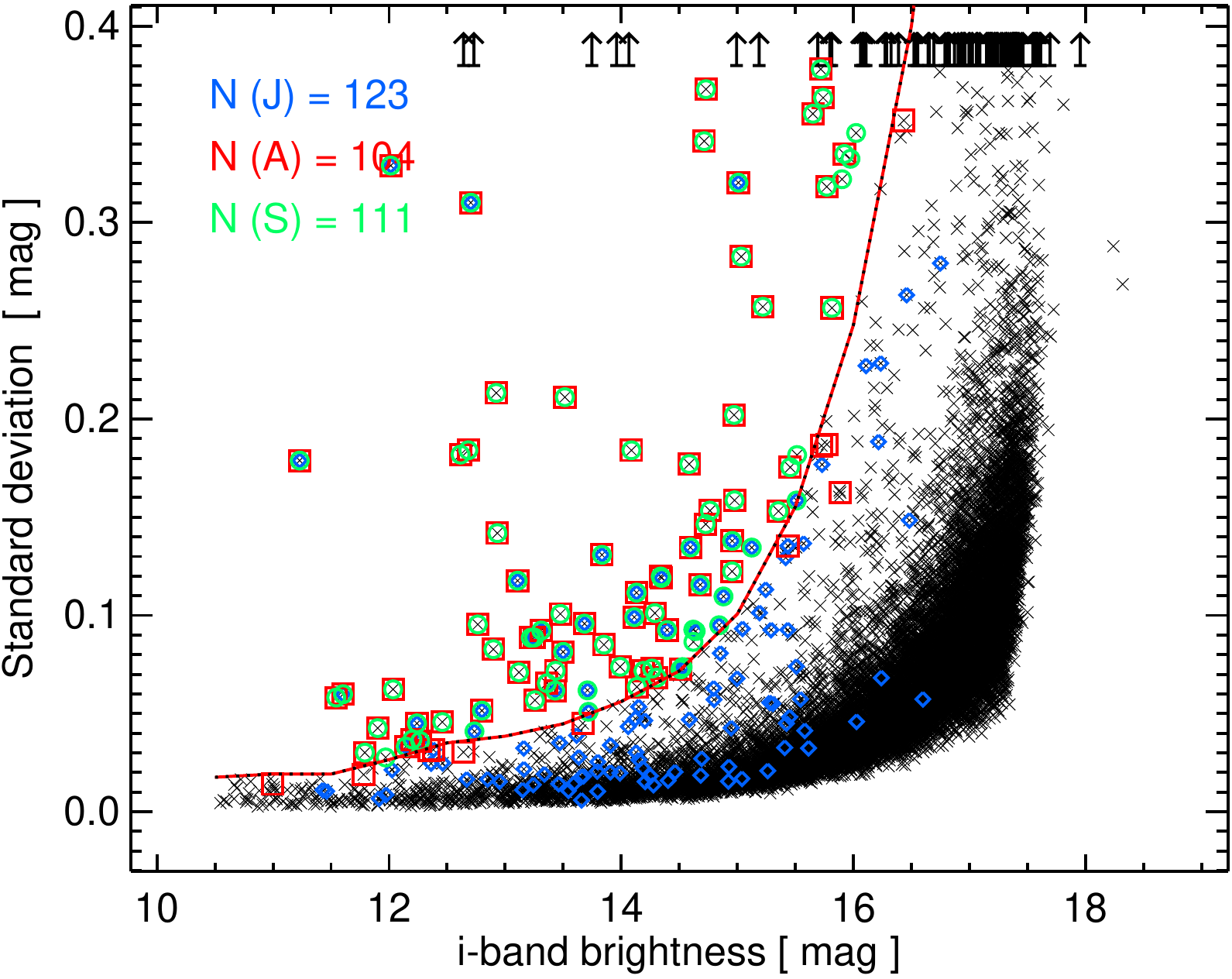}
          $\quad$
          \includegraphics[width=.3\textwidth]{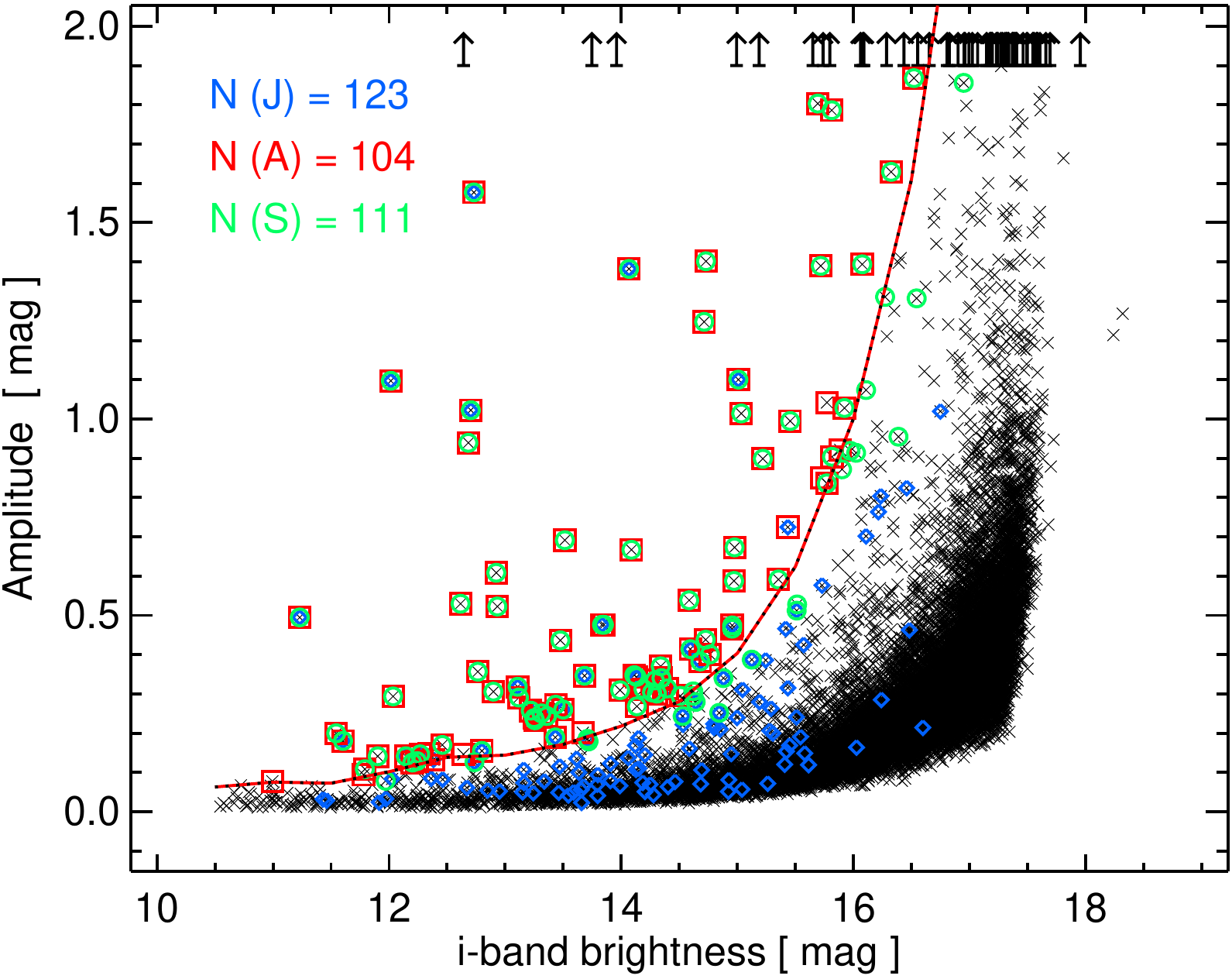}
          \\ $\quad$ \\
          \includegraphics[width=.3\textwidth]{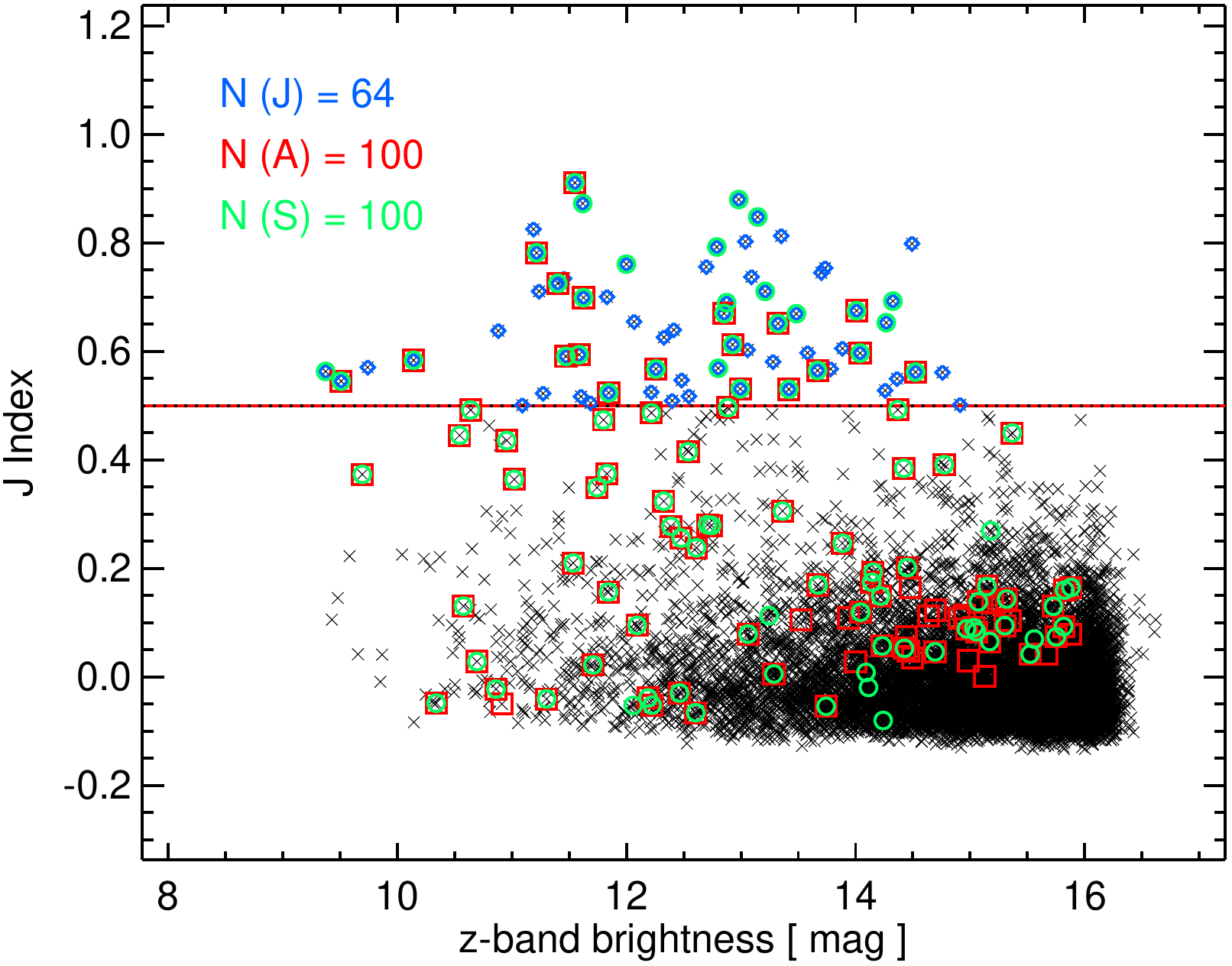}
          $\quad$
          \includegraphics[width=.3\textwidth]{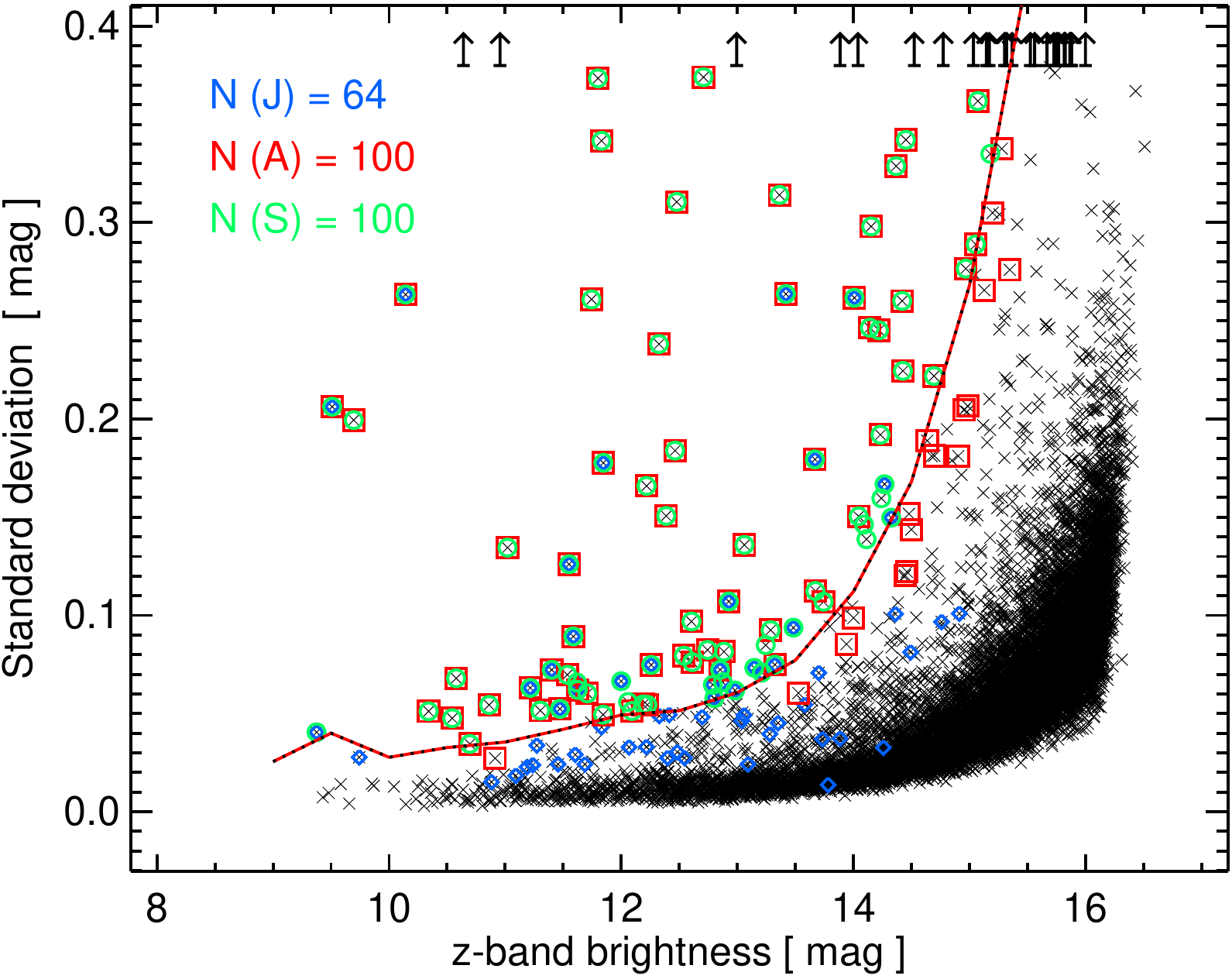}
          $\quad$
          \includegraphics[width=.3\textwidth]{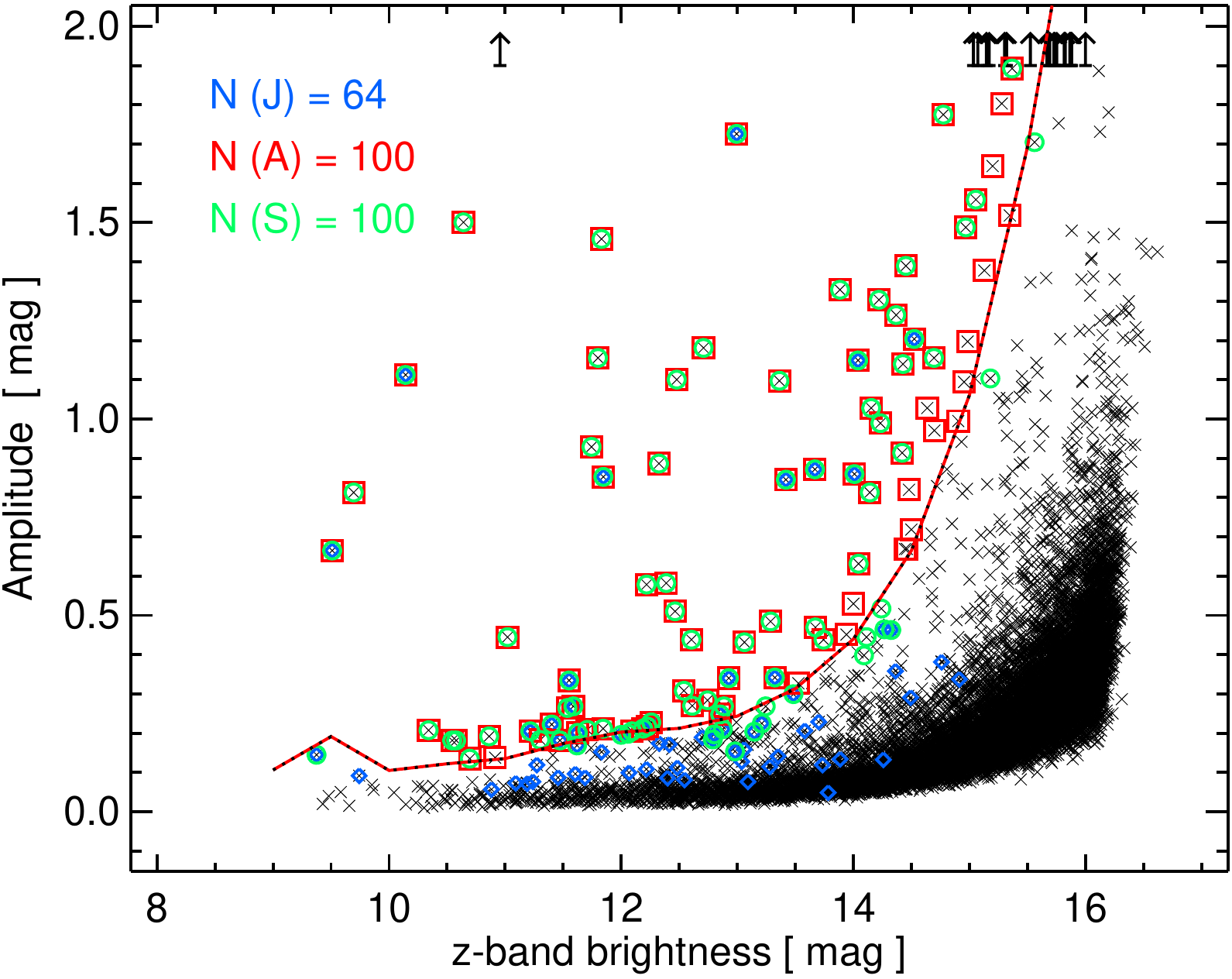}
      \caption{Distributions of the $J$-index, $S$, and $A$, which vary as a function of the brightness of the detected objects in $r$, $i$, and $z$ (from top to bottom). The variable stars identified from the $J$-index, $S$, and $A$ methods are marked as blue diamonds, green circles and red squares, respectively. $N$ is the number of variable stars. The black crosses ($\times$) denote the detected objects, and the arrows ($\uparrow$) indicate objects that were out of the display range ($S > 0.4$\,mag or $A > 2.0$\,mag).}\label{fig:select3}
\end{figure}

To be classified as a variable star, an object has to satisfy the following criteria:

\begin{equation}
\{S | S \geqslant S_{m} + 5 \sigma_S\} \cup
\{A | A \geqslant A_{m} + 5 \sigma_A\} \cup
\{J|J \geqslant 0.5\}
\end{equation}

where $S_m$ and $A_m$ are the median values of $S$ and $A$ in each bin of brightness (bin = 0.5\,mag), respectively, and $\sigma_S$ and $\sigma_A$ are the standard deviations of all $S$ and $A$ values, respectively.

Figure~\ref{fig:select3} shows the distributions of the $J$ index, $S$, and $A$, which vary with the brightness of the detected objects in the $riz$ bands. We have derived 118, 206, and 152 variable objects in the $r$, $i$, and $z$ bands, respectively. Note that all variable objects that were affected by saturation were removed. Likewise, any dubious objects that satisfied the criteria in only one band with a photometry quality flag greater than or equal to 3 (e.g., the photometric measurement of the object was significantly affected by close sources, bad pixels, and/or saturation) were also removed. This vetting stage yielded 143 objects, including 43 variable objects in $riz$, 63 variable objects in two bands ($ri$:22, $iz$:41, and $rz$:0) and 37 objects in only one band ($r$:6, $i$:17, and $z$:14).

We introduced a ranking system according to the number of eligible bands; i.e. the numbers 3, 2 and 1 indicate whether the variable object was detected in three, two, or one band(s). The amplitudes of the objects were different in each band. For the 63 variable objects found in two bands and the 37 objects in one band, if the correlation between the nonvariable and variable bands was higher than 0.8, we appended a plus sign to the main rank number, representing the possibility that it was not actually detected by the current data and methods. A six-level rank system was thus introduced to weigh the variability of the 143 objects, from high to lowest, as: 3, 2+, 2, 1++, 1+, and 1. From the 143 variable objects, 43 were classified as level 3, 20 as level 2+, 43 as level 2, 11 as level 1++, 11 as level 1+, and 15 as level 1. Variability information on the 143 variable objects is listed in Table~\ref{tab:color}.

\begin{longrotatetable}
\begin{deluxetable*}{cccLcLcLccccc}
\tablecaption{Parameters of the variable sources.\label{tab:color}}
\tablewidth{700pt}
\tabletypesize{\scriptsize}
\tablehead{
\colhead{Obs\_ID} &  \colhead{Ra(J2000.0)} & \colhead{Dec(J2000.0)} & \colhead{$m_r$} & \colhead{$A_r$} &
 \colhead{$m_i$} & \colhead{$A_i$} &  \colhead{$m_z$} & \colhead{$A_z$} &
 \colhead{Bands} & \colhead{Ranking} & \colhead{Parallax} & \colhead{Type}  \\
\colhead{} &  \colhead{(degree)} & \colhead{(degree)} & \colhead{(mag)} & \colhead{(mag)} & \colhead{(mag)} & \colhead{(mag)} & \colhead{(mag)} & \colhead{(mag)} & \colhead{} & \colhead{} & \colhead{(mas)} & \colhead{}
}
\startdata
V001$^{ }$ & 277.48768& +00.54992& 15.24(<0.01)& 0.71& 14.13(<0.01)& 0.66& 13.11(<0.01)& 0.43& $riz$ & 3   & 	0.0951(0.6678)& LP, YSO      \\
V002$^{d}$ & 277.48822& +00.55495& 16.93( 0.01)& 2.53& 15.72( 0.01)& 1.80& 14.51( 0.01)& 1.39& $riz$ & 3   & 	2.0418(0.1115)& SP, YSO      \\
V003$^{ }$ & 276.75894& +00.56156& 17.60( 0.03)& 0.43& 15.00(<0.01)& 0.23& 13.17(<0.01)& 0.17& $i	 $ & 1   & 	0.2159(0.1687)&             \\
V004$^{d}$ & 277.43388& +00.56559& 16.51( 0.01)& 0.86& 15.26( 0.01)& 0.89& 14.20( 0.01)& 0.81& $riz$ & 3   & 	2.0712(0.0917)& SP, YSO      \\
V005$^{ }$ & 276.71648& +00.56788& 16.20( 0.01)& 2.43& 13.36(<0.01)& 2.97& 10.76(<0.01)& 2.10& $riz$ & 3   & 	0.2491(0.2211)& SR\_M       \\
V006$^{ }$ & 276.72783& +00.58798& 16.39( 0.01)& 0.28& 13.71(<0.01)& 0.20& 11.88(<0.01)& 0.15& $iz $ & 2   & 	0.3410(0.1288)&             \\
V007$^{ }$ & 276.83282& +00.58941& 15.58(<0.01)& 0.40& 13.38(<0.01)& 0.24& 11.81(<0.01)& 0.14& $ri $ & 2   & 	0.1861(0.1290)& LP?         \\
V008$^{ }$ & 277.57395& +00.60300&             &     & 15.13(<0.01)& 2.47& 11.87(<0.01)& 1.46& $iz $ & 2   & -0.3016(0.3321) &SR\_M, YSO$?$ \\
V009$^{ }$ & 277.91040& +00.60008&             &     & 15.44( 0.01)& 0.31& 12.81(<0.01)& 0.19& $iz $ & 2   & 	0.4468(0.3032)& LP?         \\
V010$^{ }$ & 276.75656& +00.60338& 14.05(<0.01)& 0.19& 12.29(<0.01)& 0.12& 11.03(<0.01)& 0.06& $ri $ & 2	 & 	0.2635(0.1095)& LP          \\
V011$^{ }$ & 277.58136& +00.62222& 17.77( 0.03)& 0.46& 14.14(<0.01)& 0.34& 11.24(<0.01)& 0.20& $iz $ & 2   & 	0.5301(0.3635)&SR\_M, YSO$?$ \\
V012$^{ }$ & 277.87199& +00.63251& 13.72(<0.01)& 0.25& 13.48(<0.01)& 0.43&             &     & $ri $ & 2   & 	2.3111(0.0401)&             \\
V013$^{d}$ & 277.49323& +00.63125& 16.14( 0.01)& 0.68& 15.03(<0.01)& 0.58& 14.16(<0.01)& 0.40& $riz$ & 3   & 	2.1982(0.0728)&EW           \\
V014$^{ }$ & 277.87346& +00.63221& 13.48(<0.01)& 0.36& 12.68(<0.01)& 0.14& 11.91(<0.01)& 0.03& $ri $ & 2   & 	2.0765(0.0393)&             \\
V015$^{ }$ & 277.93818& +00.64686& 17.11( 0.02)& 0.28& 14.81(<0.01)& 0.21& 13.11(<0.01)& 0.07& $iz $ & 2+  & 	0.1907(0.1543)&             \\
V016$^{ }$ & 277.94027& +00.65329& 16.93( 0.01)& 3.91& 13.86(<0.01)& 2.18& 11.75(<0.01)& 1.15& $riz$ & 3   & 	0.2424(0.2407)& SR\_M       \\
V017$^{ }$ & 276.89269& +00.67338&             &     & 16.25( 0.02)& 0.80& 14.21( 0.01)& 0.40& $i  $ & 1+  & 	0.1095(0.1504)& SR\_M, YSO$?$ \\
V018$^{ }$ & 276.99561& +00.68410& 17.27( 0.03)& 0.45& 14.30(<0.01)& 0.30& 11.87(<0.01)& 0.21& $iz $ & 2   & 	0.0745(0.1890)& LP          \\
V019$^{c}$ & 276.72923& +00.68920& 17.59( 0.03)& 0.82& 14.97(<0.01)& 0.47& 13.21(<0.01)& 0.22& $riz$ & 3   & 	0.1068(0.1533)& LP          \\
V020$^{ }$ & 276.79820& +00.71692& 16.86( 0.01)& 0.54& 13.93(<0.01)& 0.48& 11.69(<0.01)& 0.20& $riz$ & 3   & 	0.3802(0.1503)& LP          \\
V021$^{c,d}$ & 277.52568& +00.70935& 13.69(<0.01)& 0.95& 12.76(<0.01)& 0.94& 11.92(<0.01)& 0.85& $riz$ & 3 & 	2.6038(0.1899)& SP$?$, YSO \\
V022$^{c}$ & 277.84070& +00.76939&             &     & 16.01( 0.01)& 2.05& 13.39(<0.01)& 1.10& $iz $ & 2   & 	0.0345(0.1777)& SR\_M       \\
V023$^{ }$ & 277.02053& +00.80388&             &     & 16.43( 0.02)& 0.82& 14.31( 0.01)& 0.46& $iz $ & 2   & -0.3734(0.1834) & SR\_M       \\
V024$^{ }$ & 276.86018& +00.81059&             &     & 15.54( 0.01)& 0.23& 13.40(<0.01)& 0.13& $iz $ & 2   & -0.0390(0.1721) &             \\
V025$^{c}$ & 277.01889& +00.83005&             &     & 16.26( 0.01)& 0.76& 14.17( 0.01)& 0.36& $i  $ & 1+  & -0.1202(0.1831) &SR\_M, YSO$?$ \\
V026$^{ }$ & 276.88270& +00.85169&             &     & 16.15( 0.01)& 0.70& 14.32( 0.01)& 0.46& $iz $ & 2   & 	0.0560(0.1483)& SR\_M       \\
V027$^{ }$ & 276.98476& +00.90530& 16.55( 0.01)& 0.44& 13.55(<0.01)& 0.26& 11.31(<0.01)& 0.11& $riz$ & 3   & 	0.4692(0.1843)& LP          \\
V028$^{ }$ & 276.95526& +00.90950& 15.92( 0.01)& 0.16& 13.87(<0.01)& 0.09& 12.37(<0.01)& 0.06& $ri $ & 2   & 	0.3826(0.1226)&             \\
V029$^{ }$ & 277.81701& +00.92273&             &     & 15.71( 0.01)& 1.39& 12.36(<0.01)& 0.89& $iz $ & 2   & 	0.2322(0.2855)& SR\_M       \\
V030$^{ }$ & 276.99712& +00.93316&             &     & 15.91( 0.01)& 1.79& 13.49(<0.01)& 0.84& $iz $ & 2   & -0.0564(0.2287) &SR\_M, YSO$?$ \\
V031$^{ }$ & 276.92278& +00.93857& 19.09( 0.03)& 6.70& 16.38( 0.02)& 2.10& 14.17( 0.01)& 1.15& $iz $ & 2+  & 	0.3671(0.1985)& SR\_M       \\
V032$^{ }$ & 276.77456& +00.94866& 17.62( 0.04)& 0.56& 15.56( 0.01)& 0.19& 13.92(<0.01)& 0.13& $iz $ & 2   & 	0.1366(0.1549)&             \\
V033$^{ }$ & 277.77401& +00.95637& 17.50( 0.02)& 0.68& 14.69(<0.01)& 0.38& 12.72(<0.01)& 0.19& $riz$ & 3   & 	0.0787(0.1444)&SR\_M        \\
V034$^{ }$ & 276.75242& +00.95976&             &     & 16.72( 0.02)& 1.02& 13.67(<0.01)& 0.87& $iz $ & 2   & 	0.2030(0.2094)&SR\_M        \\
V035$^{ }$ & 276.69176& +00.96266& 16.59( 0.01)& 0.60& 14.36(<0.01)& 0.33& 12.80(<0.01)& 0.18& $riz$ & 3   & 	0.0276(0.1300)& LP          \\
V036$^{ }$ & 276.84733& +00.96843& 16.98( 0.01)& 0.27& 14.64(<0.01)& 0.16& 12.92(<0.01)& 0.08& $i  $ & 1++ & 	0.1369(0.1759)&             \\
V037$^{c}$ & 276.90264& +00.97122& 15.40(<0.01)& 1.20& 12.03(<0.01)& 1.09&  9.53(<0.01)& 0.66& $riz$ & 3   & 	0.5563(0.2183)& SR\_M       \\
V038$^{a,b}$ & 276.91487& +01.00051& 13.05(<0.01)& 0.54& 12.70(<0.01)& 0.52& 12.48(<0.01)& 0.51& $riz$ & 3 & 	3.7603(0.0404)&EW         \\
V039$^{d}$ & 277.37859& +01.01860& 17.14( 0.01)& 1.09& 15.81( 0.01)& 0.90& 14.69( 0.01)& 0.71& $i  $ & 1++ & 	2.3095(0.1173)&SP, YSO      \\
V040$^{ }$ & 277.07961& +01.02547&             &     & 15.12( 0.01)& 1.01& 12.29(<0.01)& 0.58& $iz $ & 2   & -0.5477(0.2850) &SR\_M, YSO$?$ \\
V041$^{ }$ & 276.99025& +01.04211&             &     & 15.41( 0.01)& 0.46& 12.92(<0.01)& 0.27& $iz $ & 2   & -0.1658(0.2817) & LP          \\
V042$^{ }$ & 276.99627& +01.05244&             &     & 15.23( 0.01)& 0.28& 12.04(<0.01)& 0.19& $iz $ & 2   & -0.5445(0.3401) & LP          \\
V043$^{ }$ & 277.92026& +01.05220& 18.01( 0.03)& 0.66& 14.88(<0.01)& 0.33& 12.67(<0.01)& 0.16& $ri $ & 2+  & 	0.0562(0.2066)& SR\_M       \\
V044$^{ }$ & 277.87538& +01.11073&             &     & 15.71( 0.01)& 0.57& 12.95(<0.01)& 0.33& $iz $ & 2   & 	0.2208(0.1661)& LP          \\
V045$^{ }$ & 277.30429& +01.11685&             &     & 15.14( 0.01)& 0.38& 12.31(<0.01)& 0.22& $iz $ & 2   & 	1.7702(0.3288)&SR\_M, YSO    \\
V046$^{d}$ & 277.42283& +01.12714& 16.08( 0.01)& 1.27& 15.06(<0.01)& 1.10& 14.04(<0.01)& 0.86& $riz$ & 3   & 	2.4817(0.0870)&SP, YSO       \\
V047$^{ }$ & 277.32318& +01.14269& 17.60( 0.03)& 0.36& 15.32( 0.01)& 0.20& 13.30(<0.01)& 0.11& $i  $ & 1+  & 	0.5314(0.1674)&             \\
V048$^{c}$ & 277.93120& +01.15520& 16.12( 0.01)& 2.70& 14.02(<0.01)& 1.38& 12.41(<0.01)& 0.58& $riz$ & 3   & 	0.0306(0.1047)& SR\_M       \\
V049$^{ }$ & 276.81854& +01.16073& 16.87( 0.02)& 0.50& 14.84(<0.01)& 0.22& 13.31(<0.01)& 0.11& $i  $ & 1+  & 	0.0700(0.1497)&             \\
V050$^{ }$ & 278.00169& +01.21432& 18.29( 0.06)& 3.41& 15.02(<0.01)& 2.13& 12.49(<0.01)& 1.10& $iz $ & 2+  & 	0.3114(0.1762)& SR\_M       \\
V051$^{ }$ & 277.17125& +01.25486& 17.98( 0.03)& 0.47& 15.44( 0.01)& 0.11& 13.64(<0.01)& 0.05& $i  $ & 1   & 	0.1250(0.1416)&             \\
V052$^{ }$ & 277.65563& +01.26905& 18.05( 0.03)& 0.72& 15.33( 0.01)& 0.26& 13.31(<0.01)& 0.11& $iz $ & 2+  & -0.0418(0.2115) &SR\_M, YSO    \\
V053$^{ }$ & 276.91888& +01.26832& 16.71( 0.01)& 0.31& 14.15(<0.01)& 0.13& 12.60(<0.01)& 0.07& $iz $ & 2   & -0.6840(0.1429) &             \\
V054$^{d}$ & 277.51434& +01.27182& 16.13( 0.01)& 0.76& 14.98( 0.01)& 0.67& 14.05( 0.01)& 0.63& $riz$ & 3   & 	2.2286(0.1118)&E, YSO        \\
V055$^{ }$ & 277.06308& +01.27812&             &     & 15.93( 0.01)& 1.93& 12.64(<0.01)& 1.18& $iz $ & 2   & -0.1201(0.4783) & SR\_M       \\
V056$^{ }$ & 276.95821& +01.27548& 16.62( 0.01)& 0.16& 13.54(<0.01)& 0.11& 11.24(<0.01)& 0.06& $riz$ & 3   & -0.4607(0.2513) &             \\
V057$^{c}$ & 277.26533& +01.30963&             &     & 16.04( 0.01)& 2.78& 14.35( 0.01)& 1.26& $iz $ & 2   & 	0.1917(0.1992)& SR\_M       \\
V058$^{c}$ & 277.02906& +01.31650& 16.79( 0.01)& 0.83& 13.49(<0.01)& 0.69& 11.03(<0.01)& 0.44& $iz $ & 2+  & 	0.1386(0.3063)& LP          \\
V059$^{ }$ & 277.02860& +01.32279&             &     & 15.57( 0.01)& 0.42& 13.18(<0.01)& 0.20& $iz $ & 2   & -0.1006(0.3430) & SR\_M       \\
V060$^{ }$ & 278.08854& +01.33410& 18.19( 0.03)& 0.45& 15.28( 0.01)& 0.20& 13.28(<0.01)& 0.09& $i  $ & 1+  & -0.6084(0.1510) &             \\
V061$^{ }$ & 277.57212& +01.35909&             &     & 15.26( 0.01)& 0.38& 12.90(<0.01)& 0.21& $iz $ & 2   & 	0.0104(0.2213)& LP, YSO      \\
V062$^{c}$ & 276.97960& +01.35616& 16.23( 0.01)& 0.29& 14.27(<0.01)& 0.14& 12.87(<0.01)& 0.07& $ri $ & 2+  & 	0.2188(0.2152)& LP          \\
V063$^{ }$ & 277.09781& +01.36030& 15.71(<0.01)& 0.35& 13.76(<0.01)& 0.17& 12.39(<0.01)& 0.08& $i  $ & 1++ & 	0.1523(0.0979)& LP          \\
V064$^{ }$ & 277.09966& +01.37345& 17.09( 0.02)& 0.44& 15.46( 0.01)& 0.16& 14.28( 0.01)& 0.12& $iz $ & 2   & 	0.0298(0.1040)&SR\_M        \\
V065$^{c}$ & 277.79327& +01.36864& 15.73( 0.01)& 0.33& 13.74(<0.01)& 0.18& 12.25(<0.01)& 0.10& $riz$ & 3   & 	0.1347(0.1298)&SR\_M        \\
V066$^{ }$ & 277.99639& +01.38077&             &     & 16.47( 0.02)& 0.46& 13.96(<0.01)& 0.27& $i  $ & 1+  & -0.7160(0.2051) &SR\_M        \\
V067$^{c}$ & 278.03296& +01.40484& 16.83( 0.01)& 0.49& 14.44(<0.01)& 0.31& 12.75(<0.01)& 0.17& $i  $ & 1++ & -0.0590(0.1611) & SR\_M  \\
V068$^{ }$ & 277.83003& +01.41757& 14.12(<0.01)& 0.17& 13.14(<0.01)& 0.28& 11.61(<0.01)& 0.04& $ri $ & 2   & 	1.1013(0.0513)&             \\
V069$^{ }$ & 277.14764& +01.40057& 14.31(<0.01)& 0.26& 12.42(<0.01)& 0.13& 11.13(<0.01)& 0.06& $riz$ & 3   & 	0.2842(0.1250)& LP          \\
V070$^{c}$ & 277.94236& +01.41826& 17.15( 0.02)& 0.75& 14.96(<0.01)& 0.46& 13.05(<0.01)& 0.18& $i  $ & 1++ & -0.0654(0.1382) & SR\_M    \\
V071$^{ }$ & 277.83043& +01.41894& 14.07(<0.01)& 0.34& 12.91(<0.01)& 0.30& 11.61(<0.01)& 0.04& $ri $ & 2   & 	0.8434(0.0546)&             \\
V072$^{ }$ & 277.93455& +01.43036& 14.03(<0.01)& 0.48& 11.28(<0.01)& 0.49&  9.02(<0.01)& 0.31& $ri $ & 2+  & 	0.1505(0.1829)& SR\_M       \\
V073$^{ }$ & 277.27397& +01.44745& 17.56( 0.02)& 0.50& 14.89(<0.01)& 0.21& 13.08(<0.01)& 0.12& $riz$ & 3   & 	0.0629(0.1445)&             \\
V074$^{ }$ & 277.11152& +01.44332& 15.59(<0.01)& 0.40& 12.81(<0.01)& 0.35& 10.59(<0.01)& 0.18& $riz$ & 3   & 	0.7986(0.2048)& LP          \\
V075$^{ }$ & 276.75318& +01.47317& 16.73( 0.01)& 0.33& 14.57(<0.01)& 0.22& 13.09(<0.01)& 0.15& $riz$ & 3   & 	0.0207(0.1641)&             \\
V076$^{ }$ & 277.03171& +01.45608& 13.43(<0.01)& 0.18& 12.41(<0.01)& 0.17&             &     & $ri $ & 2   & 	2.3292(0.0413)&             \\
V077$^{ }$ & 276.76185& +01.50503& 14.55(<0.01)& 0.28& 12.32(<0.01)& 0.14& 10.93(<0.01)& 0.05& $riz$ & 3   & 	0.2024(0.1578)& LP          \\
V078$^{ }$ & 276.68885& +01.51194& 14.78(<0.01)& 0.24& 12.87(<0.01)& 0.15& 11.51(<0.01)& 0.08& $riz$ & 3   & -0.1072(0.1009) & LP          \\
V079$^{c,d}$ & 277.91134& +01.51511& 15.63(<0.01)& 0.35& 14.63(<0.01)& 0.30& 13.90( 0.01)& 0.26& $ri $ & 2+& 	2.1974(0.0867)&EW         \\
V080$^{ }$ & 277.01452& +01.53291& 15.28(<0.01)& 0.60& 14.60(<0.01)& 0.54& 14.13( 0.01)& 0.44& $riz$ & 3   & 	0.6636(0.0460)&EA           \\
V081$^{ }$ & 277.91146& +01.53907& 16.51( 0.01)& 0.48& 13.70(<0.01)& 0.34& 11.49(<0.01)& 0.18& $iz $ & 2+  & -1.0436(0.3172) &LP           \\
V082$^{ }$ & 277.86416& +01.56440& 16.65( 0.01)& 0.65& 14.11(<0.01)& 0.34& 12.43(<0.01)& 0.17& $iz $ & 2+  & -0.4949(0.2064) & LP          \\
V083$^{ }$ & 276.71510& +01.57916& 15.35(<0.01)& 0.11& 13.85(<0.01)& 0.06& 12.69(<0.01)& 0.05& $ri $ & 2   & -0.0195(0.1669) &             \\
V084$^{ }$ & 277.83886& +01.58508& 17.73( 0.04)& 1.51& 14.74(<0.01)& 1.40& 11.77(<0.01)& 0.93& $iz $ & 2+  & -1.0856(0.2977) & SR\_M       \\
V085$^{ }$ & 278.04609& +01.59366& 15.51(<0.01)& 1.38& 14.68(<0.01)& 1.25& 14.16(<0.01)& 1.02& $riz$ & 3   & 	1.5927(0.0712)&             \\
V086$^{ }$ & 276.97993& +01.60791& 15.01(<0.01)& 0.35& 14.24(<0.01)& 0.31& 13.70(<0.01)& 0.25& $ri $ & 2+  & 	0.8759(0.0443)&EA           \\
V087$^{ }$ & 278.04781& +01.61522& 16.72( 0.01)& 0.45& 14.39(<0.01)& 0.37& 12.67(<0.01)& 0.27& $iz $ & 2+  & -0.1946(0.1821) & LP          \\
V088$^{ }$ & 277.94803& +01.62105&             &     & 16.25( 0.01)& 0.28& 14.70( 0.01)& 0.13& $i  $ & 1   & -0.3417(0.1694) &             \\
V089$^{ }$ & 277.49932& +01.65625& 16.31( 0.01)& 0.33& 13.65(<0.01)& 0.13& 11.72(<0.01)& 0.08& $iz $ & 2+  & 	0.3161(0.1392)&             \\
V090$^{c}$ & 277.05612& +01.65435& 16.39( 0.01)& 0.49& 13.34(<0.01)& 0.24& 11.65(<0.01)& 0.16& $riz$ & 3   & 	0.1961(0.1289)& LP          \\
V091$^{ }$ & 278.02524& +01.66100& 14.81(<0.01)& 0.39& 13.87(<0.01)& 0.48& 13.31(<0.01)& 0.48& $riz$ & 3   & 	1.3318(0.0730)& EA$?$       \\
V092$^{ }$ & 277.08847& +01.68516& 14.90(<0.01)& 0.86& 12.97(<0.01)& 0.52& 11.62(<0.01)& 0.27& $riz$ & 3   & 	0.1628(0.0891)& LP          \\
V093$^{ }$ & 277.84355& +01.68106&             &     & 13.26(<0.01)& 0.23& 11.44(<0.01)& 0.22& $iz $ & 2   & -0.3132(0.2104) & LP$?$       \\
V094$^{ }$ & 278.06180& +01.69958& 16.62( 0.01)& 0.29& 14.19(<0.01)& 0.18& 12.53(<0.01)& 0.10& $riz$ & 3   & 	0.0705(0.1445)& LP          \\
V095$^{ }$ & 276.85195& +01.70252& 17.26( 0.03)& 0.71& 15.71( 0.03)& 1.39& 15.12( 0.03)& 1.70& $iz $ & 2   & 	0.8626(0.1429)&SP           \\
V096$^{ }$ & 278.12174& +01.71904& 17.79( 0.05)& 0.09& 15.41( 0.01)& 0.72& 13.37(<0.01)& 0.34& $iz $ & 2   & 	0.1897(0.1517)&             \\
V097$^{ }$ & 277.25459& +01.72288& 14.89(<0.01)& 0.20& 13.22(<0.01)& 0.10& 12.01(<0.01)& 0.06& $ri $ & 2   & 	0.0625(0.1050)&             \\
V098$^{c}$ & 278.09719& +01.73048& 17.30( 0.02)& 0.93& 14.66(<0.01)& 0.41& 12.91(<0.01)& 0.24& $iz $ & 2+  & 	0.1857(0.1464)&LP           \\
V099$^{c}$ & 277.43571& +01.74349& 16.79( 0.01)& 0.48& 14.54(<0.01)& 0.29& 13.05(<0.01)& 0.16& $i  $ & 1++ & 	0.5571(0.1649)&SR\_M     \\
V100$^{ }$ & 276.93075& +01.75038& 14.91(<0.01)& 0.17& 13.19(<0.01)& 0.32& 11.61(<0.01)& 0.33& $riz$ & 3   & 	0.0042(0.1203)& LP          \\
V101$^{ }$ & 277.29592& +01.75262& 15.28(<0.01)& 0.10& 13.51(<0.01)& 0.04& 12.28(<0.01)& 0.05& $ri $ & 2   & 	0.3596(0.1046)&             \\
V102$^{ }$ & 277.77740& +01.73918& 14.33(<0.01)& 0.32& 11.63(<0.01)& 0.18&  9.42(<0.01)& 0.14& $riz$ & 3   & 	0.0613(0.1844)&             \\
V103$^{ }$ & 276.70071& +01.76625& 14.15(<0.01)& 0.22& 12.21(<0.01)& 0.13& 10.86(<0.01)& 0.04& $ri $ & 2+  & 	0.2584(0.1070)& SR\_M       \\
V104$^{c}$ & 277.39730& +01.77845& 16.19( 0.01)& 0.24& 13.96(<0.01)& 0.12& 12.46(<0.01)& 0.08& $iz $ & 2+  & 	0.2577(0.1449)& LP          \\
V105$^{c}$ & 276.67320& +01.77073& 14.88(<0.01)& 3.77& 12.47(<0.01)& 2.65& 10.61(<0.01)& 1.50& $riz$ & 3   & 	0.1886(0.1673)& SR\_M       \\
V106$^{ }$ & 276.80925& +01.78002& 15.48(<0.01)& 0.54& 14.79(<0.01)& 0.43& 14.29( 0.01)& 0.52& $riz$ & 3   & 	1.7526(0.0589)& EW          \\
V107$^{ }$ & 277.49591& +01.77911& 15.39( 0.01)& 0.30& 14.26( 0.01)& 0.29& 13.56( 0.01)& 0.16& $ri $ & 2   & 	0.4792(0.1008)&             \\
V108$^{c}$ & 276.99586& +01.81003& 17.83( 0.04)& 1.06& 15.48( 0.01)& 0.51& 14.00( 0.01)& 0.33& $i  $ & 1++ & -0.1632(0.1614) &LP           \\
V109$^{ }$ & 276.79297& +01.79516& 14.23(<0.01)& 0.28& 11.63(<0.01)& 0.20&  9.81(<0.01)& 0.08& $riz$ & 3   & 	0.2420(0.1381)& LP          \\
V110$^{ }$ & 276.77940& +01.81539& 17.42( 0.02)& 0.58& 14.90(<0.01)& 0.25& 13.03(<0.01)& 0.15& $riz$ & 3   & 	0.4767(0.1762)& LP?         \\
V111$^{ }$ & 277.51468& +01.82385& 13.80(<0.01)& 0.10& 12.50(<0.01)& 0.07& 11.54(<0.01)& 0.05& $ri $ & 2   & 	0.2061(0.1128)&             \\
V112$^{d}$ & 277.15728& +01.82983& 12.91(<0.01)& 0.08& 12.27(<0.01)& 0.12& 11.88(<0.01)& 0.09& $i  $ & 1++ & 	2.2888(0.0362)&SP?          \\
V113$^{ }$ & 277.29339& +01.84029& 16.48( 0.01)& 0.90& 15.66( 0.01)& 0.85& 15.12( 0.02)& 1.05& $ri $ & 2   & 	0.3500(0.0823)& EA          \\
V114$^{ }$ & 277.22407& +01.81884& 14.11(<0.01)& 0.18& 11.84(<0.01)& 0.10& 10.35(<0.01)& 0.06& $ri $ & 2+  & 	0.2582(0.1005)&             \\
V115$^{ }$ & 277.24607& +01.83406& 12.63(<0.01)& 0.38& 12.10(<0.01)& 0.28& 11.77(<0.01)& 0.21& $riz$ & 3   & 	1.6174(0.0505)& EA?         \\
V116$^{ }$ & 277.38059& +01.84642& 15.47(<0.01)& 0.32& 13.49(<0.01)& 0.19& 12.15(<0.01)& 0.10& $riz$ & 3   & 	0.3657(0.1235)&LP           \\
V117$^{b,c}$ & 277.21854& +01.83793& 15.83( 0.01)& 1.11& 12.70(<0.01)& 1.02&  9.76(<0.01)& 0.81& $riz$ & 3 & 	0.2006(0.2369)& SR\_M     \\
V118$^{ }$ & 276.79390& +01.88709& 17.41( 0.02)& 0.48& 15.11( 0.01)& 0.31& 13.76(<0.01)& 0.22& $iz $ & 2+  & 	0.0094(0.1295)&LP           \\
V119$^{ }$ & 277.01326& +01.91057& 14.16(<0.01)& 0.10& 12.08(<0.01)& 0.08& 10.69(<0.01)& 0.04& $ri $ & 2   & 	0.3611(0.1102)&             \\
V120$^{ }$ & 276.94130& +01.92219& 14.91(<0.01)& 0.17& 12.80(<0.01)& 0.13& 11.29(<0.01)& 0.07& $riz$ & 3   & 	0.1301(0.1217)&LP           \\
V121$^{ }$ & 277.38283& +01.98485& 16.95( 0.02)& 0.28& 14.69(<0.01)& 0.28& 12.58(<0.01)& 0.30& $iz $ & 2   & -0.1492(0.2846) &             \\
V122$^{ }$ & 278.10518& +01.98875& 15.61( 0.01)& 0.41& 13.25(<0.01)& 0.25& 11.55(<0.01)& 0.26& $riz$ & 3   & 	0.3606(0.1537)&             \\
V123$^{c}$ & 276.77813& +01.99689& 15.65( 0.01)& 1.88& 12.77(<0.01)& 1.58& 10.19(<0.01)& 1.11& $riz$ & 3   & 	0.2219(0.2154)& SR\_M       \\
V124$^{ }$ & 278.08224& +00.58301& 15.65(<0.01)& 0.28& 13.34(<0.01)& 0.13& 11.65(<0.01)& 0.09& $r  $ & 1+  & 	0.5945(0.1478)&             \\
V125$^{ }$ & 276.67461& +01.08045& 13.01(<0.01)& 0.12& 11.04(<0.01)& 0.06&  9.62(<0.01)& 0.04& $r  $ & 1+  & 	0.6426(0.1114)&             \\
V126$^{d}$ & 277.83039& +01.11546& 14.49(<0.01)& 0.24& 13.67(<0.01)& 0.15& 13.09(<0.01)& 0.18& $r  $ & 1++ & 	2.2120(0.0347)& SP          \\
V127$^{ }$ & 276.73488& +01.13430& 16.40( 0.01)& 0.21& 14.76( 0.01)& 0.10&             &     & $r  $ & 1   & 	0.0280(0.1124)&             \\
V128$^{ }$ & 276.85291& +01.59634& 14.95(<0.01)& 0.18& 13.33(<0.01)& 0.07& 12.15(<0.01)& 0.05& $r  $ & 1++ & 	0.1292(0.0782)&             \\
V129$^{ }$ & 276.67759& +01.73506& 16.32( 0.01)& 0.46& 15.56( 0.01)& 0.42& 15.00( 0.01)& 0.43& $r  $ & 1   & 	0.7313(0.0892)&EW           \\
V130$^{ }$ & 277.29091& +00.57929&             &     &             &     & 15.29( 0.02)& 1.10& $z  $ & 1   & 	0.3998(0.5975)&SR\_M, YSO    \\
V131$^{ }$ & 277.70348& +00.63292&             &     & 17.15( 0.04)& 2.35& 13.92( 0.01)& 1.33& $z  $ & 1+  & -0.6644(0.5244) & SR\_M, YSO$?$\\
V132$^{ }$ & 277.84742& +00.77975&             &     & 15.99( 0.01)& 0.18& 13.77(<0.01)& 0.11& $z  $ & 1   & -0.1951(0.1932) &             \\
V133$^{ }$ & 277.65091& +00.79869&             &     &             &     & 14.52( 0.01)& 0.28& $z  $ & 1   & 	0.0878(0.3746)&             \\
V134$^{ }$ & 277.31297& +00.86997& 17.35( 0.04)& 0.83&             &     & 12.39(<0.01)& 0.17& $z  $ & 1   & -0.0063(0.1719) &YSO          \\
V135$^{ }$ & 277.14578& +00.96323&             &     & 16.13( 0.02)& 0.56& 13.52(<0.01)& 0.29& $z  $ & 1   & -0.4595(0.2251) &             \\
V136$^{ }$ & 276.85562& +01.06284& 17.43( 0.04)& 0.86&             &     & 14.35( 0.01)& 0.36& $z  $ & 1+  & -0.2429(0.1698) & LP          \\
V137$^{ }$ & 277.14016& +01.07997&             &     & 17.12( 0.04)& 2.03& 14.62( 0.01)& 1.20& $z  $ & 1+  & -0.1782(0.4169) & SR\_M       \\
V138$^{ }$ & 277.43467& +01.08149&             &     &             &     & 14.88( 0.01)& 1.77& $z  $ & 1   & 	1.6704(0.3561)& SR\_M, YSO   \\
V139$^{ }$ & 277.78025& +01.14533&             &     &             &     & 14.42( 0.01)& 0.91& $z  $ & 1   & -0.4059(0.2649) & LP          \\
V140$^{ }$ & 277.72387& +01.18904&             &     &             &     & 14.76( 0.01)& 0.38& $z  $ & 1   & 	0.0811(0.2307)& LP, YSO      \\
V141$^{c}$ & 277.65789& +01.66850&             &     & 15.45( 0.01)& 0.45& 12.76(<0.01)& 0.28& $z  $ & 1   & 	0.2749(0.1855)& SR\_M       \\
V142$^{c}$ & 277.15479& +01.82846&             &     &             &     & 13.04(<0.01)& 1.73& $z  $ & 1   & -0.1102(0.1832) & SR\_M, YSO$?$\\
V143$^{ }$ & 277.69838& +01.85889& 17.03( 0.01)& 0.50& 15.97( 0.01)& 0.32& 15.15( 0.02)& 0.37& $z  $ & 1++ & 	0.1414(0.0880)& SP          \\
\enddata
\tablecomments{The bands in which a star satisfied the criteria of a variable star are indicated in the column `Bands'. The parallaxes and parallax errors of the variable objects are from Gaia\,DR2. $^a$, $^b$, and $^c$ denote known variable objects from GCVS, VSX, and Gaia\,DR2, respectively. $^d$ The variable object is a member of Serpens Main.}
\end{deluxetable*}
\end{longrotatetable}

\subsection{Period Determination}\label{subsec:period}
The methods of multiband periodogram (MP; \citealt{2015ApJ...812...18V}), multiband analysis of variance (AoV; \citealt{2015ApJ...811L..34M}), and quadratic mutual information (QMI; \citealt{2018ApJS..236...12H}) were used to derive the periods of the suspected variables in this study, all of which are suited for searching for periodic signals in nonuniformly sampled time-domain data.

\begin{figure}[!h]
 \centering
        \includegraphics[width=.9\textwidth]{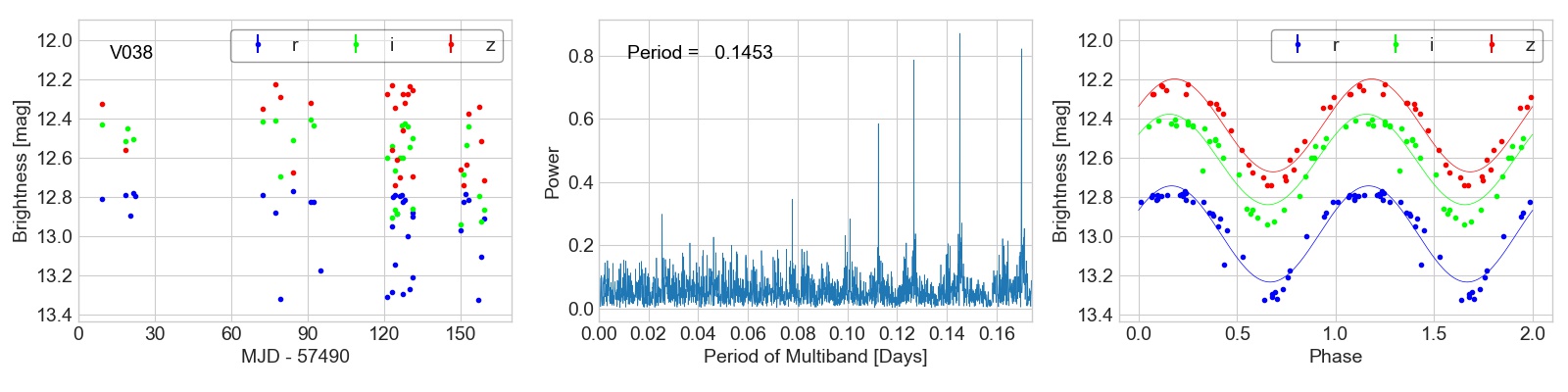}
        \includegraphics[width=.9\textwidth]{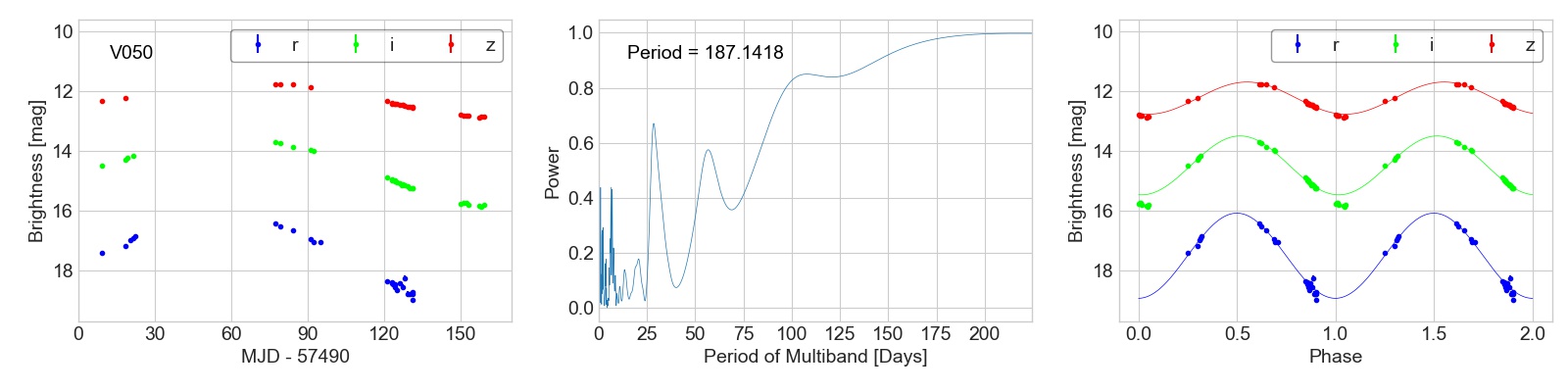}
        \includegraphics[width=.9\textwidth]{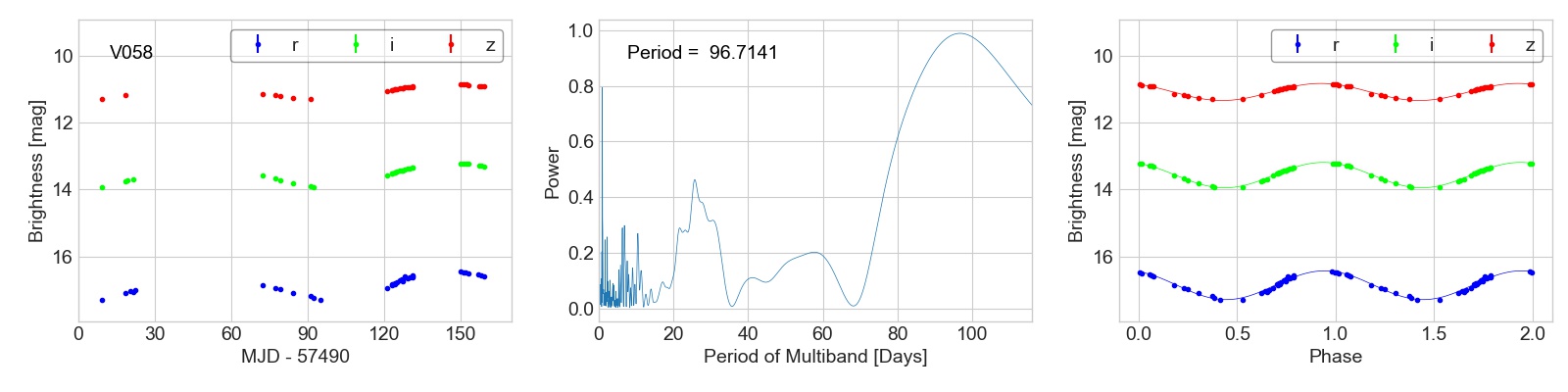}
        \includegraphics[width=.9\textwidth]{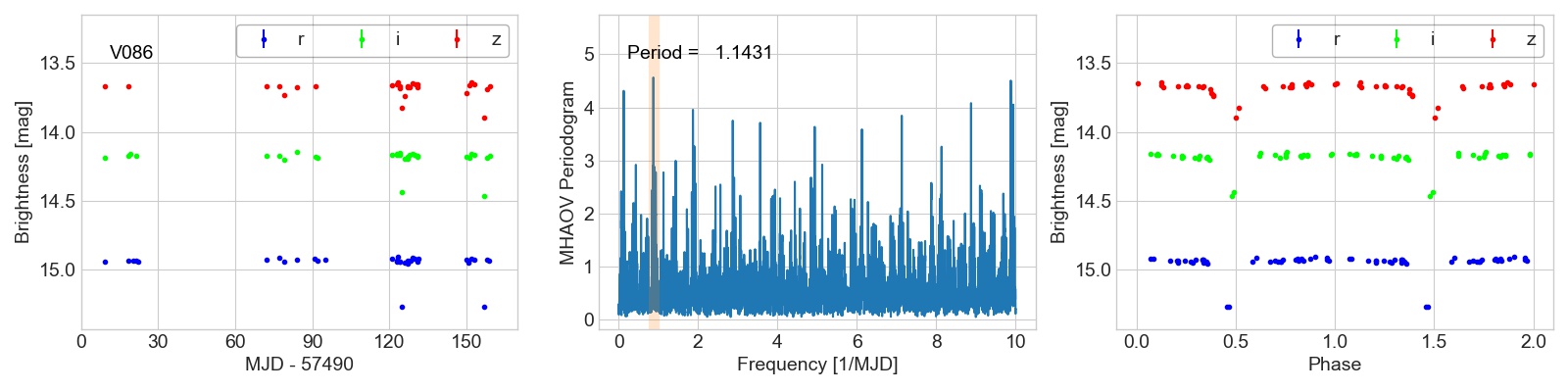}
      \caption{Light curves, periodograms, and folded light curves of four variables. The light curve of V086 was obtained with the AoV method, and those of V038, V050, and V058 were obtained with the MP method. The typical photometric error is less than 0.01\,mag.}\label{fig:select}
\end{figure}

\begin{longrotatetable}
\begin{deluxetable*}{crrrcrrr}
\tablecaption{Periods of the variable sources.\label{tab:period}}
\tablewidth{0pt}
\tabletypesize{\scriptsize}
\tablehead{
\colhead{Obs\_ID} & \multicolumn{3}{c}{Period}&\colhead{}  & \multicolumn{3}{c}{FWHM}  \\
 \cline{2-4}
 \cline{6-8}
\colhead{} & \colhead{MP(d)} & \colhead{AoV(d)} & \colhead{QMI(d)} &\colhead{} & \colhead{MP(d$^{-1}$)} & \colhead{AoV(d$^{-1}$)} & \colhead{QMI(d$^{-1}$)}
}
\startdata
V001 & {   0.9654}& {\bf 62.50}& {   0.1101}& & 0.0063& 0.0042& 0.0183\\
V002 & {     23.53}& {    138.1}& {    20.41}& & 0.0062& 0.0010& 0.0224\\
V003 & {   0.9981}& {   0.9901}& {   0.5022}& & 0.0118& 0.0055& 0.0247\\
V004 & {   0.1014}& {   0.1841}& {   0.1018}& & 0.0206& 0.0028& 0.0233\\
V005 & {    187.1}& {\bf 250.6}& {    1.010}& & 0.0021& 0.0024& 0.0299\\
V006 & {   0.5025}& {    147.3}& {   0.5028}& & 0.0111& 0.0163& 0.0241\\
V007 & {   0.4996}& {\bf 95.79}& {    20.41}& & 0.0052& 0.0021& 0.0285\\
V008 & {   0.9936}& {\bf 245.1}& {   0.5040}& & 0.0079& 0.0040& 0.0238\\
V009 & {\bf 65.08}& {    130.5}& {    66.40}& & 0.0035& 0.0015& 0.0214\\
V010 & {\bf 41.66}& {    41.67}& {    1.020}& & 0.0050& 0.0055& 0.0210\\
V011 & {\bf 164.9}& {   0.9927}& {    75.19}& & 0.0030& 0.0035& 0.0145\\
V012 & {   0.3070}& {   0.2502}& {   0.1001}& & 0.0049& 0.0036& 0.0257\\
V013 & {\bf 9.012}& {    9.034}& {    9.026}& & 0.0062& 0.0012& 0.0062\\
V014 & {   0.2497}& {   0.1666}& {   0.2436}& & 0.0047& 0.0040& 0.0665\\
V015 & {   0.9909}& {   0.9921}& {   0.9841}& & 0.0065& 0.0051& 0.0199\\
V016 & {    164.9}& {\bf 186.6}& {    1.011}& & 0.0058& 0.0032& 0.0205\\
V017 & {   1.0019}& {\bf 192.3}& {    200.0}& & 0.0061& 0.0030& 0.0121\\
V018 & {\bf 62.54}& {    63.86}& {   0.5020}& & 0.0041& 0.0021& 0.0142\\
V019 & {   0.4960}& {\bf 114.2}& {   0.9859}& & 0.0067& 0.0024& 0.0187\\
V020 & {\bf 77.57}& {    83.33}& {    77.34}& & 0.0044& 0.0020& 0.0126\\
V021 & {\bf0.9904}& {    61.84}& {   0.9836}& & 0.0264& 0.0027& 0.0144\\
V022 & {    187.1}& {    980.4}& {    1.010}& & 0.0016& 0.0035& 0.0212\\
V023 & {\bf 154.3}& {    158.0}& {   0.5041}& & 0.0031& 0.0017& 0.0269\\
V024 & {   0.5028}& {   0.3342}& {    1.014}& & 0.0118& 0.0033& 0.0181\\
V025 & {   0.9936}& {    190.5}& {   0.9920}& & 0.0059& 0.0016& 0.0141\\
V026 & {    128.8}& {\bf 200.0}& {   0.9907}& & 0.0041& 0.0036& 0.0292\\
V027 & {\bf 96.46}& {    148.6}& {    86.28}& & 0.0074& 0.0028& 0.0269\\
V028 & {    1.011}& {    1.008}& {    1.013}& & 0.0071& 0.0052& 0.0236\\
V029 & {    187.1}& {\bf 222.2}& {    85.47}& & 0.0064& 0.0054& 0.0210\\
V030 & {\bf 187.1}& {   0.9921}& {   0.9852}& & 0.0005& 0.0113& 0.0271\\
V031 & {\bf 174.5}& {    196.9}& {   0.1689}& & 0.0034& 0.0013& 0.0105\\
V032 & {    1.019}& {    125.0}& {    1.019}& & 0.0102& 0.0029& 0.0141\\
V033 & {\bf 169.0}& {    198.0}& {    71.43}& & 0.0030& 0.0019& 0.0219\\
V034 & {   0.9934}& {\bf 250.0}& {    83.47}& & 0.0008& 0.0051& 0.0261\\
V035 & {    44.50}& {\bf 89.77}& {    73.31}& & 0.0068& 0.0012& 0.0218\\
V036 & {    44.21}& {   0.9756}& {    1.020}& & 0.0057& 0.0025& 0.0205\\
V037 & {    121.5}& {\bf 173.3}& {    82.37}& & 0.0040& 0.0012& 0.0260\\
V038 & {\bf0.1453}& {   0.1453}& {   0.1453}& & 0.0081& 0.0027& 0.0115\\
V039 & {   0.9038}& {   0.1665}& {\bf 13.38}& & 0.0074& 0.0024& 0.0154\\
V040 & {    119.9}& {\bf 200.0}& {   0.9868}& & 0.0047& 0.0025& 0.0269\\
V041 & {\bf 92.52}& {    88.89}& {   0.1004}& & 0.0049& 0.0018& 0.0365\\
V042 & {\bf 148.8}& {   0.9911}& {    74.40}& & 0.0078& 0.0019& 0.0214\\
V043 & {\bf 152.8}& {    155.3}& {    1.010}& & 0.0067& 0.0014& 0.0135\\
V044 & {\bf 112.0}& {    286.5}& {    1.009}& & 0.0045& 0.0028& 0.0239\\
V045 & {\bf 170.5}& {    159.2}& {   0.5018}& & 0.0062& 0.0008& 0.0136\\
V046 & {\bf 1.004}& {    1.004}& {    66.67}& & 0.0197& 0.0009& 0.0140\\
V047 & {   0.9746}& {    86.73}& {    1.009}& & 0.0110& 0.0049& 0.0224\\
V048 & {    171.3}& {    153.1}& {    76.92}& & 0.0060& 0.0043& 0.0228\\
V049 & {    1.020}& {   0.5046}& {   0.3350}& & 0.0074& 0.0031& 0.0244\\
V050 & {\bf 187.1}& {    222.2}& {    1.010}& & 0.0078& 0.0091& 0.0174\\
V051 & {   0.5013}& {   0.3316}& {   0.3316}& & 0.0242& 0.0085& 0.0207\\
V052 & {    1.018}& {\bf 153.8}& {    1.018}& & 0.0108& 0.0054& 0.0152\\
V053 & {   0.9816}& {    75.36}& {   0.9838}& & 0.0201& 0.0030& 0.0297\\
V054 & {   0.2571}& {\bf 1.061}& {    13.16}& & 0.0179& 0.0017& 0.0132\\
V055 & {    187.1}& {\bf 191.6}& {    92.85}& & 0.0070& 0.0011& 0.0281\\
V056 & {   0.9900}& {    500.0}& {    69.83}& & 0.0105& 0.0079& 0.0280\\
V057 & {    1.002}& {\bf 302.1}& {    1.009}& & 0.0012& 0.0084& 0.0292\\
V058 & {\bf 96.71}& {    198.8}& {    1.009}& & 0.0042& 0.0012& 0.0331\\
V059 & {    1.012}& {\bf 159.7}& {    75.82}& & 0.0047& 0.0036& 0.0271\\
V060 & {   0.5048}& {    119.2}& {   0.5040}& & 0.0167& 0.0033& 0.0165\\
V061 & {   0.3315}& {\bf 130.9}& {    79.94}& & 0.0163& 0.0006& 0.0140\\
V062 & {\bf 41.82}& {    80.58}& {    1.022}& & 0.0070& 0.0020& 0.0133\\
V063 & {    1.003}& {\bf 92.08}& {   0.5018}& & 0.0123& 0.0033& 0.0270\\
V064 & {\bf 150.3}& {    215.5}& {    74.46}& & 0.0093& 0.0066& 0.0080\\
V065 & {   0.9926}& {    242.1}& {   0.5018}& & 0.0071& 0.0044& 0.0062\\
V066 & {\bf 187.1}& {    1.004}& {   0.5040}& & 0.0048& 0.0013& 0.0158\\
V067 & {   0.9939}& {   0.9930}& {    1.009}& & 0.0057& 0.0036& 0.0237\\
V068 & {   0.0822}& {   0.1411}& {   0.2491}& & 0.0703& 0.0027& 0.0241\\
V069 & {\bf 58.07}& {   0.9796}& {   0.1001}& & 0.0123& 0.0043& 0.0234\\
V070 & {\bf 187.1}& {   0.9921}& {    1.010}& & 0.0054& 0.0026& 0.0136\\
V071 & {   0.1112}& {   0.1247}& {   0.2491}& & 0.0912& 0.0042& 0.0218\\
V072 & {\bf 158.3}& {    166.7}& {    73.53}& & 0.0070& 0.0012& 0.0179\\
V073 & {    1.010}& {    156.3}& {    71.43}& & 0.0062& 0.0053& 0.0204\\
V074 & {\bf 112.7}& {    179.5}& {   0.1004}& & 0.0041& 0.0018& 0.0619\\
V075 & {    1.003}& {    197.6}& {    1.010}& & 0.0061& 0.0027& 0.0141\\
V076 & {   0.2493}& {   0.1995}& {   0.1662}& & 0.1373& 0.0038& 0.0137\\
V077 & {\bf 76.31}& {    148.6}& {    80.58}& & 0.0101& 0.0015& 0.0240\\
V078 & {\bf 140.2}& {    139.5}& {    69.78}& & 0.0031& 0.0069& 0.0269\\
V079 & {\bf 3.865}& {    3.861}& {    3.861}& & 0.0055& 0.0025& 0.0128\\
V080 & {\bf 1.444}& {   0.1603}& {    2.801}& & 0.0100& 0.0034& 0.0409\\
V081 & {   0.9744}& {\bf 42.12}& {   0.5042}& & 0.0063& 0.0024& 0.0222\\
V082 & {\bf 82.81}& {    83.26}& {    81.90}& & 0.0036& 0.0021& 0.0227\\
V083 & {    1.010}& {    72.62}& {    79.49}& & 0.0068& 0.0054& 0.0133\\
V084 & {\bf 178.2}& {    166.7}& {    74.46}& & 0.0057& 0.0022& 0.0205\\
V085 & {    1.010}& {   0.5017}& {    85.11}& & 0.0069& 0.0049& 0.0229\\
V086 & {   0.5347}& {\bf 1.143}& {    1.142}& & 0.0184& 0.0047& 0.0310\\
V087 & {\bf 99.54}& {    191.9}& {    1.010}& & 0.0047& 0.0028& 0.0203\\
V088 & {   0.9911}& {   0.5022}& {    1.010}& & 0.0163& 0.0052& 0.0128\\
V089 & {   0.9762}& {   0.9835}& {   0.9829}& & 0.0215& 0.0052& 0.0232\\
V090 & {\bf 77.89}& {    156.5}& {    78.43}& & 0.0089& 0.0020& 0.0137\\
V091 & {\bf0.6285}& {   0.2406}& {   0.1379}& & 0.0166& 0.0096& 0.0888\\
V092 & {   0.4943}& {\bf 114.3}& {    90.91}& & 0.0108& 0.0010& 0.0238\\
V093 & {   0.9834}& {\bf 62.66}& {    71.94}& & 0.0139& 0.0039& 0.0195\\
V094 & {\bf 62.17}& {    76.92}& {    72.05}& & 0.0092& 0.0063& 0.0214\\
V095 & {\bf0.1995}& {   0.1662}& {   0.1662}& & 0.0218& 0.0029& 0.0096\\
V096 & {    1.005}& {    163.9}& {    72.05}& & 0.0198& 0.0329& 0.0243\\
V097 & {   0.9780}& {   0.9782}& {   0.1996}& & 0.0165& 0.0026& 0.0702\\
V098 & {    106.3}& {\bf 87.41}& {   0.9863}& & 0.0108& 0.0037& 0.0225\\
V099 & {    41.94}& {    91.07}& {   0.9461}& & 0.0046& 0.0018& 0.0233\\
V100 & {\bf 140.4}& {    144.5}& {    71.43}& & 0.0084& 0.0035& 0.0132\\
V101 & {   0.5027}& {    1.014}& {   0.4955}& & 0.0084& 0.0055& 0.0263\\
V102 & {   0.9921}& {    159.2}& {   0.9833}& & 0.0067& 0.0036& 0.0207\\
V103 & {\bf 171.3}& {    126.6}& {    1.019}& & 0.0035& 0.0010& 0.0272\\
V104 & {    1.012}& {\bf 142.5}& {    67.84}& & 0.0133& 0.0028& 0.0148\\
V105 & {    187.1}& {    302.1}& {    72.99}& & 0.0017& 0.0037& 0.0221\\
V106 & {   0.1477}& {\bf0.2955}& {   0.1477}& & 0.0071& 0.0013& 0.0080\\
V107 & {    1.007}& {    1.004}& {   0.3334}& & 0.0110& 0.0042& 0.0209\\
V108 & {\bf 142.6}& {    143.9}& {   0.1118}& & 0.0059& 0.0014& 0.0320\\
V109 & {    1.023}& {\bf 83.06}& {   0.5047}& & 0.0060& 0.0017& 0.0216\\
V110 & {    1.013}& {    1.013}& {\bf 66.67}& & 0.0123& 0.0017& 0.0145\\
V111 & {   0.9921}& {   0.9918}& {    1.010}& & 0.0065& 0.0046& 0.0137\\
V112 & {\bf 2.456}& {    53.76}& {    1.677}& & 0.0057& 0.0194& 0.0089\\
V113 & {   0.4352}& {\bf0.8701}& {   0.3029}& & 0.0060& 0.0038& 0.0206\\
V114 & {    1.038}& {    2.076}& {    1.010}& & 0.0061& 0.0014& 0.0074\\
V115 & {\bf0.8848}& {   0.1095}& {   0.1424}& & 0.0121& 0.0033& 0.1069\\
V116 & {\bf 137.6}& {    64.35}& {    71.43}& & 0.0085& 0.0045& 0.0136\\
V117 & {    1.001}& {    500.0}& {   0.5018}& & 0.0051& 0.0042& 0.0222\\
V118 & {\bf 115.0}& {    111.1}& {   0.1252}& & 0.0064& 0.0042& 0.0293\\
V119 & {   0.5023}& {    166.7}& {    1.012}& & 0.0080& 0.0044& 0.0167\\
V120 & {\bf 123.5}& {    127.9}& {   0.5043}& & 0.0044& 0.0013& 0.0295\\
V121 & {   0.9918}& {   0.9916}& {    1.009}& & 0.0107& 0.0154& 0.0214\\
V122 & {   0.4943}& {    1.006}& {    1.014}& & 0.0137& 0.0019& 0.0190\\
V123 & {    187.1}& {    219.8}& {    76.92}& & 0.0050& 0.0028& 0.0238\\
V124 & {   0.9861}& {   0.9911}& {    1.011}& & 0.0132& 0.0046& 0.0082\\
V125 & {   0.9537}& {   0.9829}& {   0.9294}& & 0.0065& 0.0023& 0.0184\\
V126 & {\bf 1.028}& {   0.1059}& {   0.2571}& & 0.0061& 0.0053& 0.0119\\
V127 & {    46.55}& {    90.91}& {    1.021}& & 0.0066& 0.0030& 0.0078\\
V128 & {   0.0836}& {    72.83}& {    14.57}& & 0.0107& 0.0016& 0.0095\\
V129 & {\bf0.2133}& {   0.2133}& {   0.2134}& & 0.0388& 0.0014& 0.0076\\
V130 & {   0.9932}& {\bf 283.3}& {    83.19}& & 0.0086& 0.0032& 0.0269\\
V131 & {   0.9930}& {\bf 200.0}& {    1.010}& & 0.0072& 0.0020& 0.0206\\
V132 & {   0.5027}& {   0.2504}& {   0.1248}& & 0.0170& 0.0041& 0.0113\\
V133 & {    145.6}& {    1.005}& {    73.21}& & 0.0082& 0.0008& 0.0172\\
V134 & {   0.9921}& {    1.004}& {   0.4960}& & 0.0068& 0.0044& 0.0170\\
V135 & {   0.4964}& {   0.4961}& {   0.3315}& & 0.0094& 0.0046& 0.0294\\
V136 & {\bf 109.3}& {    333.3}& {   0.1113}& & 0.0063& 0.0052& 0.0239\\
V137 & {\bf 187.1}& {    934.6}& {    71.33}& & 0.0001& 0.0070& 0.0220\\
V138 & {   0.9940}& {\bf 354.6}& {   0.9919}& & 0.0063& 0.0031& 0.0150\\
V139 & {\bf 115.3}& {    216.9}& {   0.9867}& & 0.0059& 0.0028& 0.0226\\
V140 & {\bf 59.25}& {    117.8}& {    61.73}& & 0.0123& 0.0052& 0.0212\\
V141 & {   0.9906}& {\bf 333.3}& {   0.9867}& & 0.0081& 0.0067& 0.0211\\
V142 & {    140.1}& {    1.005}& {    71.43}& & 0.0029& 0.0004& 0.0211\\
V143 & {\bf 17.40}& {    17.49}& {    1.058}& & 0.0109& 0.0039& 0.0109\\
\enddata
\tablecomments{The period of V038 is 0.2905\,d in GCVS and VSX. The period of V117 in VSX and Gaia\,DR2 is 437\,d and 528.5\,d, respectively. The period of V079 in the Gaia\,DR2 and this work is 89.1\,d and 7.73\,d, respectively. The other objects that have Gaia\,DR2 periods (in units of d) are V022(230.1), V025(355.2), V048(164.1), V065(399.9), V067(436.8), V099(329.5), V105(248.2), V123(297.7), and V142(213.9). The full-width half max (FWHM) of the periodogram peaks is obtained from the three methods (MP, AoV and QMI).}
\end{deluxetable*}
\end{longrotatetable}

Based on the Lomb--Scargle (LS) approach \citep{1982ApJ...263..835S}, \citet{2015ApJ...812...18V} developed the MP method to detect periodic signals in the time-domain data. The main improvement of the MP method was that the multiband light curves were modeled as an arbitrary truncated Fourier series, where Tikhonov regularization was used to decrease the model complexity. Since the MP method can efficiently determine a period using randomly sampled multiband light curves, it was employed as the main method to derive the periods of the variable candidates. As the MP method may not be sensitive to nonsinusoidal or sharp signals, the AoV and QMI methods were also considered to find any missing signals. The multiband AoV method was used here based on the Lafler--Kinman (LK) method \citep{1965ApJS...11..216L}, which employs periodic orthogonal polynomials and variance statistics to search for periodic signals, and it has excellent performance in detecting nonsinusoidal signals. The QMI method estimates a period using QMI in different bands. Compared with the MP and the AoV methods, it is robust to non-Gaussian noise and outliers, and it does not require a particular model to describe the light curve and its underlying probability density.

The three aforementioned methods were used to obtain the periods of all 143 variable candidates, where the sampling time of the MP method was about 0.1--150\,d and the sampling frequencies of the AoV and QMI methods were both about 0.0001--10\,d$^{-1}$. These periods were derived from the maximum powers of MP, AoV, and QMI under a frequency spacing of $\Delta f = 1.3\times10^{-6}$, $1.0\times10^{-5}$, and $1.0\times10^{-5}$\,day$^{-1}$, respectively, which are listed in the `Period' column in Table~\ref{tab:period}, and the FWHMs of the periodogram peaks for the corresponding periods are listed in the `FWHM' column.

By examining the light curves of all 143 variables, we obtained 99 variables with good periodic folded light curves, as shown in Figures~\ref{fig:sp_lc}--\ref{fig:sr_lc} in the Appendix \ref{app:figs}. However, all of the periods obtained from the three methods are not quite consistent, although the light curves of the 57, 30, and 2 variables obtained from the MP, AoV, and QMI methods, respectively, showed clear periodicities. The corresponding period results were used to classify the variables, which are shown in bold in Table~\ref{tab:period}, while the periods of the 10 Gaia reference variables are also given for reference. As shown in Figure~\ref{fig:select}, the results from the MP and AoV methods of four variables are presented.

\section{Variable Stars toward Serpens Main}
\label{sec:vs}

\subsection{Known Variables}
\label{subsec:catalog}

The General Catalog of Variable Stars (GCVS; \citealt{2017ARep...61...80S}), International Variable Star Index (VSX; \citealt{2006SASS...25...47W}), and Gaia DR2 \citep{2018A&A...618A..58M} were queried to determine which of our 143 identified variables were already known.

\begin{figure}[!h]
 \centering
          \includegraphics[width=.45\textwidth]{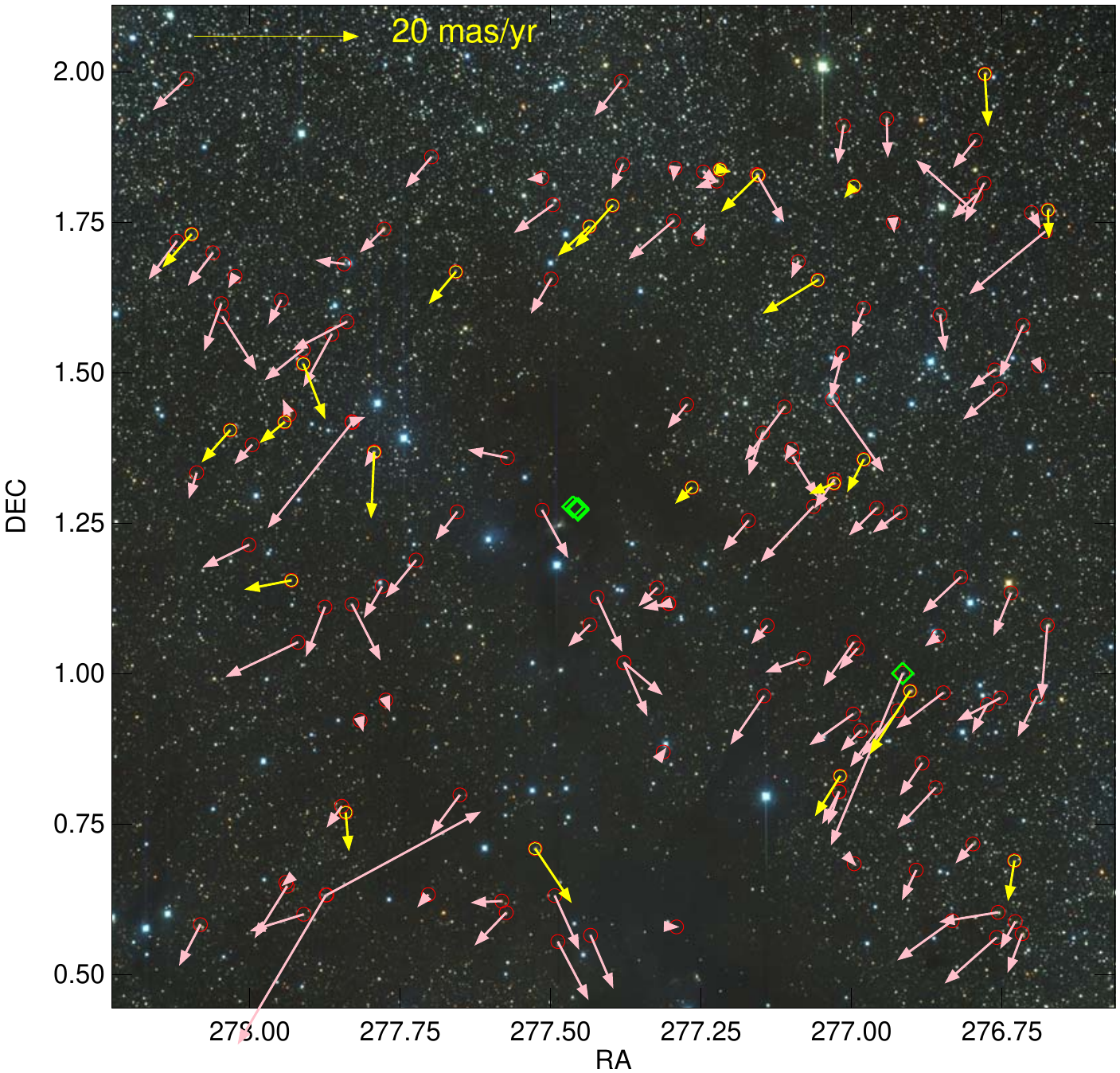}
      \caption{Color composite image of our photometric field (blue: $r$, green: $i$, and red: $z$). The known variables of GCVS are marked by green diamonds. The matched sources of Gaia\,DR2 are marked by red circles, and their proper motions are shown with arrows. The yellow arrows indicate LP variable stars to distinguish them from the pink ones (non-variable stars).}
      \label{fig:gcvs}
\end{figure}

In the observation field, there were four variable stars from GCVS, as shown by green diamonds in Figure~\ref{fig:gcvs}, of which three variables, OO\,Ser, V0370\,Ser, and V0371\,Ser were not detected in $riz$. The variable V0623\,Ser from GCVS (labeled V038 here) is an eclipsing variable (W\,Ursae\,Majoris type; EW), with a period of 0.2905\,days, an amplitude of $13.15 - 13.9$\,mag in the $V$ band, and a proper motion of (0.001, 0.023)\,arcsec yr$^{-1}$. The period in the GCVS catalog is twice as long as our result (see V038 in Figure~\ref{fig:select}); this difference may result from an overlap of periodic signals caused by primary and secondary eclipses.

The observed area also contains 43 variables from the VSX catalog, of which two variables, V0623\,Ser and Mis\,V0424, were identified within $2\arcsec$. The period of Mis\,V0424 (labeled V117) is uncertain; VSX lists 437\,days while Gaia DR2 gives 528.5\,days. Our result is about 500\,days.

The 143 variable stars from our survey coincide within 2$\arcsec$ with objects from Gaia\,DR2. However, three objects from our observations each match with two objects in the Gaia\,DR2 database with highly accurate positions. In Figure~\ref{fig:gcvs}, the proper motions of 123 objects from Gaia\,DR2 are shown. Here 23 objects are long-period (LP) variables; the photometric periods of 11 objects are given in Gaia\,DR2. As our survey covers only 150\,days, we used the results from Gaia\,DR2 to determine the periods of 10 objects (V022, V025, V048, V065, V067, V099, V105, V117, V123, and V142). Interestingly, V079 shows a period of 89.1\,days in the Gaia\,DR2 database, whereas we find a period of 7.73\,days. Both the folded light curves from Gaia\,DR2 and those from our data show consistent periodicities. However, in our folded light curves, there are obvious sinusoidal variations, while the original light curves show the characteristic of short periods (SPs). Therefore, we suggest that V079 may be an eclipsing system with a period of 7.73\,days.

\subsection{Classification}
\label{subsec:class}

Based on the classification method used by GCVS, the detected variable stars were classified according to the shape, period, and amplitude of the light curves. The 143 variables contain 44 aperiodic variables (classified as unknow type VAR) and 99 periodic variables. As the data points are not abundant, it was difficult to determine the variability types for most of the 99 periodic variables in our sample. Therefore, we divided the 99 periodic variables basically into SP variables (periods $<$ 30\,d) and LP (periods $\geqslant$ 30\,days) variables. The light curves of most LPs are more sinusoidal than those of SPs. For the LPs, if the amplitudes $\Delta i >$ 1\,mag or the periods $>$ 150\,days, the variables were classified as semiregular (SR) variables and Miras (SR\_M). The Miras have periods $>$ 100\,days and amplitudes of $\Delta I > 0.8$\,mag \citep{2009AcA....59..239S, 2017ARep...61...80S}, whereas the light curves of SRs exhibit more irregular behavior, with shorter periods and smaller amplitudes. As our sample is not very large, the SRs and Miras could not be clearly distinguished. Therefore, both were classified as SR\_Ms. For the SPs in our sample, eight eclipsing binaries were classified by their light-curve behavior: eclipsing (E) binaries, (semi)detached eclipsing binary systems or Algol type (EA), and contact binaries, i.e. EW type. The classification analysis is presented in Appendix~\ref{app:model}.

\begin{figure}[!h]
 \centering
      \includegraphics[width=.8\textwidth]{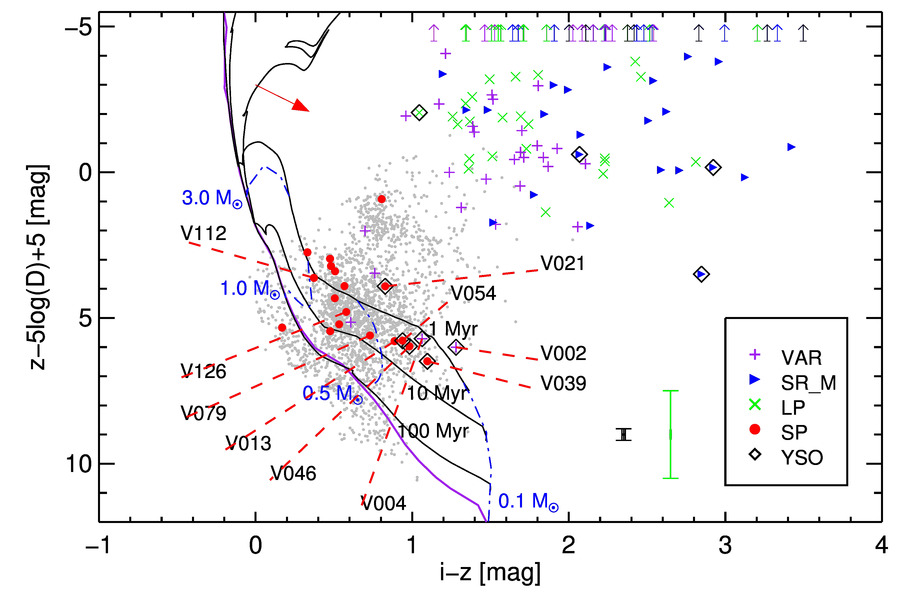}
      \caption{The CMD for all stars in the FOV. The gray points denote stars detected in all three bands in the Serpens Main field whose $\varpi/\sigma(\varpi)$ values of their parallaxes are greater than 5. The potential variable sources (132 sources detected in $i$ and $z$) are marked by colored symbols: 23 blue triangles (SR\_M), 24 green crosses (LP), 18 red circles (SP), 28 purple plus signs (VAR), and 39 colored arrows (distant variables brighter than -5 mag; 16 SR\_Ms in blue, 12 LPs in green, and 11 VARs in purple). The black diamonds denote 10 YSOs and YSO candidates, and the black arrows denote seven distant YSO candidates. The purple curve marks the MS for a distance of 10\,pc, the black curves present isochrones of 1, 10, and 100\,Myr, and the dashed-dotted blue curves represent evolutionary tracks of 3.0, 1.0, 0.5, and 0.1\,M$_{\sun}$, obtained from CMD\,3.3 \citep{2017ApJ...835...77M,2019MNRAS.485.5666P}. The red arrow denotes the extinction vector for $A_V = 2$\,mag (we used the extinction values in the SDSS bands from \citealt{2019ApJ...877..116W}). The distances, $D$, of the variable stars are obtained by the corresponding parallaxes in Gaia\,DR2. The Obs\_IDs and the red dashed lines indicate the positions of 10 cluster members of Serpens Main. The typical photometric error is less than 0.01\,mag. The error bar in black indicates the typical errors of stars around the PMS area, and the error bar in green indicates the distant stars (some VARs, SR\_Ms, LPs, and some YSOs) in the upper giant branch.}\label{fig:cmd1}
\end{figure}

A CMD is a useful tool for investigating the properties and evolutionary stage of a star. Driven by various physical mechanisms, a star's location in the CMD depends, among various physical properties, on its variability type \citep{2019A&A...623A.110G}. For the 143 variables found in this work, 11 variables were not detected in the $i$ or $z$ bands (including three LP and three SR\_M-type variables); the remaining 132 variables are shown in different colors in Figure~\ref{fig:cmd1} and the absolute magnitude of 39 distant variables (including 16 SR\_Ms, 12 LPs, and 11 VARs) is brighter than -5\,mag. From Figure~\ref{fig:cmd1}, the LPs populate the upper red part of the diagram ($1.2 < i - z < 3.2$), where the SR\_Ms are also found. Both types of variables are primarily stars on the upper giant branch, including a sequence of pulsating stars from Miras to SRs, and red giant branch (RGB) stars. As most of these objects are also quite distant, their intrinsically red $i-z$ colors are enhanced by interstellar extinction along the line of sight. The SPs are found in the range $0.2 < i - z < 1.4$ and show a spread of about 5\,mag in absolute brightness. Most SPs are located in the PMS area with ages between 1 and 10\,Myr. The 10 variables marked with Obs\_IDs are identified as cluster members of Serpens Main in Section \ref{sec:cluster}. The variables associated with YSOs are described in Section \ref{sec:color}.


\section{Cluster Members of Serpens Main}\label{sec:cluster}
During the last decade, a mean distance of $436.0 \pm 9.2$\,pc to Serpens Main wass well constrained by measuring the parallaxes of young stars in the region \citep{2011RMxAC..40..231D, 2017ApJ...834..143O}. \citet{2018ApJ...869L..33O} compared the mean parallaxes of YSOs in Serpens determined in Gaia\,DR2 with those from Very Long Baseline Array (VLBA) observations, and their results were consistent.

\begin{figure}[!h]
 \centering
          \includegraphics[width=.48\textwidth]{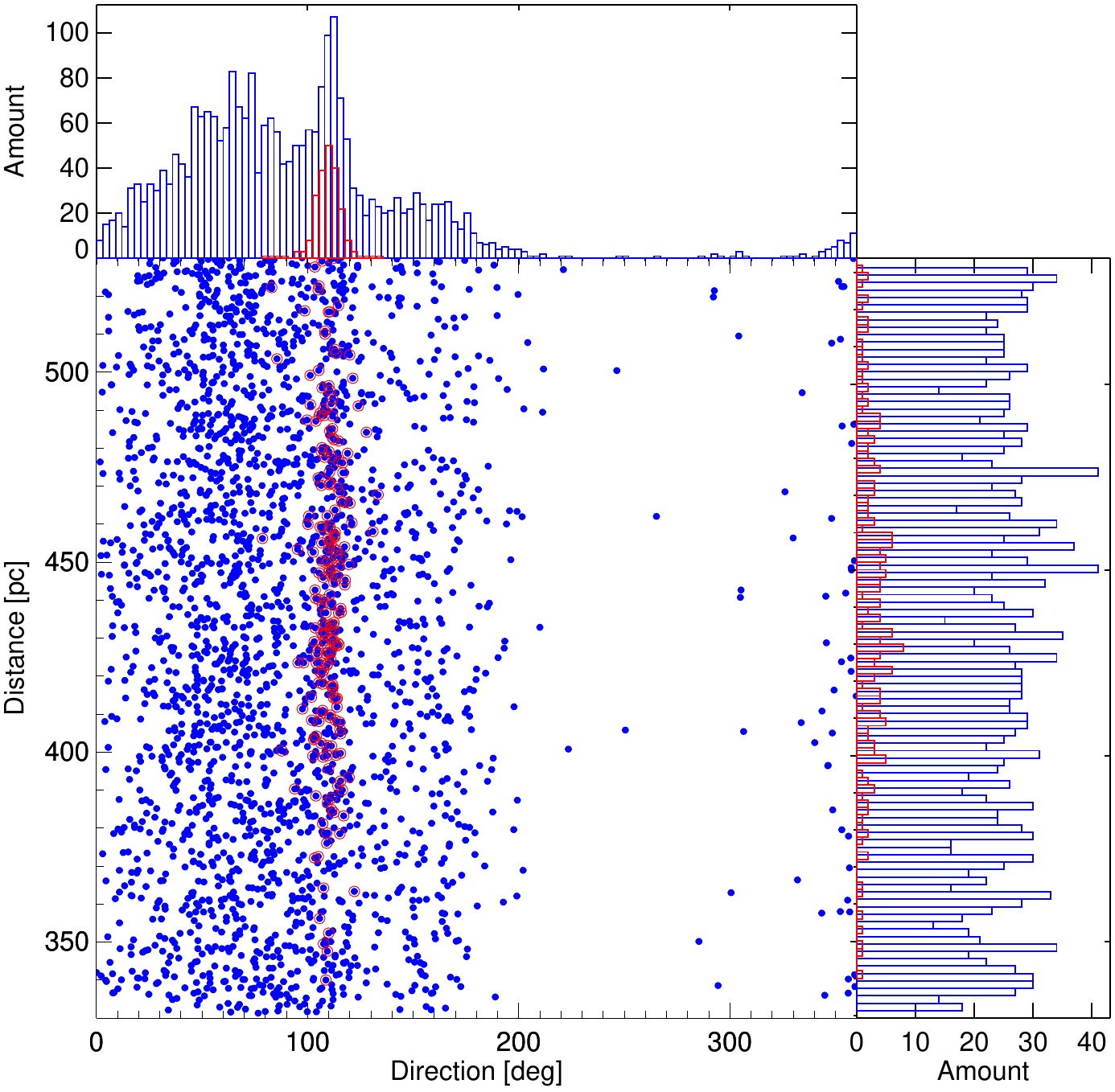}
          \includegraphics[width=.49\textwidth]{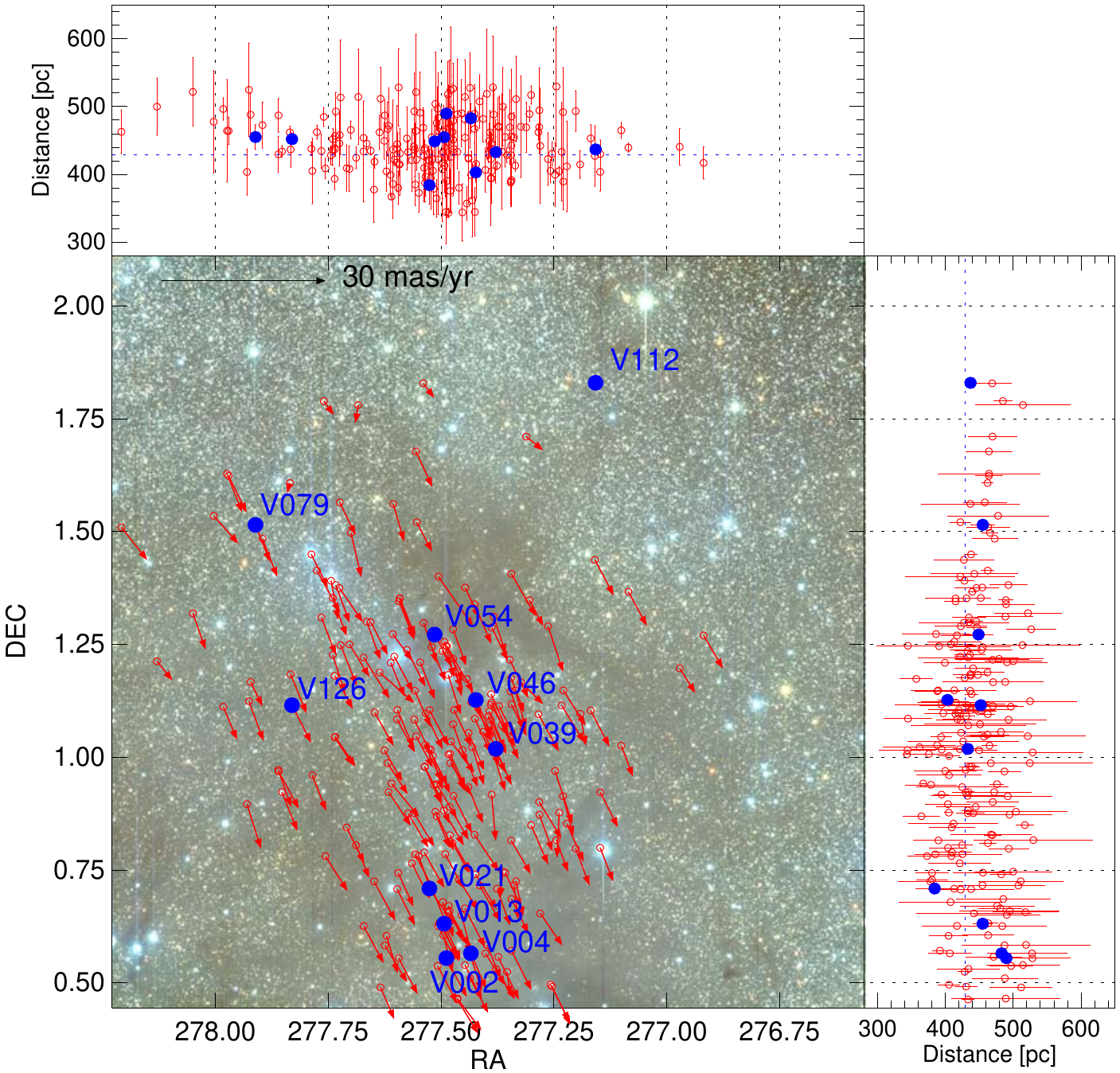}
      \caption{Identified cluster members in Serpens Main. The left panel shows the relationship between distance and proper motion direction of the stars. The top and right panels show the statistics of proper motion direction and distance, respectively. The direction of proper motion is a counter-clockwise orientation with zero to the east. The blue dots and blue statistical bars denote the stars obtained from \emph{Gaia} DR2 while the red circles and bars show the cluster members, i.e. the results of our clustering analysis. The right panel shows the spatial distribution of the cluster members with proper motion. The background of the inner panel is a color composite image (blue: $r$, green: $i$, and red: $z$). The top and right panels show the distances as functions of R.A. and Dec., respectively. The black arrow at the top left of the inner panel indicates a proper motion of 30\,mas/yr. The red circles and arrows denote the cluster members with their proper motions. The red circles and error bars show the distances of members using the errors shown in the top and right panels. The 10 variable members are shown as blue dots and labeled with their Obs\_ID in Table~\ref{tab:ysos}.}\label{fig:gaia-pm}
\end{figure}

To identify variables that are likely members of the Serpens Main cluster, the method of Hierarchical Density-Based Spatial Clustering of Applications with Noise (HDBSCAN; \citealt{2017JOSS....2..205M}) was used to analyze the clustering of stars in this region. As the locations, distances, and proper motions of cluster members are usually consistent, five parameters from Gaia\,DR2, i.e. R.A., decl., parallax and proper motion in R.A. and decl., were used to identify cluster members. When measuring the clustering, all of the parameters were fixed. All of the Gaia\,DR2 stars within the FOV of Figure~\ref{fig:comp} with $\varpi/\sigma(\varpi) > 5$ were considered, and a range window of $330-530$\,pc covering the mean distance to Serpens Main was used. In Figure~\ref{fig:gaia-pm}, 212 potential cluster members are shown with red circles. The distance distribution is nearly flat for all of the Gaia\,DR2 stars, while there is a weak concentration in the $400-490$\,pc range found for the identified cluster members. It should be noted that two significant statistical peaks appeared in the distribution of proper-motion directions. We attribute this to the identified cluster members being concentrated in a narrow spread toward a proper-motion direction of about 110$\degr$, while the distribution of field stars had a large spread range and concentrated at 70$\degr$. Furthermore, we compared these 212 potential cluster members with the recently published catalog of Serpens cluster members by \citet{2019ApJ...878..111H}, who identified the young cluster of Serpens Main in a search of the 5D phase space of location, proper motion, and distance from Gaia\,DR2, and considered stars that had higher probabilities of parallaxes, proper motions, and locations near the centroid of a cluster as members. There are 270 sources in the FOV of our data and in the $330-530$\,pc distance range in the catalog of \citet[Table\,7]{2019ApJ...878..111H}, including 84 members of Serpens Main (listed as "Main"), 39 members of distributed Serpens star formation (listed as "Distrib"), and 147 nonmembers (listed as "N/A"). We matched 189 sources of our cluster members with their catalog, including 72, 35, and 82 sources listed as "Main", "Distrib" and "N/A", respectively.

For the 143 variable stars found in this work, an object was considered a member of Serpens Main when it was close to any member within $2\arcsec$. The right side of Figure~\ref{fig:gaia-pm} shows the spatial distributions of the cluster members and the 10 variable members. It should be noted that V112 was not identified by the HDBSCAN method, as it was isolated from other members; however, see the work of \citet{2019ApJ...878..111H}, who suggested that it is a cluster member. The distances of the 10 variables, spread about 100\,pc, are larger than their projected distances, which are about $0\fdg7$, corresponding to 5.3\,pc at 430\,pc. The proper motions are about 3.0\,mas yr$^{-1}$ in R.A. and -8.0\,mas yr$^{-1}$ in decl. All the 10 variables are brighter than $G = 17$\,mag, and their measured parallaxes are within $\varpi/\sigma(\varpi) > 10$. Table~\ref{tab:ysos} lists the properties of the 10 variable cluster members.

\begin{deluxetable*}{llrclccccll}
\tablecaption{Parameters of the variable cluster members \label{tab:ysos}}
\tablewidth{0pt}
\tablehead{
\colhead{Obs\_ID} & \colhead{2MASS ID} & \colhead{Period$^a$} & \colhead{$A_r^a$} & \colhead{Sp. Ty.} & \colhead{log($L$)} & \colhead{log($T$)} & \colhead{$M$} & \colhead{log(age)} & \colhead{Info.} & \colhead{Ref.}\\
\colhead{} & \colhead{} & \colhead{(days)} & \colhead{(mag)} & \colhead{} & \colhead{($L_{\sun}$)} & \colhead{(K)} & \colhead{($M_{\sun}$)} & \colhead{(yr)} & \colhead{} & \colhead{}
}
\startdata
V002 & J18295715+0033182 & 23.53 & 2.53 & G2.5 & 1.30  & 3.77 & 2.47 & 6.50 & Class II & 1, 10\\
V004 & J18294410+0033559 & 0.1841 & 0.86 & M0    & 0.04  & 3.59 & 0.96 & 6.16 &  Class III & 1, 10\\
V013 & J18295836+0037529 & 9.012 & 0.68 & 	     & 		&		&		 & 		& X-ray 	& 8 \\
V021 & J18300616+0042336 & 0.9904 & 0.95 & F3     &    & $\sim$3.80 & $\sim$1.3 &  & Class II  & 2, 3\\
V039 & J18293084+0101071 & 13.38 & 0.93 &           & -0.28 &  3.49 &  		& 		&  Class II & 4, 5\\
V046 & J18294147+0107378 & 1.004 & 1.27 & K5     & -0.05 & 3.64 & 0.69 & 6.14 & Class II & 6, 7, 10\\
V054 & J18300341+0116191 & 1.061 & 0.76 & K5     & -0.16 & 3.64 & 0.75 & 6.35 & Class II & 6, 7, 10\\
V079 & J18313870+0130546	 & 3.865  & 0.35  &      &  &  &  &  &  & 9\\
V112 & J18283776+0149471 & 2.456  & 0.10 &      &  &  &  &  &  & 9\\
V126 & J18311927+0106557 & 1.028  & 0.24 &   		&      	& 		& 		& 		&  & \\
\enddata
\tablecomments{ $^a$Period and amplitude $A_r$ in the $r$ band are calculated in this work. The spectral type, luminosity, temperature, mass, age, and other information are quoted from the cited references. The mass of V021 is from Ser-Bolo 33 of \citet{2009ApJ...692..973E} estimated by 1.1\,mm emission. In the catalog of \citet{2019ApJ...878..111H}, V002 and V021 are listed as "N/A", V004 and V046 are listed as "Distrib", and the others are listed as "Main".}
{References. (1) \citet{2013ApJ...762..128O}, (2) \citet{2003AJ....126.2971V}, (3) \citet{2009ApJ...692..973E}, (4) \citet{2009ApJS..184...18G}, (5) \citet{2009ApJS..181..321E}, (6) \citet{2015AJ....149..103E}, (7) \citet{2014AaA...572A..62A}, (8) \citet{2016PASA...33...52F}, (9) \citet{2019ApJ...878..111H} and (10) \citet{2007ApJ...663.1149H}. }
\end{deluxetable*}

\section{Color Variability}
\label{sec:color}

Color variability can be caused by various processes like extinction, starspots, disk rotation, accretion, eclipses, and pulsations. Extinction variations caused by structural inhomogeneities of the ambient molecular cloud and the velocity field of the molecular gas \citep{2001AJ....121.3160C} are common for variables, especially in the high-extinction regions of the Serpens cloud.

\citet{2019A&A...623A.110G} visualized the variability-induced motions of stars in a time-dependent CMD and summarized five different motions. The motions of pulsating stars show loop-like shapes similar to a pulsation cycle, while the magnitudes and colors appear linearly dependent; the corresponding variability amplitudes are large. The motions of eclipsing binary systems are linear trajectories, but their slopes are steeper than those of pulsating stars. For rotationally induced variables, the motions are rather horizontal with small amplitudes in magnitude. For eruptive stars, the slopes of motions have the opposite sign with respect to the sign of pulsating stars. Cataclysmic variables show significant color variability toward bluer values.

We produced CMDs and C-CDs to study the variability-induced motions of the identified variables. Like the extinction vector in the CMD, we represented the variability-induced motions of the variables in terms of vectors, where the vector slope or vector angle indicates the direction of motion, and the length of the vector indicates the distance of motion. In this study, the colors of the variables were obtained from the difference of two bands at adjacent observation times, with time intervals less than 0.03\,days for $r-i$ and $i-z$. For the motions of most LPs, linear relationships are seen between the magnitudes and colors (most correlation coefficients are greater than 0.8); the LP analysis is given in Appendix~\ref{subsec:lps}. The motions of most SPs are scattered, and their amplitudes of variation of magnitude and color are smaller than those of the LPs.

\begin{figure}[!h]
 \centering
          \includegraphics[width=.9\textwidth]{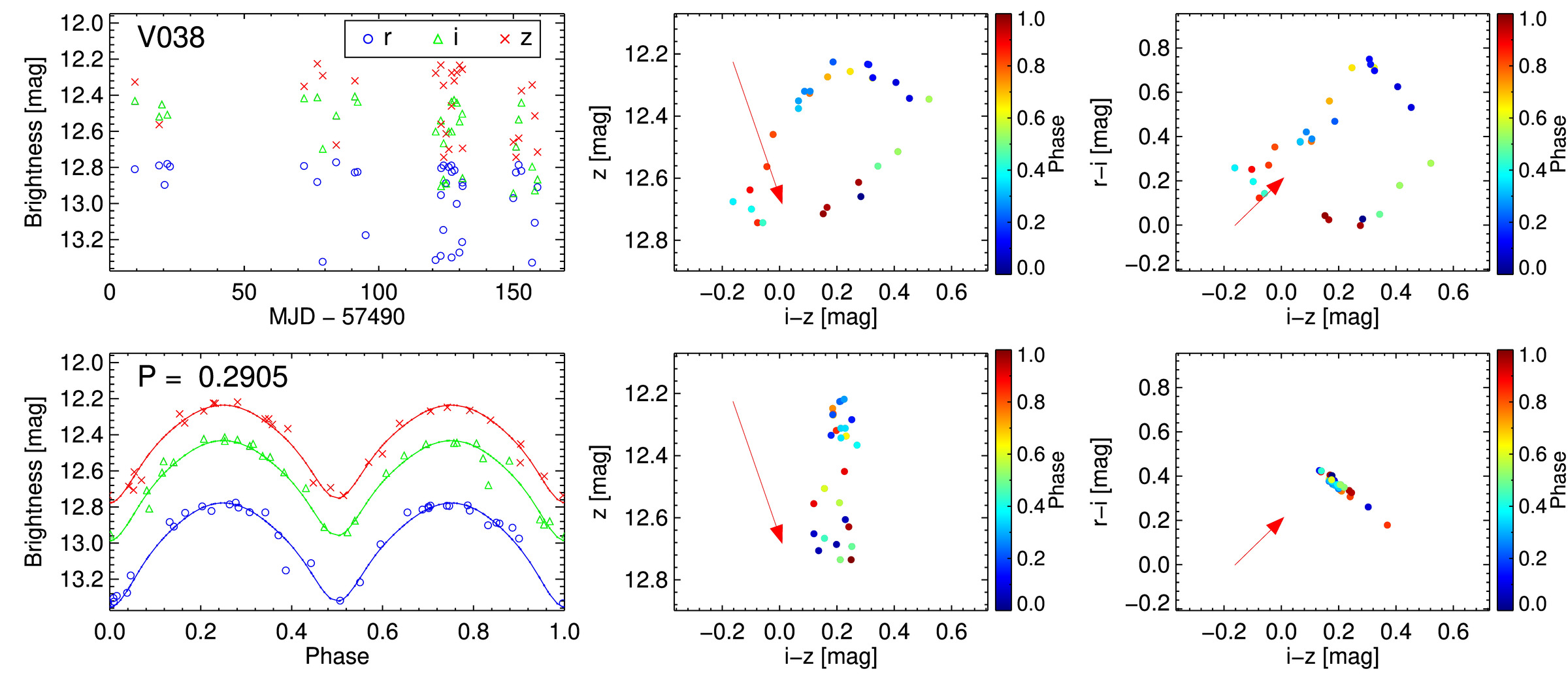}
      \caption{Example of the effect of non-simultaneous observations on color variability. The first column shows the original, folded, and fitted light curves of V038 in $r$ (blue), $i$ (green), and $z$ (red). The second column shows the CMDs constructed from our observations and the best-fitting results, where phase is indicated by color. The third column shows the C-CDs made from our observations and the corresponding best-fitting results; again, phase is indicated by color. The colors of the second row have been corrected using the best-fitting results from PHOEBE. The arrows indicate an extinction vector of $A_V = 1$\,mag. The typical photometric error is less than 0.01\,mag.}\label{fig:cmd46}
\end{figure}

The motions of several SPs appear to be circularly shaped, which is not in line with expectations. These abnormal motions could be caused by nonsimultaneous observations, especially for SPs with periods less than 1.5\,days. This may produce a periodic color bias or color-phase difference, as shown in Appendix~\ref{subsec:phase}. For example, the motion of V038 varied from linear to approximately ellipsoidal, as shown in Figure~\ref{fig:cmd46}. To correct these colors, the best-fitting results of the PHysics Of Eclipsing BinariEs (PHOEBE; \citealt{2005ApJ...628..426P}), shown in Appendix~\ref{app:model}, were used. The corrected motions thus become linear in the CMDs, which agrees with the expected shapes of eclipsing binaries. The corrected linear motions in the C-CDs are inaccurate, because there is an obvious linear correlation in the linear motions obtained using light curves in two selected bands from PHOEBE. Color-phase differences of the motions are prevalent for V106, V129, and for some aperiodic variables; it was difficult to correct the motions of the aperiodic SPs.

\begin{figure}[!h]
 \centering
          \includegraphics[width=.9\textwidth]{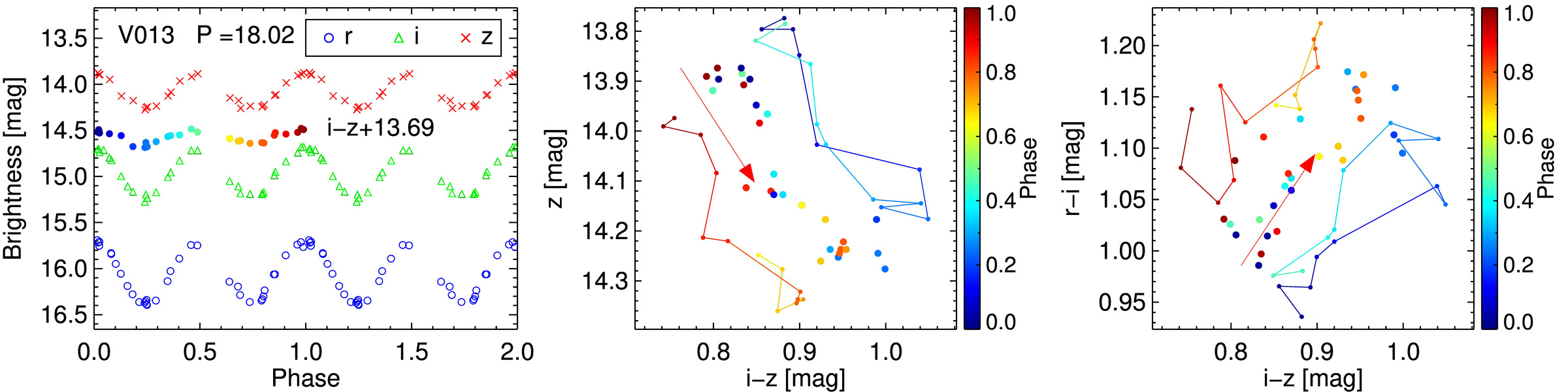}
      \caption{Folded light curves ($r$: blue, $i$: green, and $z$: red), and the CMDs and C-CDs of V013. The dots in the left column show the colors ($i-z$ + constant) as they vary with phase. The red and blue curves in the CMD and C-CD, respectively represent the motions of the first and second half phases. The phase is indicated by color, and the arrows indicate the extinction vector $A_V = 0.5$\,mag. The typical photometric error is less than 0.01\,mag.}\label{fig:cmd17}
\end{figure}

For SPs with periods larger than 1.5\,days, the color-phase difference can be neglected. Figure~\ref{fig:cmd17} presents the light curves and the motions of V013. It has a period of 18.02\,days, and its motion is not affected by color phase difference; it is a cluster member of Serpens Main. The variations of motion in the CMD are almost linear, and the directions are the same as the reddening vector.

There are 22 YSOs and YSO candidates identified through cross-matching our observed variables and the YSO catalog (as mentioned in Section~\ref{sec:axldata}) within 2$\arcsec$. Five YSOs and YSO candidates (V130, V134, V138, V140, and V142) were detected only in the $z$ band. As shown in Figure~\ref{fig:cmd1}, 11 YSOs and YSO candidates are located in the area of asymptotic giant branch (AGB) stars, and six YSOs are located in the PMS area. For these YSOs, the variability-induced motions were used to confirm the misclassified nonmember YSOs (Section~\ref{sec:nmysos}) and constrain the properties of the member YSOs (Section~\ref{sec:mysos}).

\subsection{YSOs}
\label{sec:nmysos}

The 11 YSOs and YSO candidates outside the Serpens cluster detected in the $i$ and $z$ bands are presented in this section. As listed in Table~\ref{tab:lysos}, the YSOs and YSO candidates are LP variables with distances larger than 1800\,pc, except V045, which has a distance of 565\,pc. The Class\,II source V001 presents the typical SEDs of an optically thick accretion disk \citep{2007ApJ...663.1149H,2013ApJ...762..128O}, and which has a related X-ray counterpart in the 3XMM catalog \citep{2016A&A...590A...1R}. The YSOs V045, V052, and V061 were identified by \citet{2007ApJ...669..493W}. They are transition disk objects due to the excesses observed at 8.0 and/or 24\,$\mu$m but not at shorter IRAC wavelengths. The YSO candidates V008, V011, V017, V025, V030, V040, and V131 were identified from MIPS data and Two Micron All Sky Survey (2MASS) photometry \citep{2007ApJ...663.1149H}. The YSOs, YSO candidates, and post-AGB stars are mixed in the CMD and cannot be distinguished by IR excess alone \citep{2013A&A...557A..51C}, although in a few cases, classifications could be made on the basis of light-curve morphology \citep{2016AJ....152...74B}. The periodic YSOs and YSO candidates, especially the Mira-like objects, are likely to be late-type evolved stars. This latter classification is usually based on the shape of the light curves, which resemble SR variables and Mira stars.

\begin{deluxetable*}{llccrccc}
\tablecaption{YSOs and YSO candidates outside the cluster \label{tab:lysos}}
\tablewidth{0pt}
\tablehead{
\colhead{Obs\_ID} & \colhead{2MASS ID} & \colhead{Period$^a$} & \colhead{$A_i^a$} & \colhead{Parallax$^b$} & \colhead{Info.} & \colhead{Type$^c$} & \colhead{Ref.}\\
\colhead{} & \colhead{} & \colhead{(days)} & \colhead{(mag)} & \colhead{(mas)} & \colhead{} & \colhead{} & \colhead{}
}
\startdata
V001 & J18295702+0033000 & 62.50 & 0.66 & 0.0951 & Class II   & Irr & 1, 3, 4   \\
V008 & J18301774+0036108 & 245.1 & 2.47 & -0.3016 & YSO?  & Mira & 1   \\
V011 & J18301951+0037203 & 164.9 & 0.34 & 0.5301 & YSO?  & SR  & 1   \\
V017 & J18273423+0040252 & 192.3 & 0.80 & 0.1095 & YSO?  & Mira & 1   \\
V025 & J18280452+0049483 & 355.2$^d$ & 0.76 & -0.1202 & YSO?  & Mira & 1   \\
V030 & J18275931+0055595 & 187.1 & 1.79 & -0.0564 & YSO?  & Mira & 1   \\
V040 & J18281908+0101319 & 200.0 & 1.01 & -0.5477 & YSO? & Mira  & 1   \\
V045 & J18291302+0107007 & 170.5 & 0.38 & 1.7702 & YSO  & Mira & 1, 2   \\
V052 & J18303733+0116086 & 153.8 & 0.26 & -0.0418 & YSO  & SR & 1, 2   \\
V061 & J18301730+0121325 & 130.9 & 0.38 & 0.0104 & Class III   & SR & 1, 2, 3   \\
V131 & J18304882+0037590 & 200.0 & 2.35 & -0.6644 & YSO? & Mira  &  1  \\
\enddata
\tablecomments{ $^a$The Period and amplitude, $A_i$, in the $i$ band are calculated in this work. The YSO classifications are quoted from the cited references. $^b$The parallaxes are from Gaia\,DR2. $^c$The variability types are classified in this work, including irregular (Irr), SR, and Mira types. $^d$The period of V025 is from Gaia\,DR2. }
{References. (1) \citet{2007ApJ...663.1149H}, (2) \citet{2007ApJ...669..493W}, (3) \citet{2009ApJ...692..973E}, (4) \citet{2013ApJ...762..128O}. }
\end{deluxetable*}

\begin{figure}[!h]
 \centering
          \includegraphics[width=.45\textwidth]{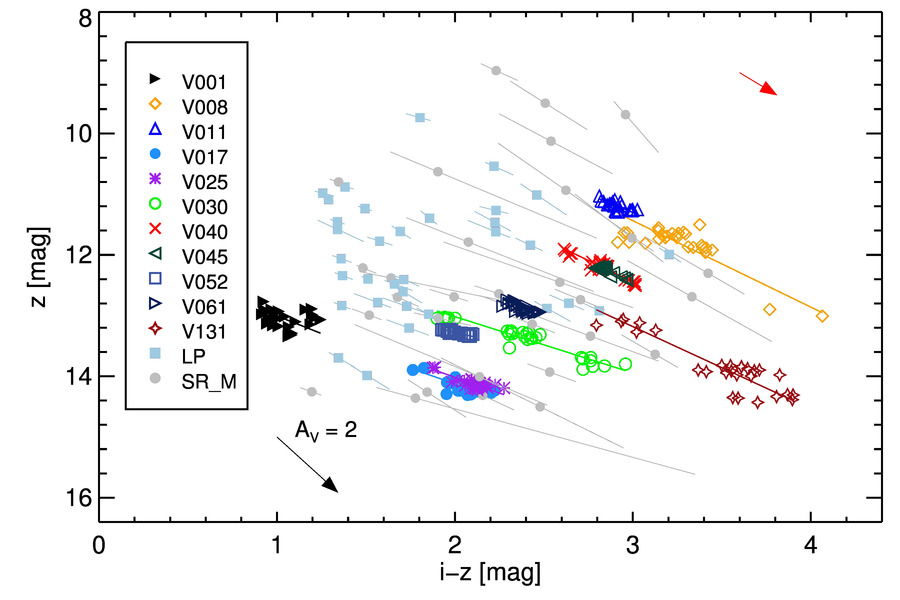}
          \includegraphics[width=.45\textwidth]{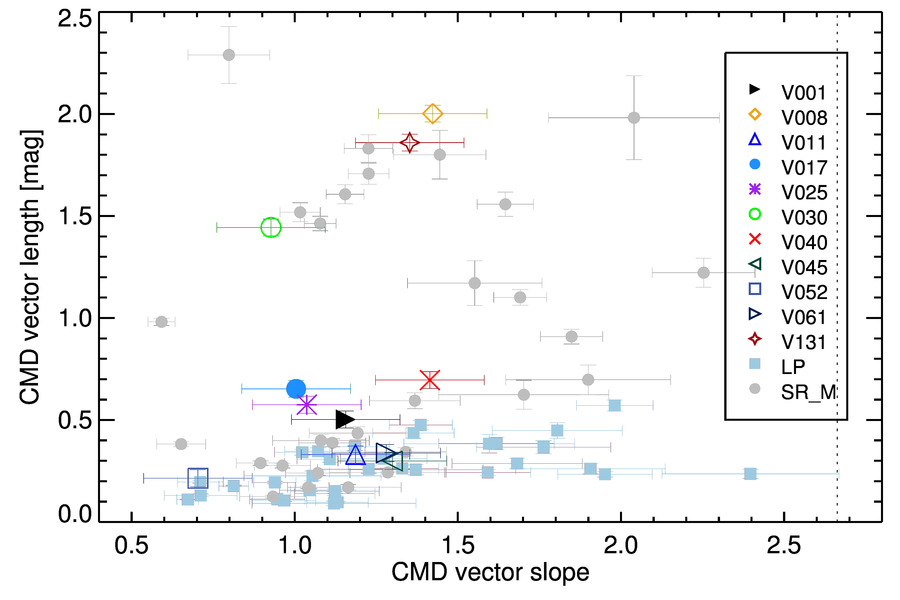}
      \caption{Motions of the variables in the CMD and their vectors of motion. The motions of the YSOs are presented in different colors. The gray dots and lines indicate the median position and the linear fitted results of the LP motions, and the light blue squares and lines indicate those of the SR\_Ms. The black arrow indicates an extinction vector of $A_V = 2$\,mag. The red arrow indicates the vector of a pulsating star (as described in Appendix~\ref{subsec:lps}). The lengths and slopes of the vectors with error bars are displayed in the right panel, and the dotted line is the slope of extinction. The typical photometric error is less than 0.01\,mag.}\label{fig:ysos}
\end{figure}

To understand their color variabilities, the YSOs and YSO candidates are shown in the CMD in Figure~\ref{fig:ysos}. From the left panel of Figure~\ref{fig:ysos}, the motions of young objects are linear, depending on their magnitudes and colors; the same holds for the motions of LPs and SR\_Ms. The median angles of the vectors of the YSOs and YSO candidates, LPs, and SR\_Ms are 52$^{\circ}\pm6^{\circ}$, 49$^{\circ}\pm8^{\circ}$, and 50$^{\circ}\pm9^{\circ}$, respectively. The angles of their vectors are between 30$^{\circ}$ and 70$^{\circ}$, which is less than the angle of the extinction vector (about 70$^{\circ}$), suggesting that the variability-induced motions of these objects in the CMD are not caused by extinction. However, the color variabilities of the YSOs and YSO candidates in the CMD are not obviously different from those of the LPs and SR\_Ms.

The YSOs are expected to show different levels of variability caused by a variety of physical processes; indeed, YSO variability should be rather irregular, with timescales ranging from a few days up to a decade \citep{2019MNRAS.482.5567M}. For example, in the outbursts of FU Orionis stars, abrupt changes in the accretion rate from the disk can lead to an increase in brightness with long periods \citep{2012ApJ...744...56H}. From our light curves, only V001 showed quasiperiodic variations and a $z$-band excess that led to irregular motions in the CMDs, which may be associated with the accretion disk. Objects V011, V052, and V061 show SR variations with periods of about 150\,days, and V008, V017, V025, V030, V040, V045, and V131 have Mira-like variations with periods greater than 180\,days. The variability classifications of the YSOs and YSO candidates are listed in Table~\ref{tab:lysos}.

\subsection{Cluster Members}
\label{sec:mysos}

Rotation studies of YSOs and their starspots suggest that optical observations are dominated by events and features near the stellar photosphere \citep{2018AJ....155...99W}. Due to magnetic activity or material accreting on the stellar surface, the starspots may be cooler or hotter than the photospheric temperature and thus change the brightness of the star as it rotates. The timescale of photometric variability caused by starspots is determined by the period of stellar rotation.

To investigate the color variability of YSOs, a simple spot model was used. In this model, we assume that a spot is a single-temperature blackbody ($T_{\rm spot}$) and covers a fraction, $f$, of the stellar photosphere with an effective temperature ($T_*$). The variation of photospheric emission can be expressed as $\Delta m(\lambda) = -2.5 {\rm log} \{1 - f[1.0 - B_{\lambda}(T_{\rm spot})/B_{\lambda}(T_{*})]\}$, where $B_{\lambda}(T)$ is the Planck function \citep{2001AJ....121.3160C}. \citet{2001AJ....121.3160C} used this model to calculate the photometric amplitudes expected from both cold and hot stellar spots and presented the results of low-mass PMS stars in the near-IR. In their CMDs and C-CDs, the photometric amplitudes showed linear variations caused by hot spots and cold spots, where those of hot spots were larger; in addition, the directions of the variation were different from that of the extinction vector.

\begin{figure}[!h]
 \centering
          \includegraphics[width=.46\textwidth]{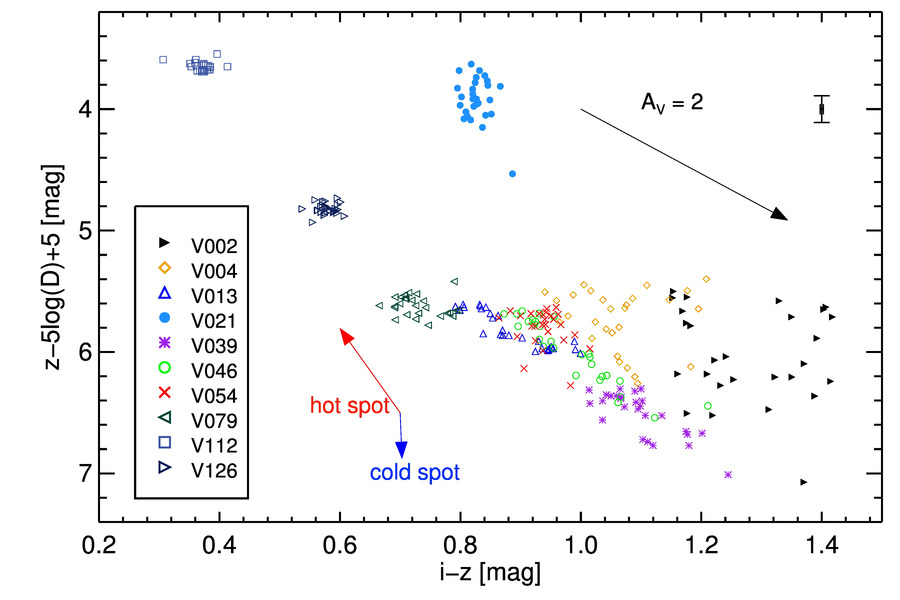}
          \includegraphics[width=.46\textwidth]{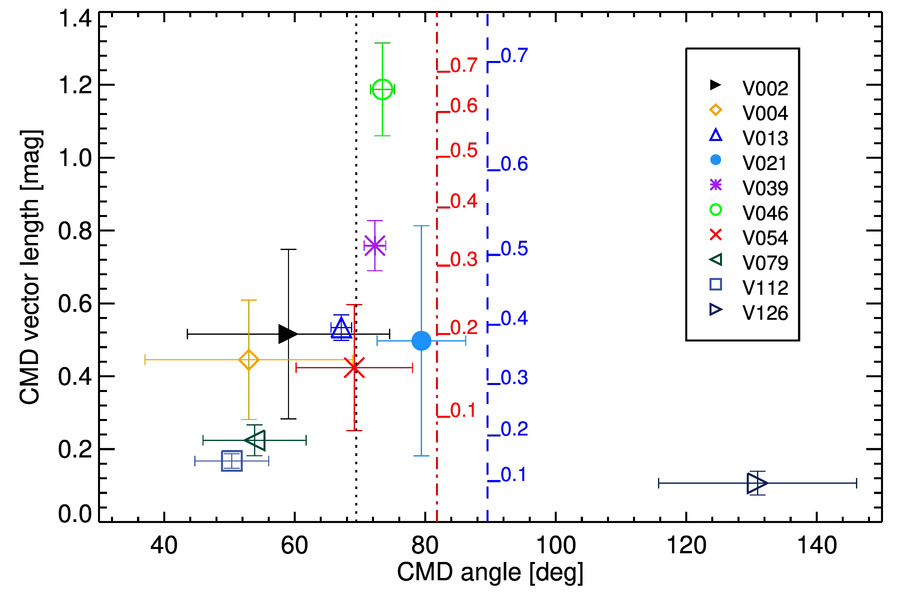}
      \caption{Motions of variable cluster members in the CMD and the vectors of motion. The members are represented in different colors. The error bar in the left panel indicates the typical errors of the members (0.003 and 0.11\,mag). The red, blue, and black arrows indicate hot spots, cold spots, and an extinction vector of $A_V = 2$\,mag, respectively. The angles and lengths of the motions in the CMD are indicated in the right panel. The black dotted line, red dotted-dashed line, and blue dashed line indicate the angles of the extinction vector, negative hot-spot vector, and cold-spot vector, respectively. The lengths of different spots range are from 0.1 (10\%) to 0.7 (70\%), with an interval of 0.1 (10\%).}\label{fig:member}
\end{figure}

To calculate the vectors of the hot and cold spots (as shown in Figure~\ref{fig:member}), the photospheric temperature was set to 4000\,K, the maximum spot coverage to 30\%, and the temperatures of the hot and cold spots to 6000 and 2000\,K, respectively. Hot spots make the star brighter and bluer, while cold spots make the star fainter and redder. Indeed, hot spots, cold spots, and extinction cause the stars to move in different directions in the CMD. The more detailed model given by \citet{2017MNRAS.468..931M} showed a range of variations of starspot vectors, where the amplitudes of both the magnitudes and color variations mostly depend on the spot coverage. The variations in the slopes mostly depend on the ratio of the temperature of the spotted photosphere to the unspotted photosphere.

Figure~\ref{fig:member} shows the color variabilities of the cluster members in the CMD, including six YSOs (V002, V004, V021, V039, V046, and V054). The angles of the vectors in the left panel in Figure~\ref{fig:member} were converted to a range spanning 0$^{\circ}$--180$^{\circ}$ to provide a comparison with their directions of motion. According to the light curves of V013 and V079, they are binary systems; the corresponding analysis is presented in Appendix~\ref{app:model}. In terms of the directions and lengths of the vectors, the left eight objects can be divided into four types: (i) objects with horizontal motions and small variations in brightness (V112 and V126), (ii) objects with chaotic motions and small vector angles (V002 and V004), (iii) objects with linear motions similar to the extinction vector (V039, V046, and V054), and (iv) an object with a loop-like motion corresponding to a hot-spot vector (V021).

The diskless YSOs (Class III) show periodic variations generally attributed to spot modulation and have smaller variability amplitudes in the mid-IR than those of YSOs with disks (Class I, Class F, and Class II; \citealt{2015AJ....150..118P}). From our spot model, cold spots should present small variability at the observed wavelengths. Objects V112 and V126 showed quasiperiodic signals with small variability, which may be caused by spot modulation and unstable extinction. As displayed in the left panel in Figure~\ref{fig:member}, the two objects with small variabilities may be related to diskless YSOs, and the lengths of their vectors indicate a cold-spot coverage of about 0.1 (10\%).

For the remaining types, the variations of brightness are larger than 0.5\,mag in the $i$ band. The light curves show quasiperiodic variations with a dipper-like morphology \citep{2014AJ....147...82C}, which can be attributed to stellar occultations by an inhomogeneous dusty disk \citep{2011ApJ...733...50M}. The three types of objects were distinguished by the angles of their vectors, and the differences are associated with the variability mechanisms of accretion spots and disk structure. In Table~\ref{tab:ysos}, the YSOs are Class II/III objects with a surrounding disk. According to the ratio of the disk luminosity to stellar luminosity ($L_{\rm d/s}$), V002 is a Class II YSO with a disk ($L_{\rm d/s} \sim 0.003$) close to a debris disk ($L_{\rm d/s} < 0.001$; \citealt{2013ApJ...762..128O,2007ApJ...663.1149H}). Class III YSO V004 and Class II YSO V021 have optically thin disks ($L_{\rm d/s} \sim 0.02$; \citealt{2013ApJ...762..128O,2009ApJ...692..973E}). Class II YSOs V039, V046, and V054 have optically thick disks ($L_{\rm d/s} \sim 0.1$; \citealt{2009ApJS..184...18G, 2014AaA...572A..62A}).

Class II objects are hybrid systems including an accretion disk and a star with both bright and dark spots \citep{2015AJ....150..145W}; in this scenario, variability can be dominated by one mechanism or a mixture of multiple mechanisms. If an accretion event occurs, the hot spots can be monitored. Considering the random fluctuations and the occultation of the inhomogeneous disk, the vectors of YSOs are dispersed. The V021 vector is close to the hot-spot vector, which implies a hot-spot coverage of about 0.2. The vectors of V039, V046, and V054 are close to the extinction vector, which may be mainly caused by structural inhomogeneities in a circumstellar disk. Objects V002 and V004 show chaotic variability caused by inhomogeneous disks and starspots. For V004, the phase difference (about 0.3$\pi$, in Appendix~\ref{subsec:phase}) has a certain contribution to the chaotic variability.

A high variability fraction of YSOs can be obtained in the YSO clusters, about 80\% of Class I/F/II YSOs are variable at 3.6 and 4.6\,$\mu$m bands, and about 20\% of Class III objects are variable \citep{2018AJ....155...99W}. In Serpens Main, 57 cluster members coincide with YSOs identified by previous studies (as mentioned in Section \ref{sec:cluster}). The variability criteria for identifying variability may be conservative for YSOs, especially for YSOs with low amplitudes, and only six YSO members are identified as variables. For the nonvariable YSO members, we inspected the periodogram and light curves and found that most members vary periodically or aperiodically with low amplitude. If these members are considered as variables, then the variability fraction of YSOs is about 60\% (35/57). These likely YSO variables require more observations to confirm, and we will study the variability fraction of YSOs in future work.

\section{Summary}
\label{sec:summary}

The BEST\,II telescope was used to search for stellar variability in the Serpens SFR. The observations were obtained in the Pan-STARRS1 $r$, $i$, and $z$ bands from spring to fall 2016, spanning about 30 nights. In the observed field, 143 variable stars were identified, of which 119 are new stars. From this sample, 99 objects with clear periodicity were chosen, and their periods and variability classifications were determined. Based on the periods and the amplitudes of their light curves, 18, 39, and 42 variables were classified as SP, LP, and SR\_M, respectively. Eclipsing binaries and LP variables were located in each region in the CMDs.

From the distribution of proper-motion directions, the locations of the peaks of cluster members and field stars were different. By matching the observed variables and the cluster members obtained from the clustering of Gaia\,DR2 stars in the observed field, the variables in the Serpens Main cluster could be identified. We found 10 variable stars belonging to the Serpens Main cluster, including six known YSOs; their brightnesses are brighter than $G = 17$\,mag, and their proper motions are about 3.0\,mas yr$^{-1}$ in R.A. and -8.0\,mas yr$^{-1}$ in decl.

We studied the color variability of these objects to distinguish different types of variables by means of their location in CMDs and C-CDs. The motions of the SPs and LPs in these diagrams were induced by color variability, showing expected linear or loop-like shapes. The motions of some SPs were affected by phase differences (larger than 2\% of their period), likely caused by nonsimultaneous observations. YSOs and AGBs with LPs could not be distinguished by means of their IR excess alone; instead, a variability classification was made by inspecting their light-curve behavior. The motions of the cluster members turned out to be different, which may be caused by accreting spots and circumstellar disks.

Moreover, we presented the light curves of eight eclipsing binary systems identified with the PHOEBE model, including the known V0623\,Ser. The parameters of the binaries were obtained, such as the effective surface temperature of the secondary star, mass ratio, inclination, semi-major axis, and the equivalent radius of the primary.

\acknowledgments
We thank the anonymous referee, whose useful comments improved this paper. We wish to thank Michael Ramolla for operating the BEST\,II telescope for the acquisition of the data used in this paper. We acknowledge support from the DFG grant CH 71/33-1, the general grants 11973004, 11903083, and 11873063 of the National Natural Science Foundation of China, and the IdP II 2015 0002 64 and DIR/WK/2018/09 grants of the Polish Ministry of Science and Higher Education. This work has made use of data from the European Space Agency (ESA) mission Gaia (\url{https://www.cosmos.esa.int/gaia}), processed by the Gaia Data Processing and Analysis Consortium (DPAC; \url{https://www.cosmos.esa.int/web/gaia/dpac/consortium}). Funding for the DPAC has been provided by national institutions, in particular the institutions participating in the Gaia Multilateral Agreement.

%






\appendix

\section{Light Curves of the Variables}
\label{app:figs}
The light curves of different types of variable stars were obtained, and according to the classifications of variable stars in Section \ref{subsec:class}, the light curves of SPs, LPs, and SR\_Ms are shown in Figures \ref{fig:sp_lc}, \ref{fig:lp_lc}, and \ref{fig:sr_lc}, respectively.

\setcounter{figure}{0}
\renewcommand{\thefigure}{A\arabic{figure}}

\begin{figure}[!h]
 \centering
\includegraphics[width=.24\textwidth]{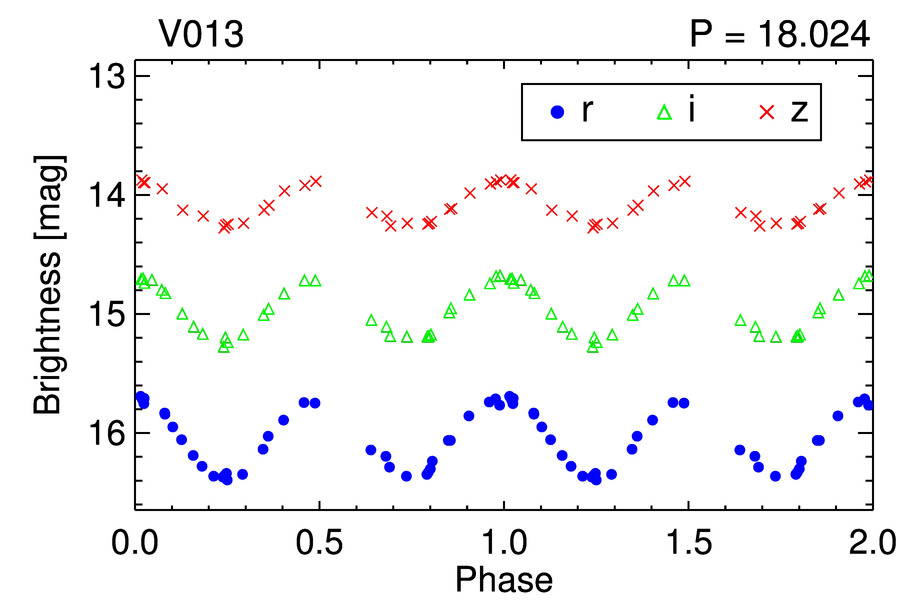}
\includegraphics[width=.24\textwidth]{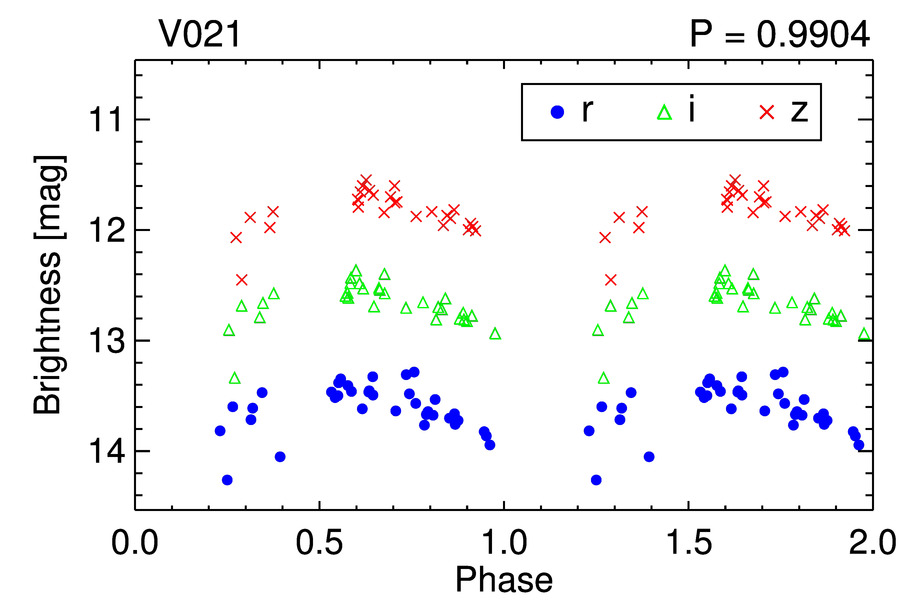}
\includegraphics[width=.24\textwidth]{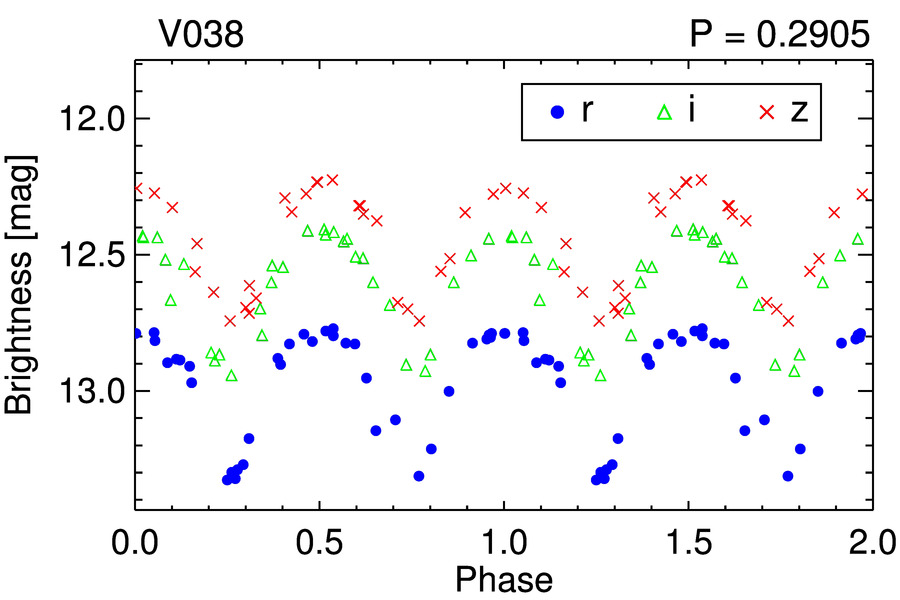}
\includegraphics[width=.24\textwidth]{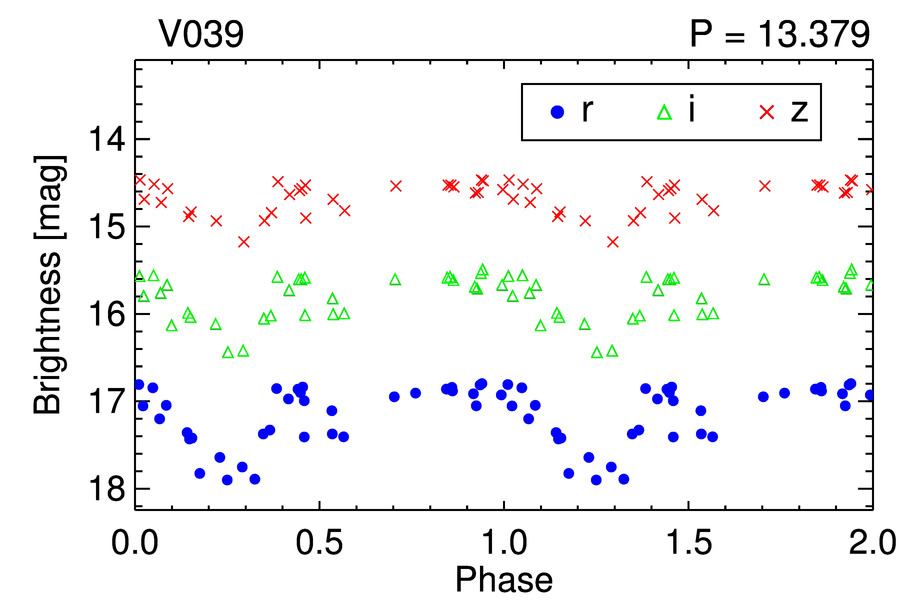}
\includegraphics[width=.24\textwidth]{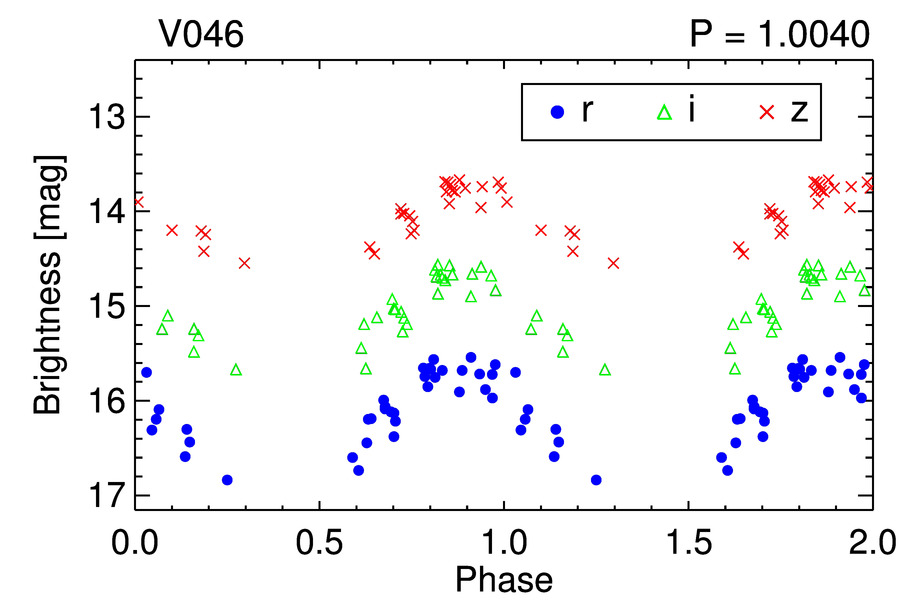}
\includegraphics[width=.24\textwidth]{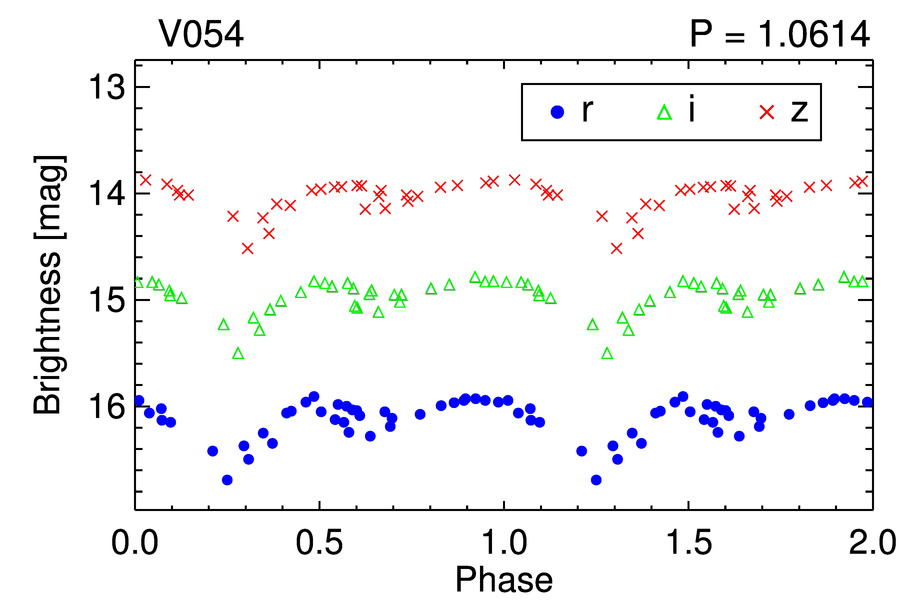}
\includegraphics[width=.24\textwidth]{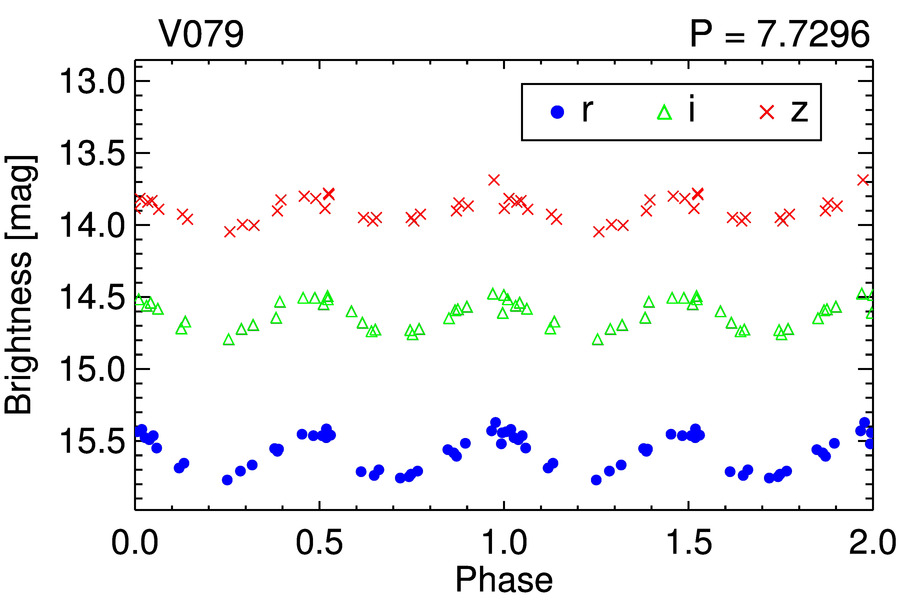}
\includegraphics[width=.24\textwidth]{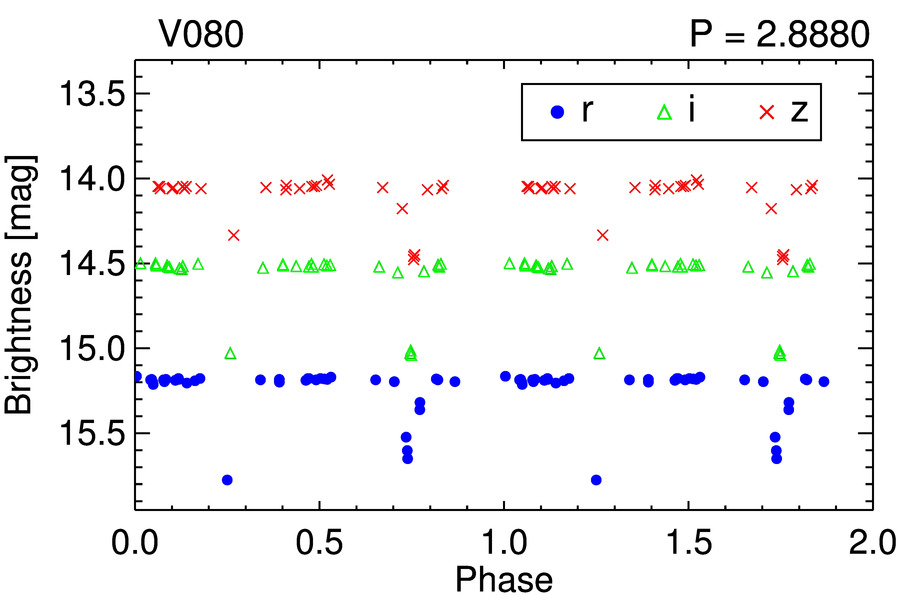}
\includegraphics[width=.24\textwidth]{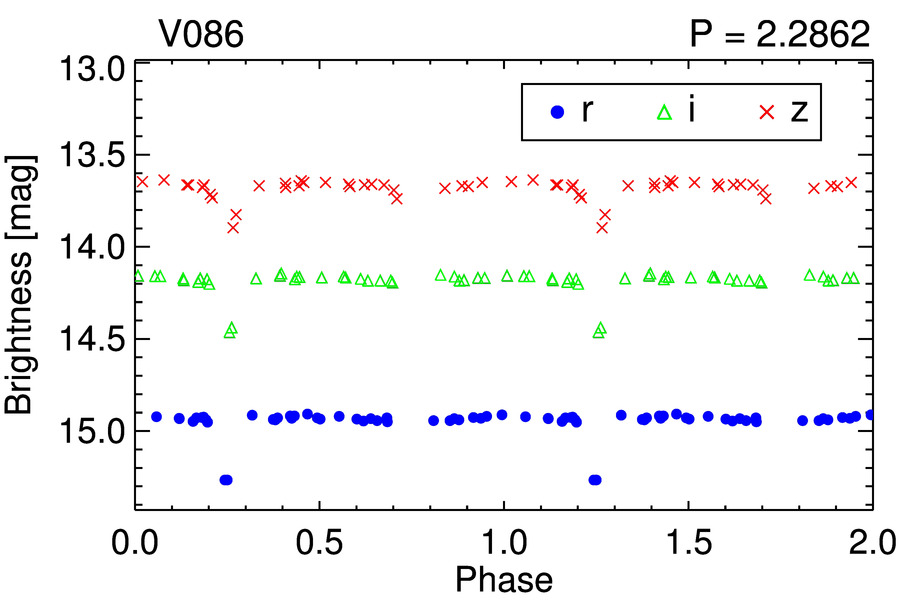}
\includegraphics[width=.24\textwidth]{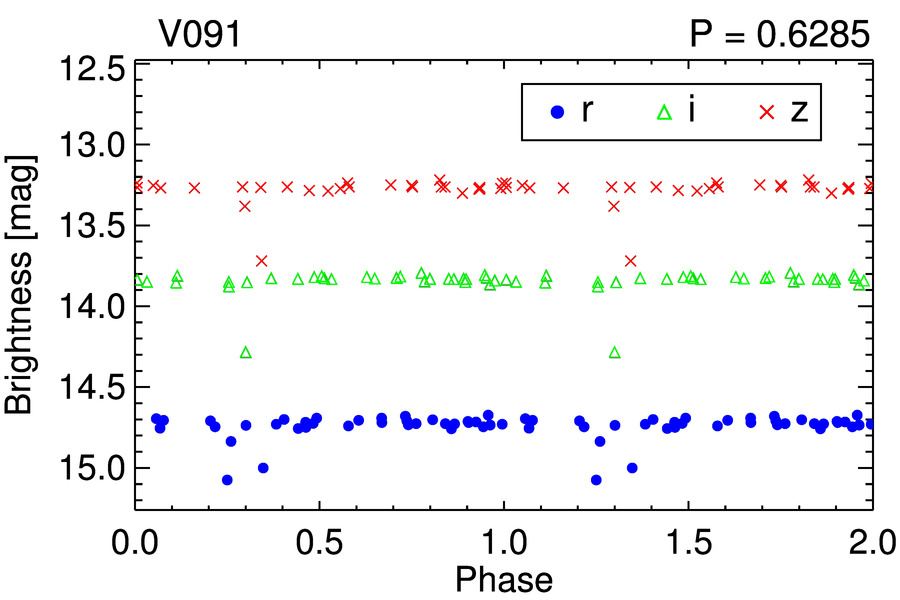}
\includegraphics[width=.24\textwidth]{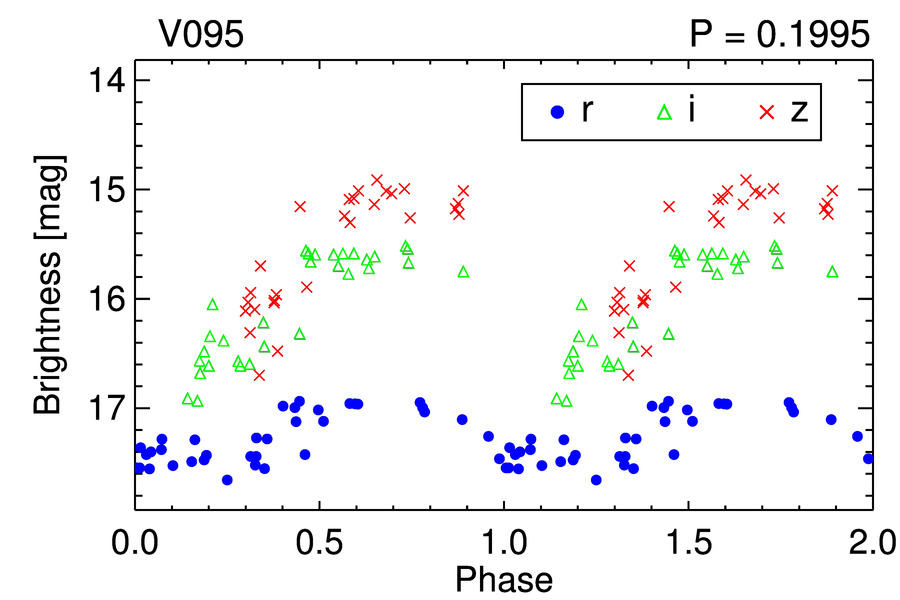}
\includegraphics[width=.24\textwidth]{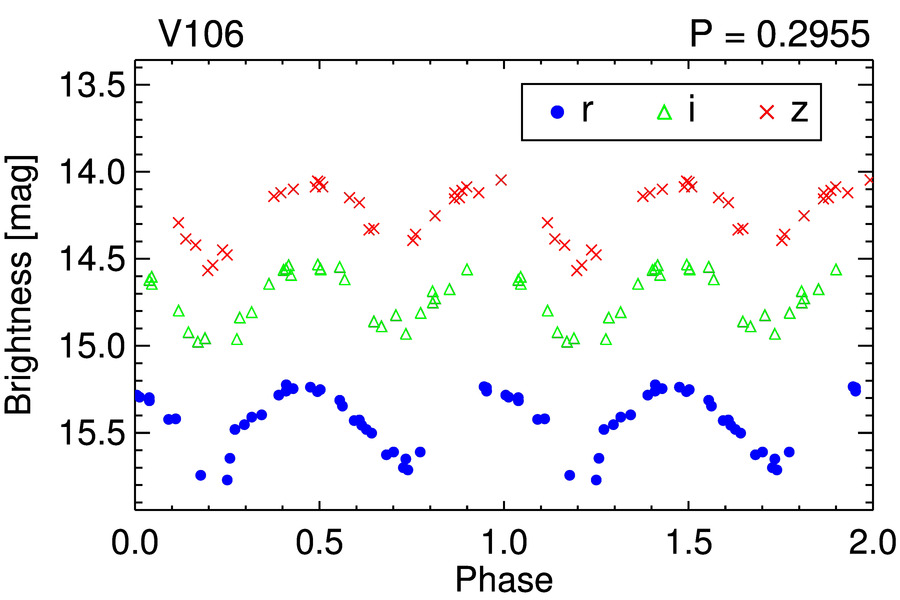}
\includegraphics[width=.24\textwidth]{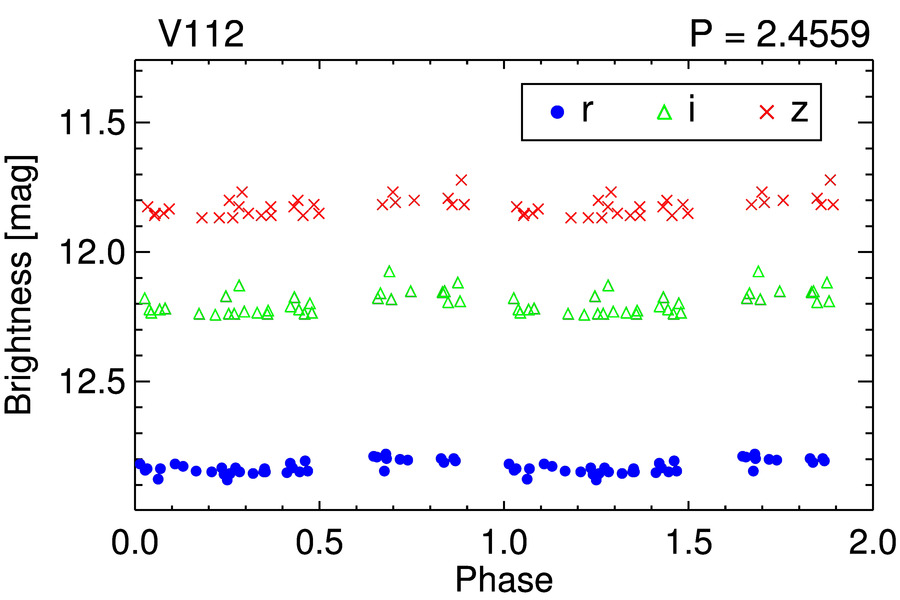}
\includegraphics[width=.24\textwidth]{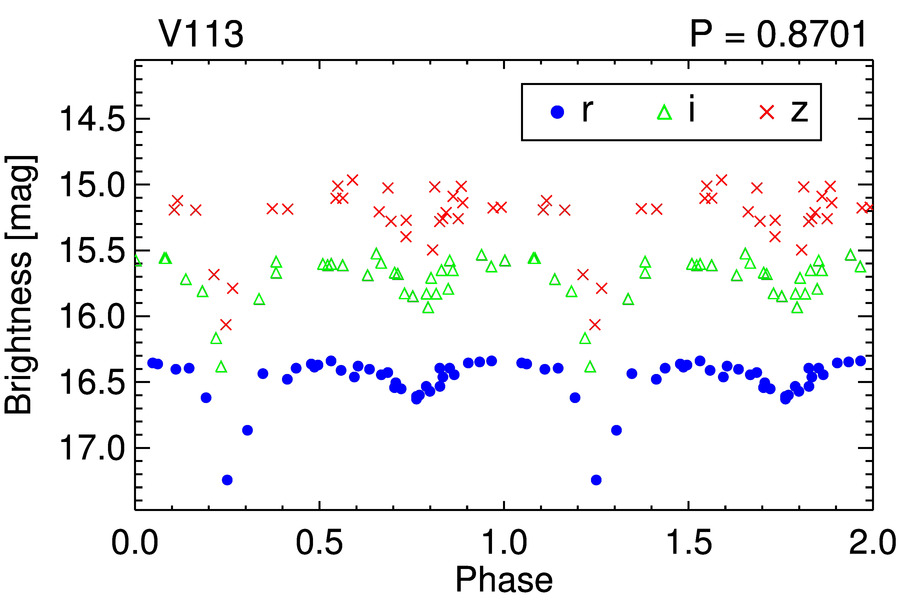}
\includegraphics[width=.24\textwidth]{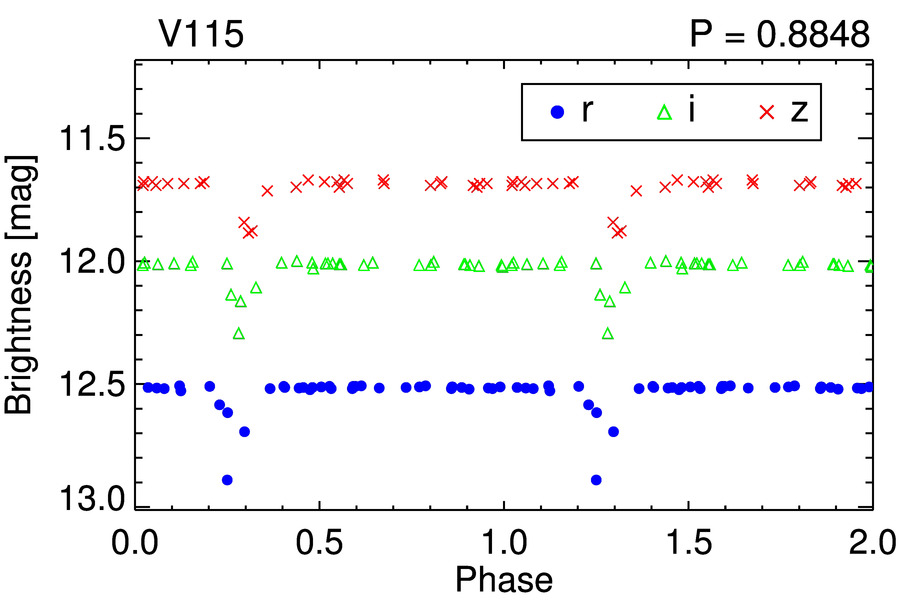}
\includegraphics[width=.24\textwidth]{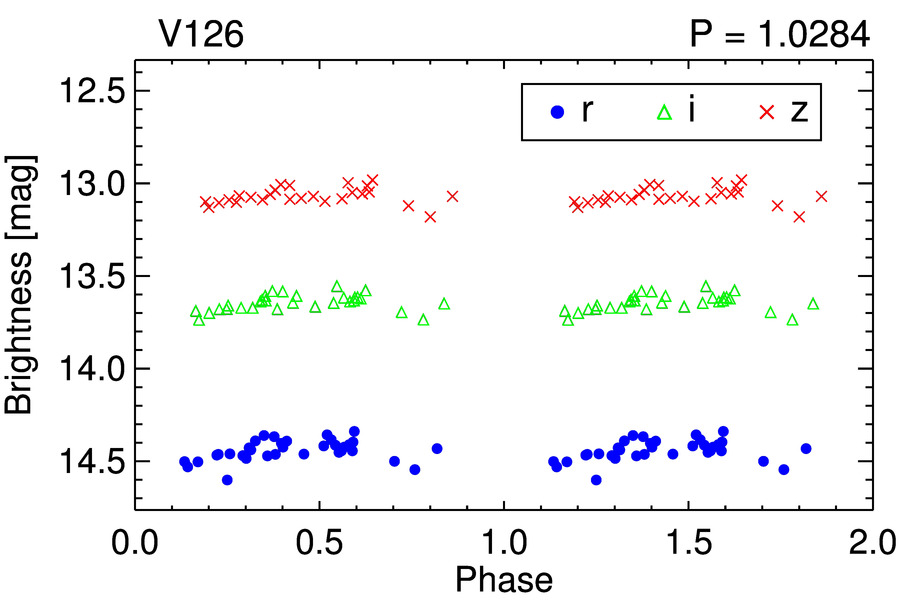}
\includegraphics[width=.24\textwidth]{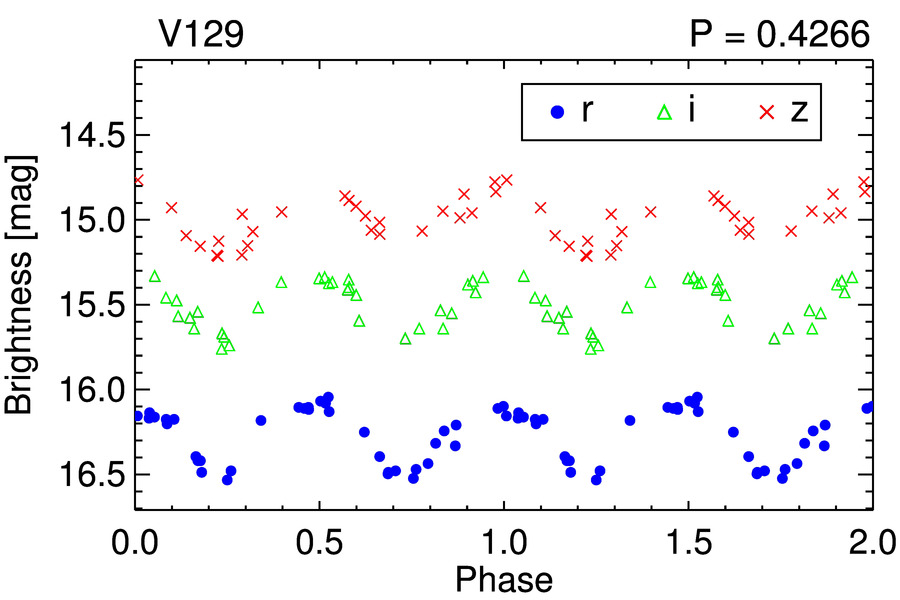}
\includegraphics[width=.24\textwidth]{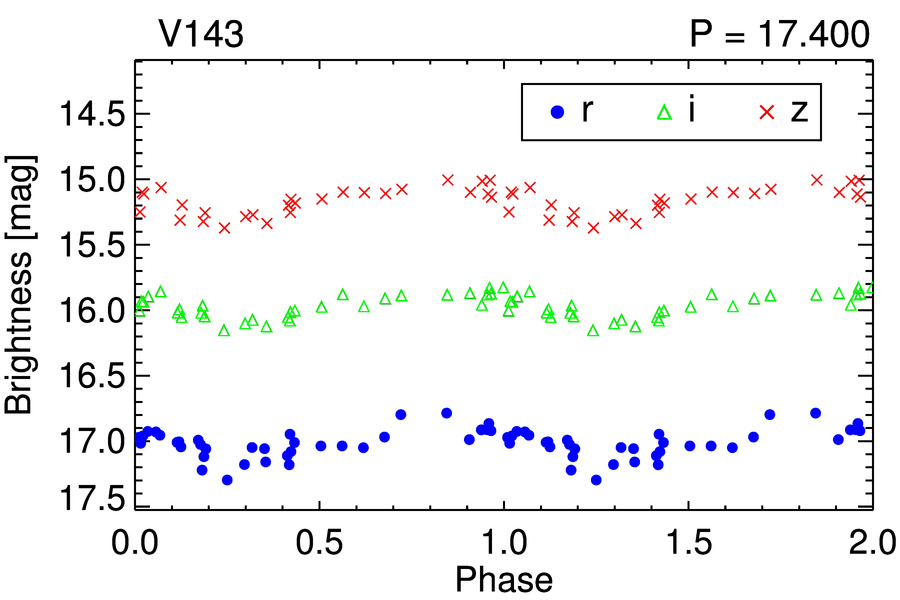}
      \caption{Light curves of the SP variables. The typical photometric error is less than 0.01\,mag.}\label{fig:sp_lc}
\end{figure}

\clearpage
\begin{figure}[!h]
 \centering
      \includegraphics[width=.24\textwidth]{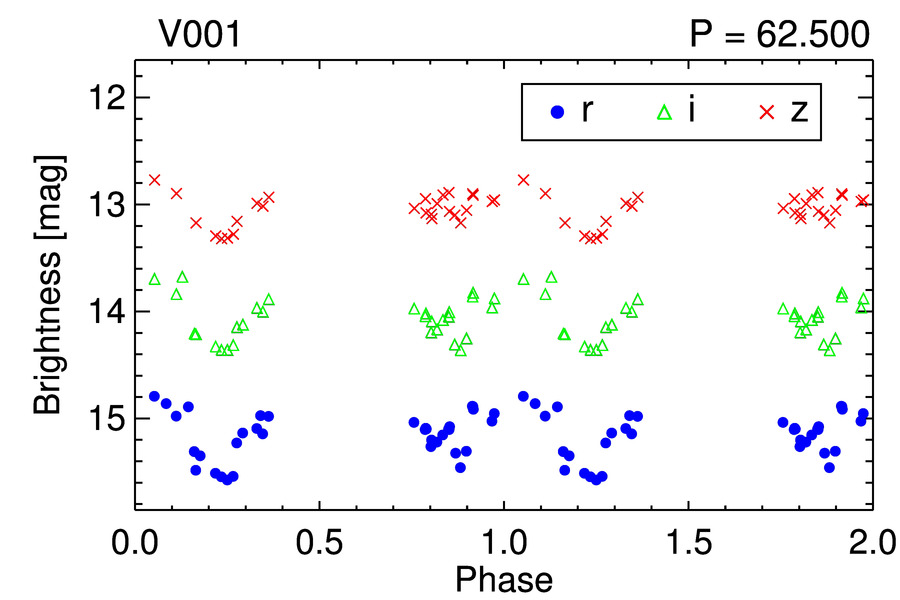}
\includegraphics[width=.24\textwidth]{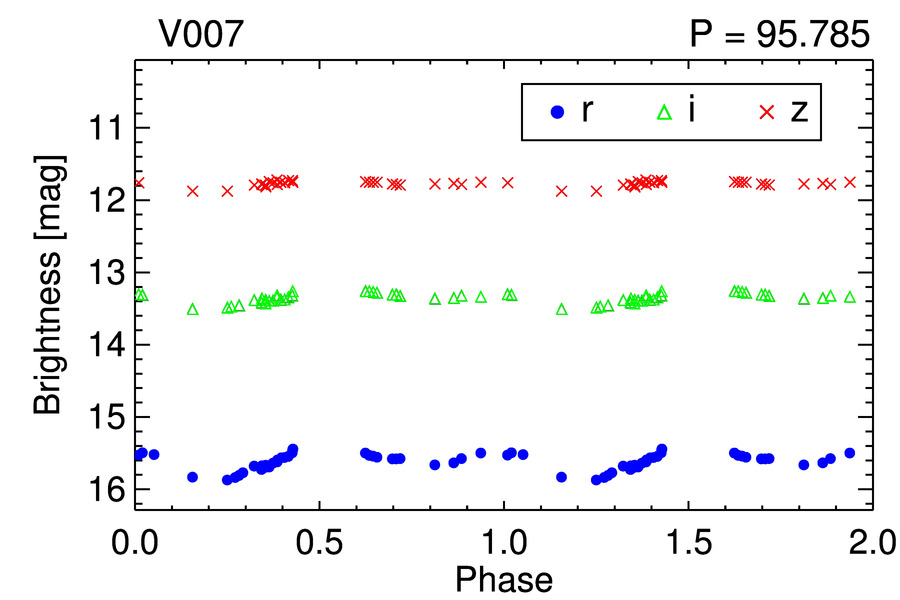}
\includegraphics[width=.24\textwidth]{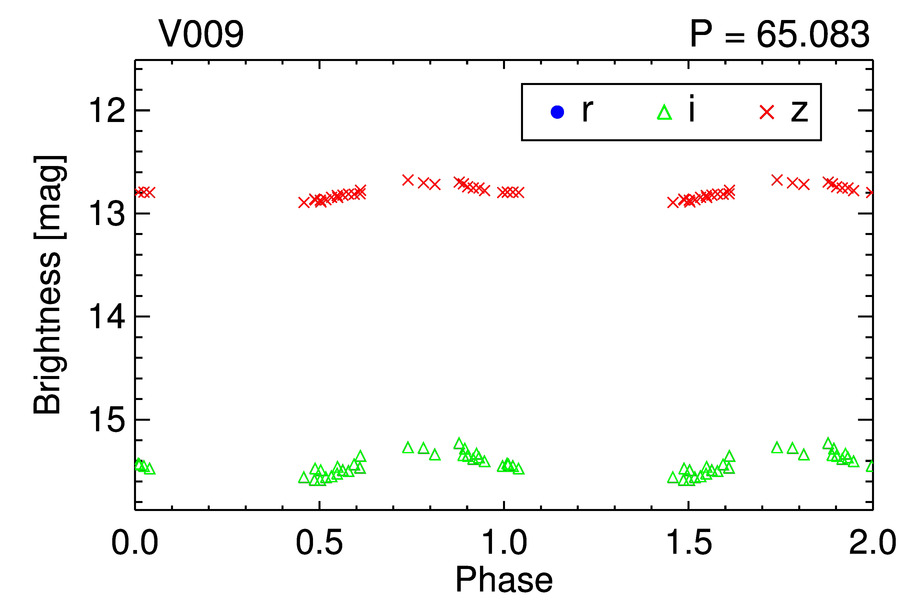}
\includegraphics[width=.24\textwidth]{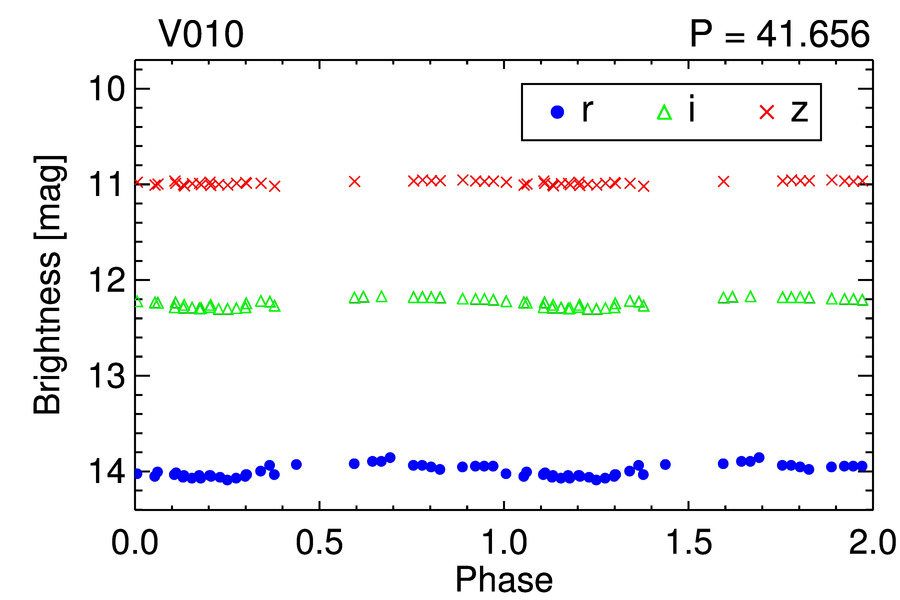}
\includegraphics[width=.24\textwidth]{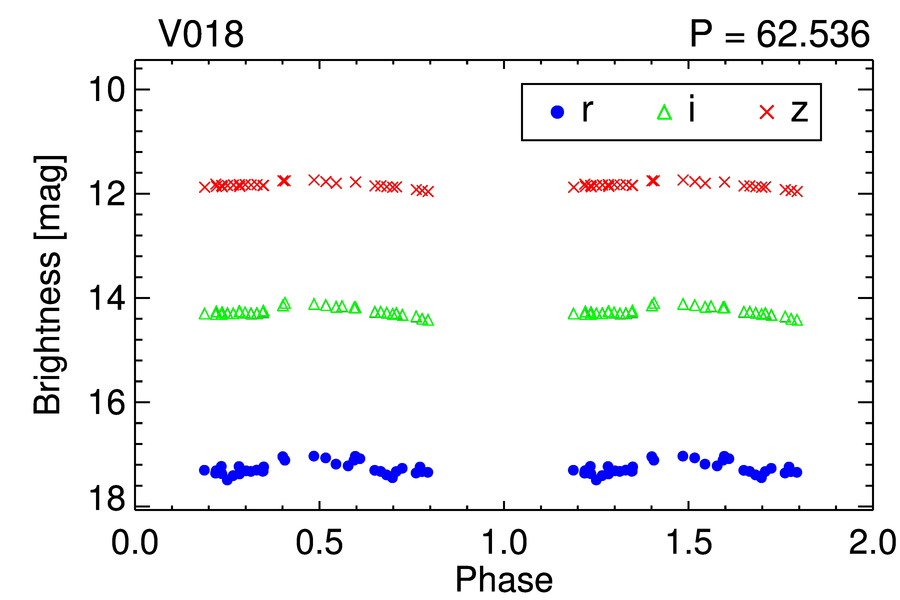}
\includegraphics[width=.24\textwidth]{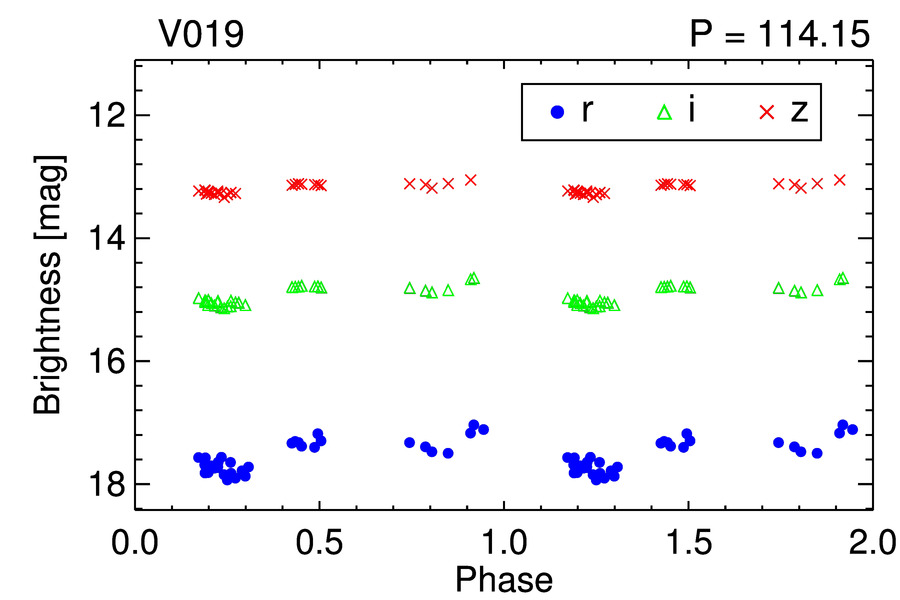}
\includegraphics[width=.24\textwidth]{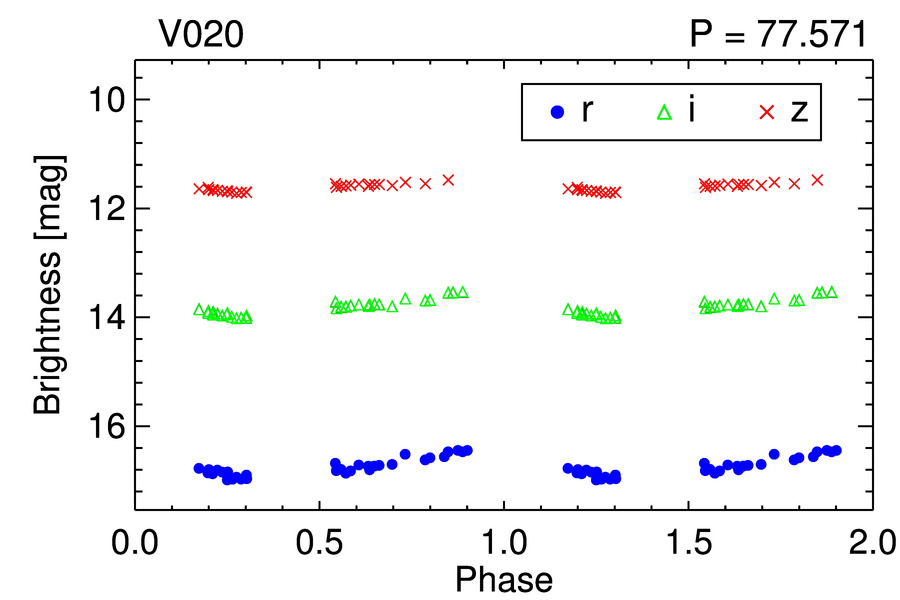}
\includegraphics[width=.24\textwidth]{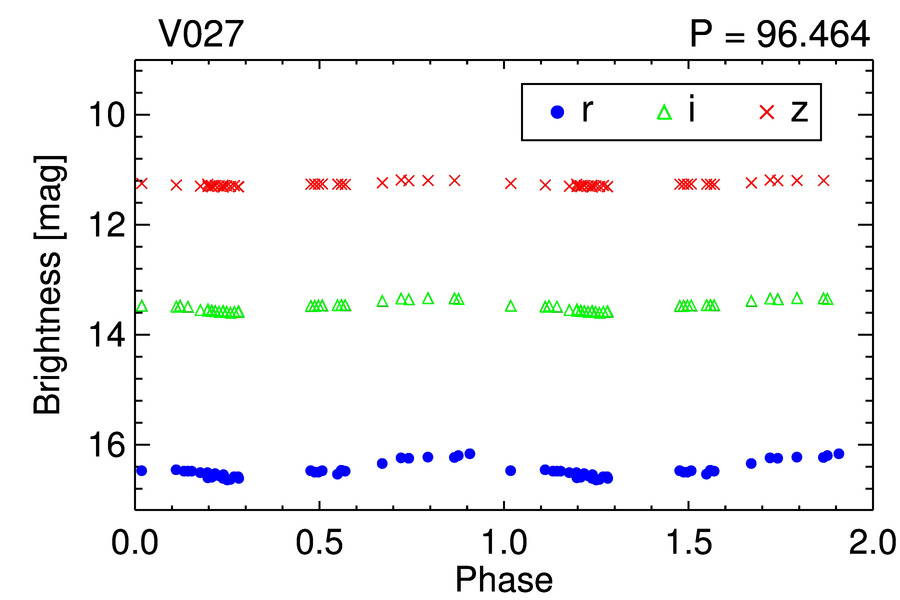}
\includegraphics[width=.24\textwidth]{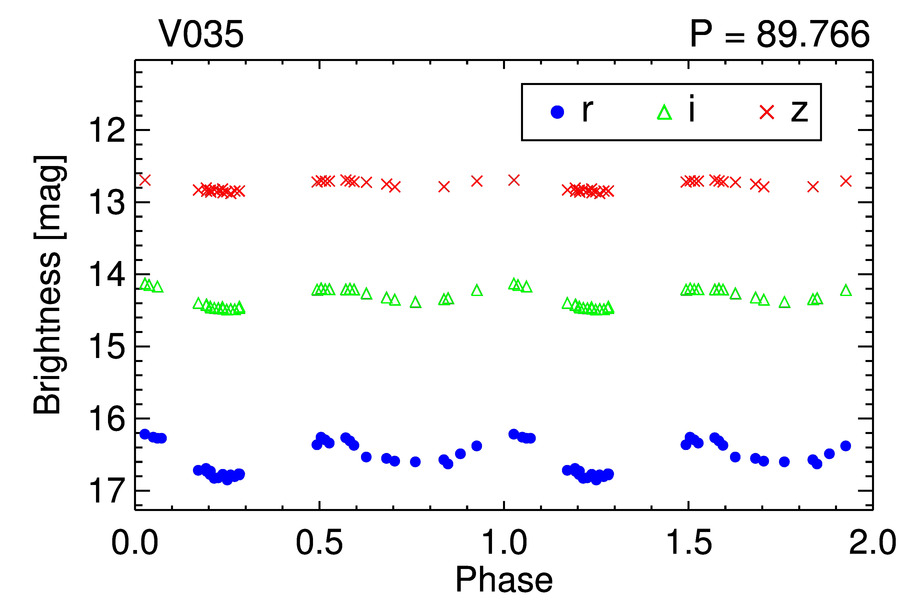}
\includegraphics[width=.24\textwidth]{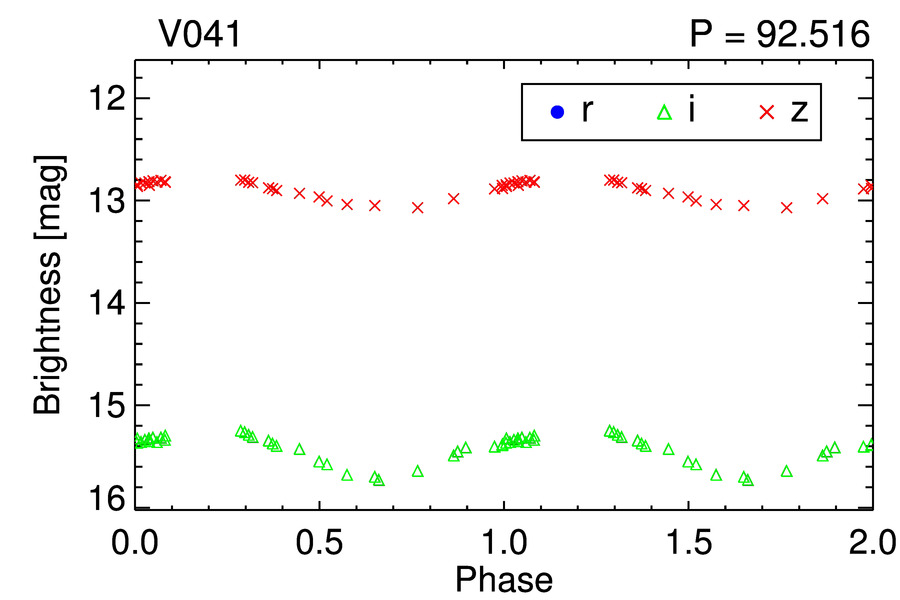}
\includegraphics[width=.24\textwidth]{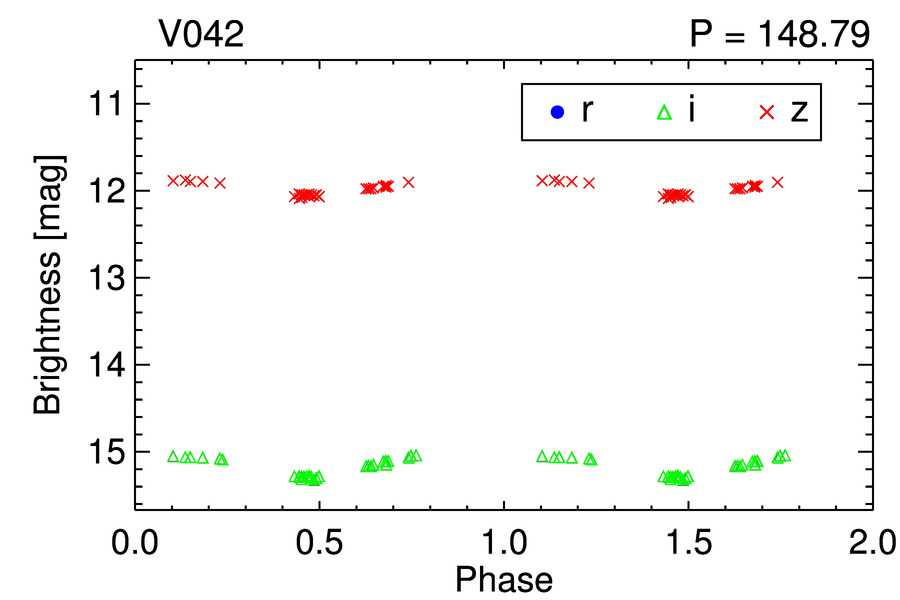}
\includegraphics[width=.24\textwidth]{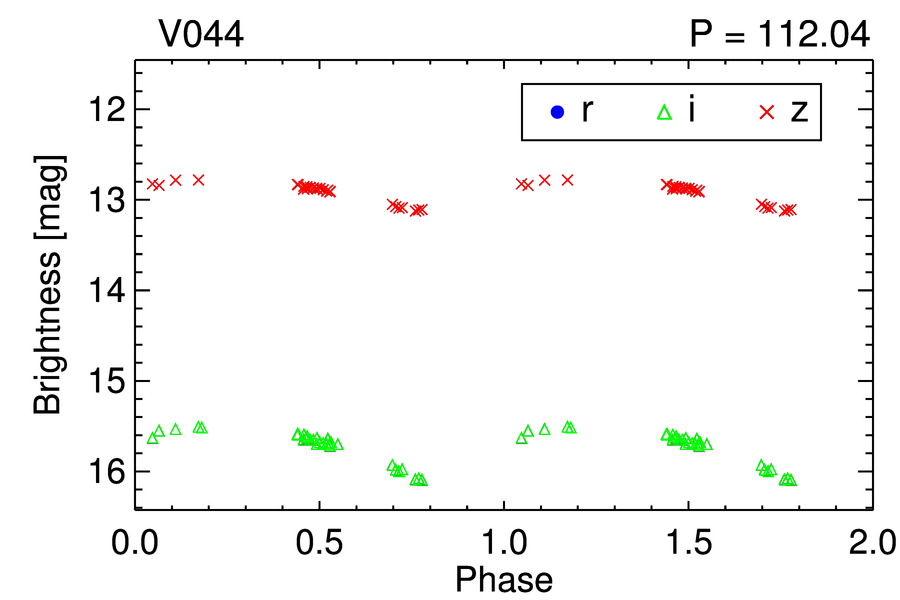}
\includegraphics[width=.24\textwidth]{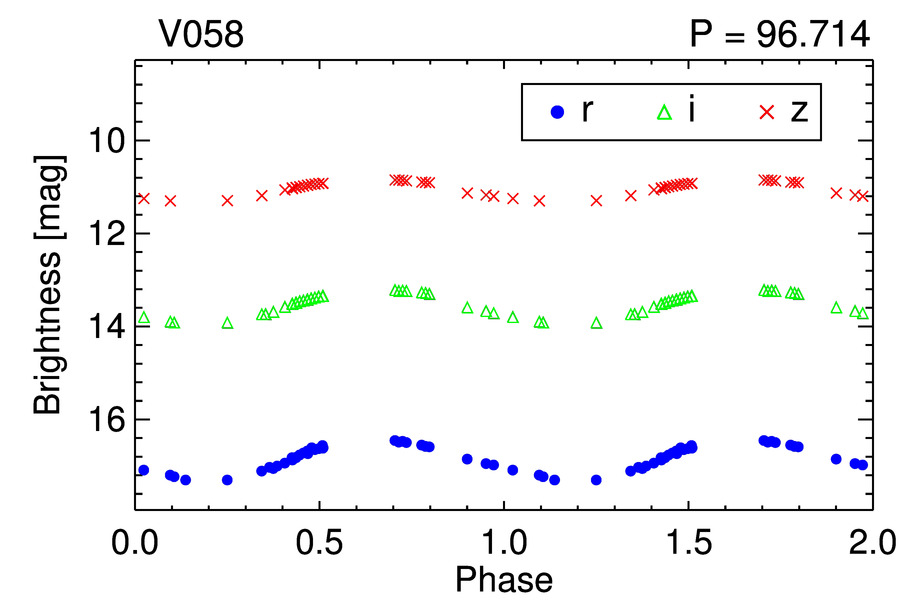}
\includegraphics[width=.24\textwidth]{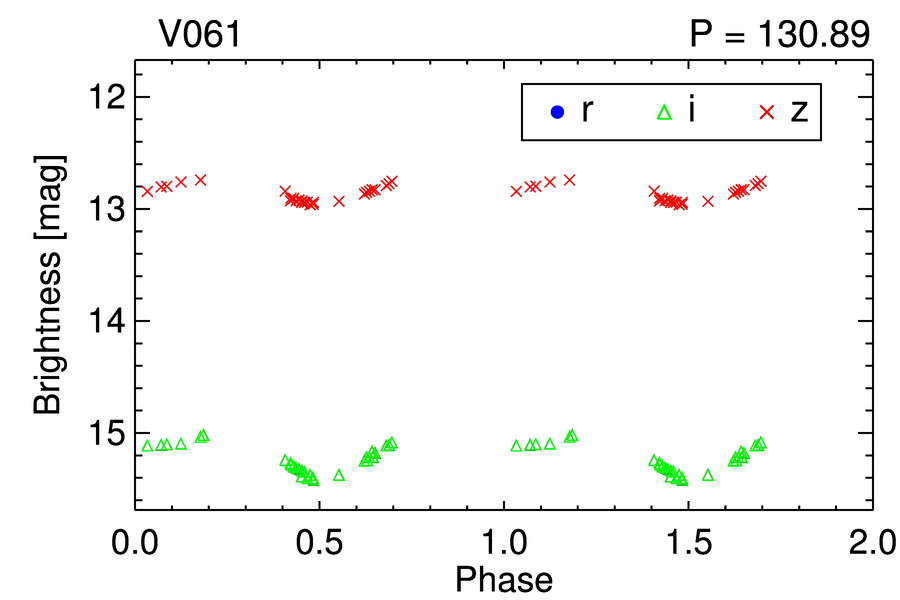}
\includegraphics[width=.24\textwidth]{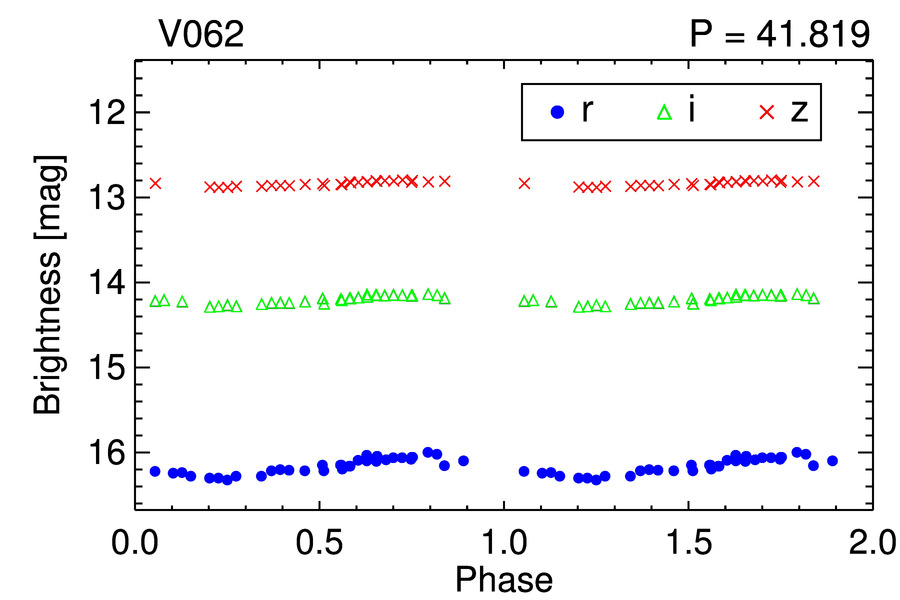}
\includegraphics[width=.24\textwidth]{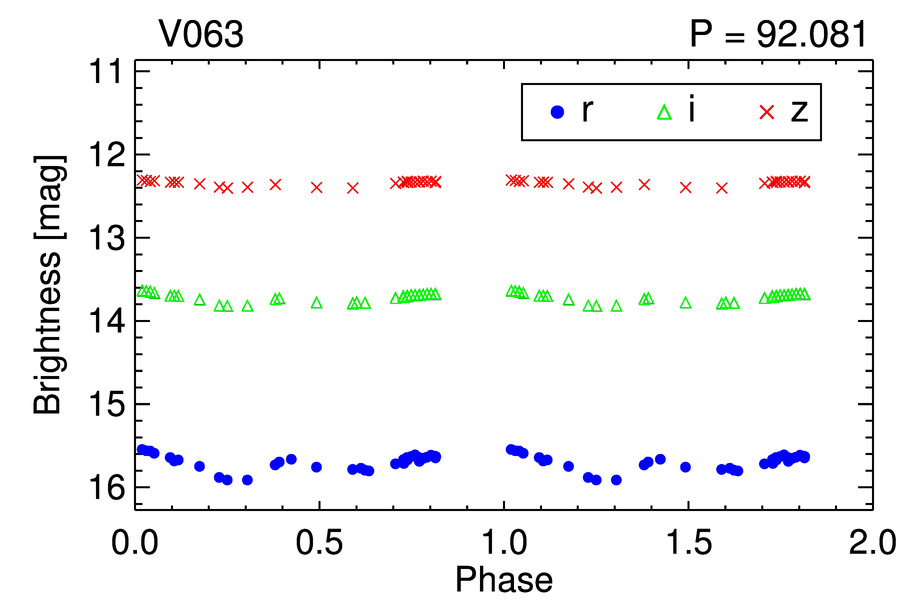}
\includegraphics[width=.24\textwidth]{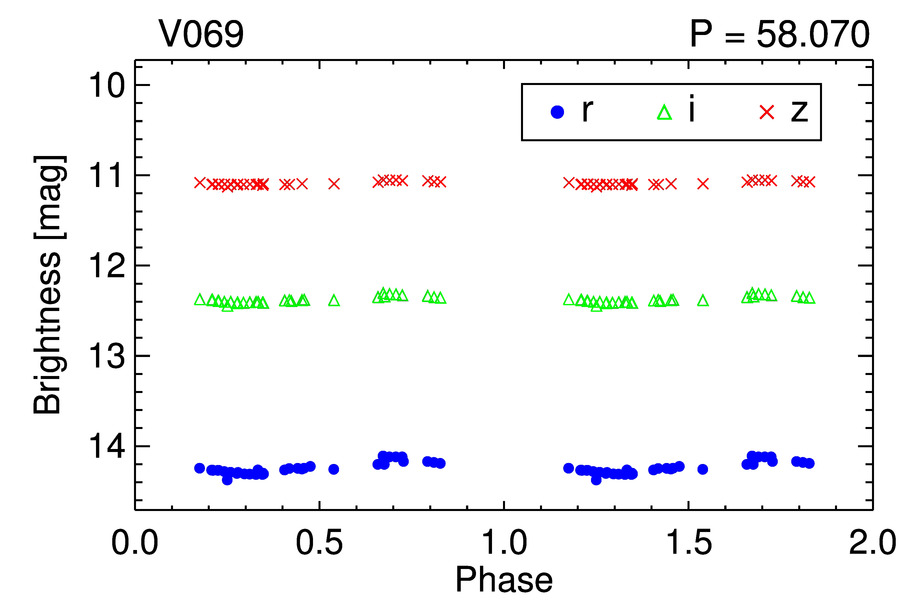}
\includegraphics[width=.24\textwidth]{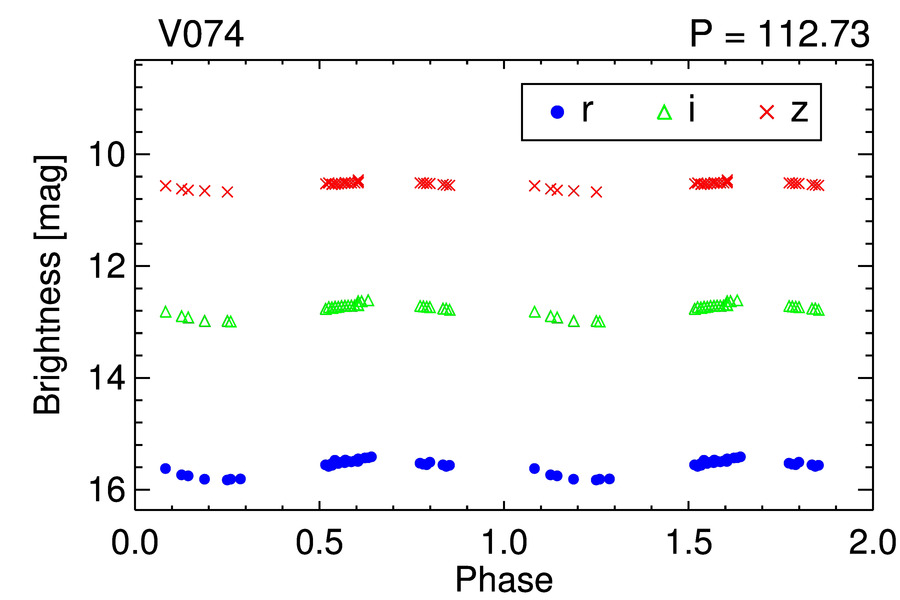}
\includegraphics[width=.24\textwidth]{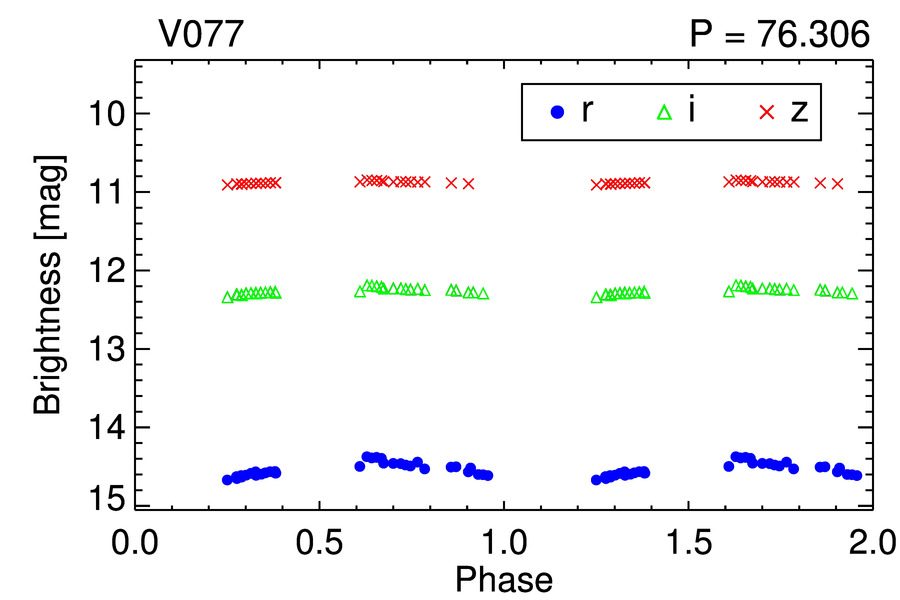}
\includegraphics[width=.24\textwidth]{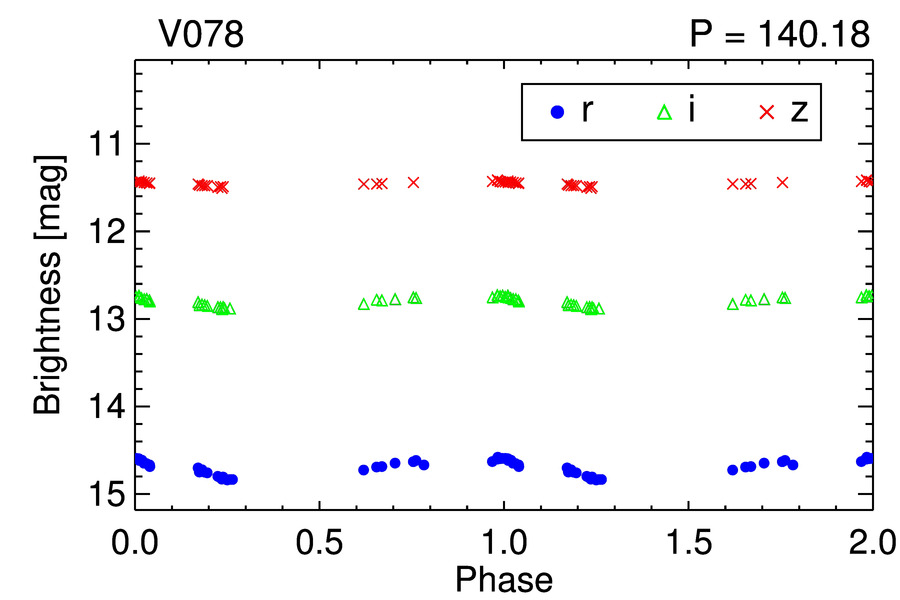}
\includegraphics[width=.24\textwidth]{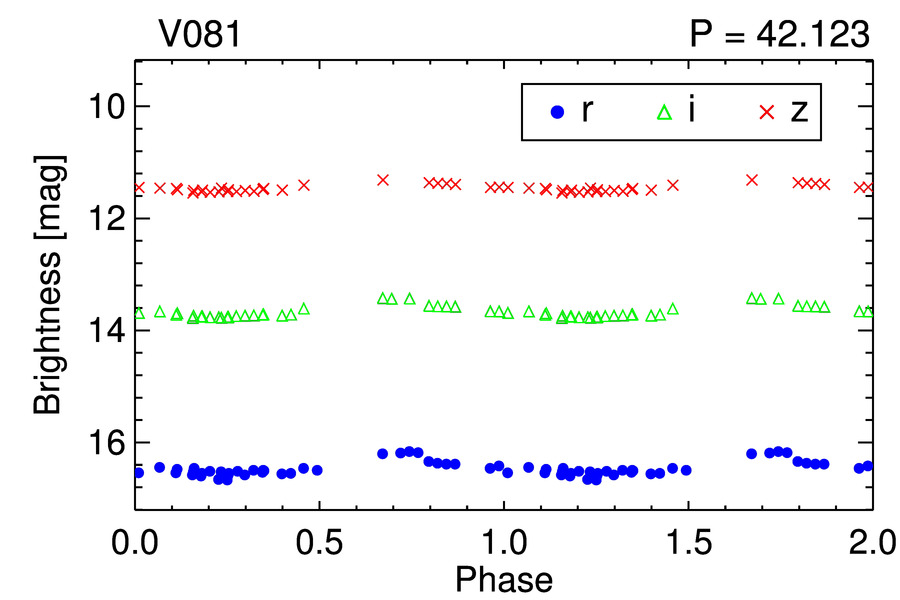}
\includegraphics[width=.24\textwidth]{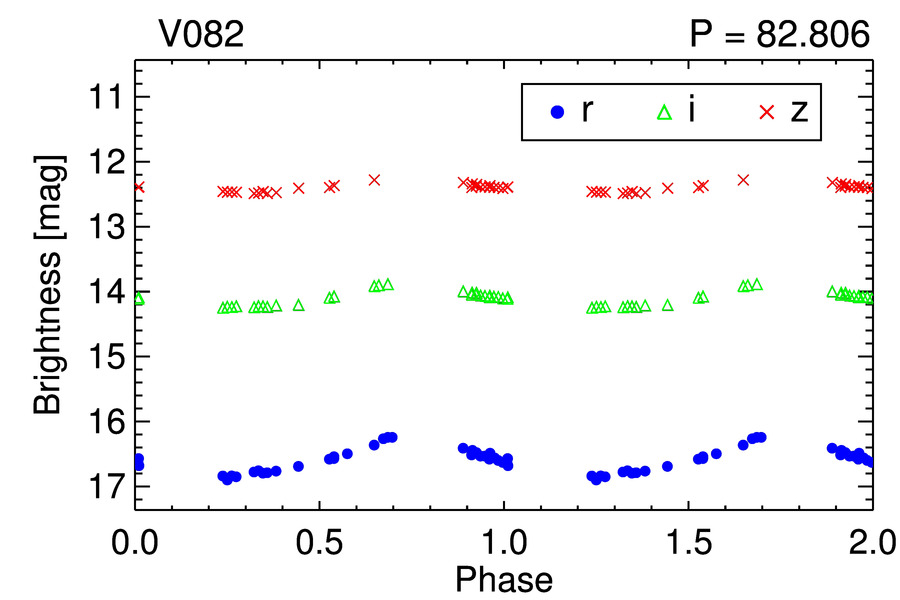}
\includegraphics[width=.24\textwidth]{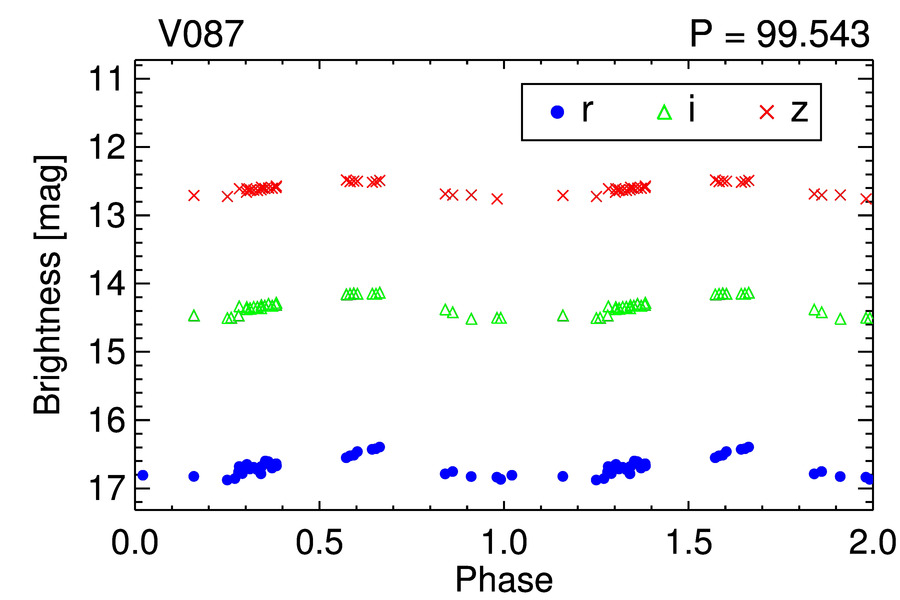}
\includegraphics[width=.24\textwidth]{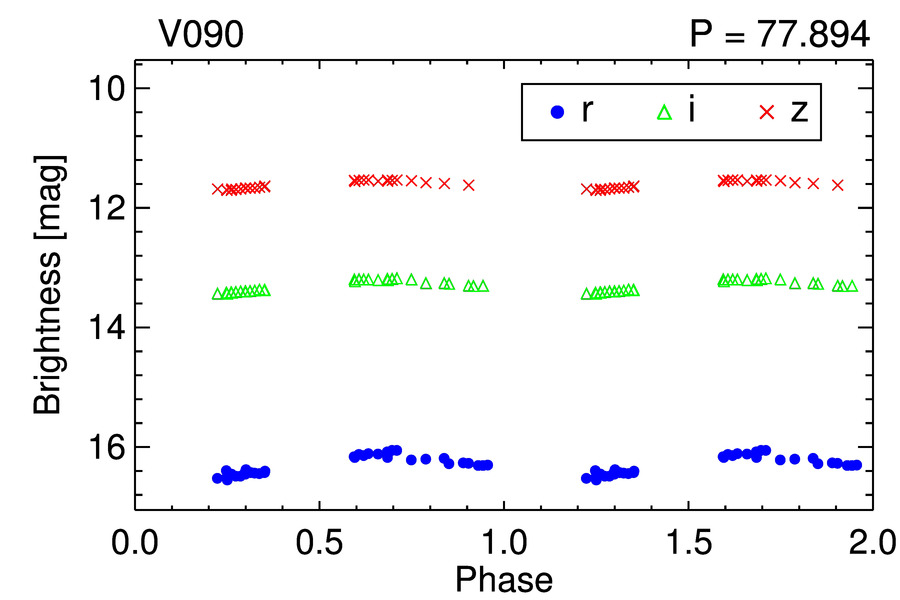}
\includegraphics[width=.24\textwidth]{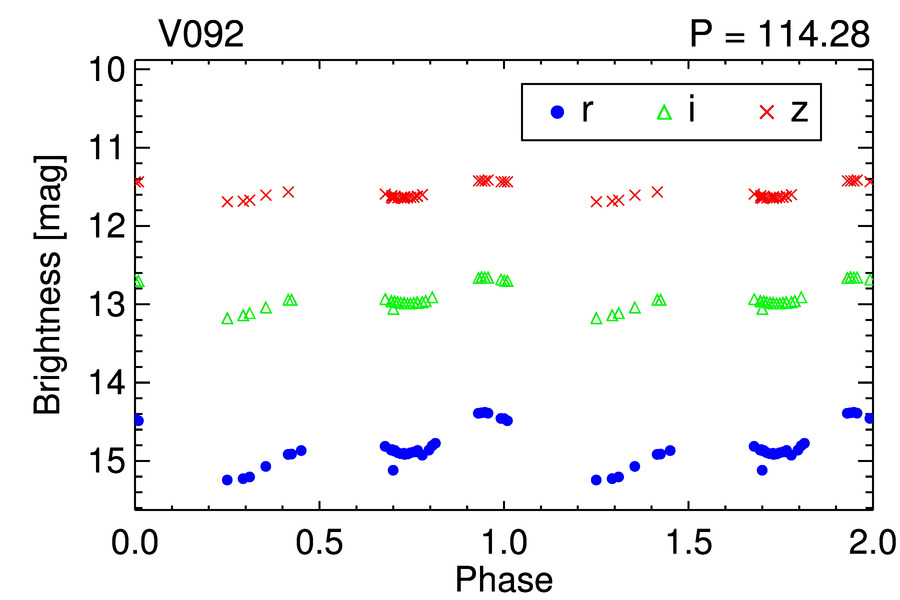}
\includegraphics[width=.24\textwidth]{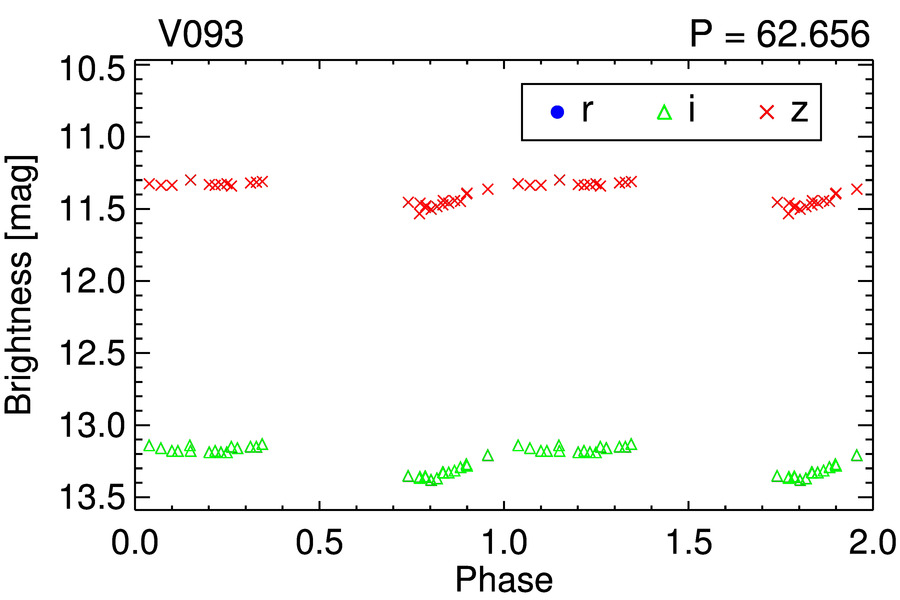}
\includegraphics[width=.24\textwidth]{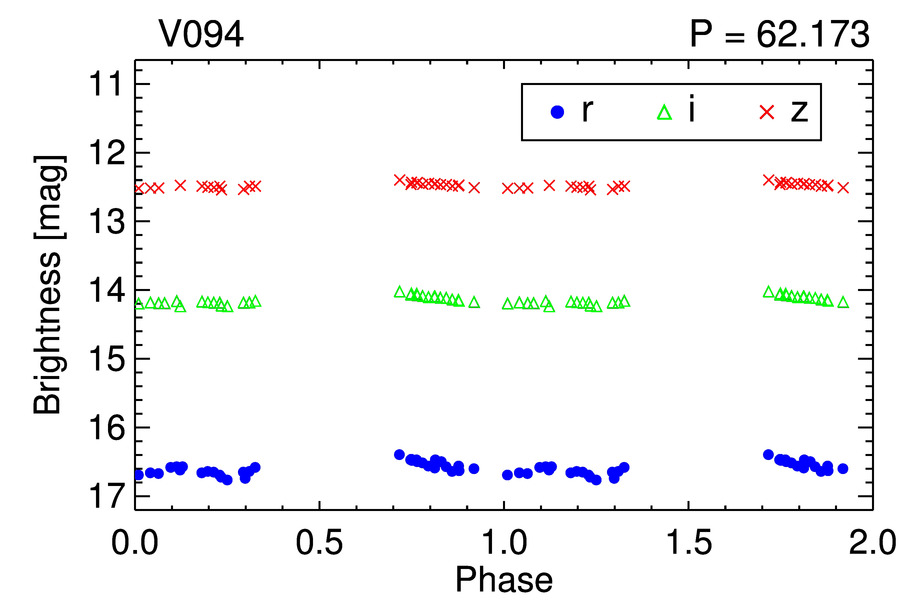}
\includegraphics[width=.24\textwidth]{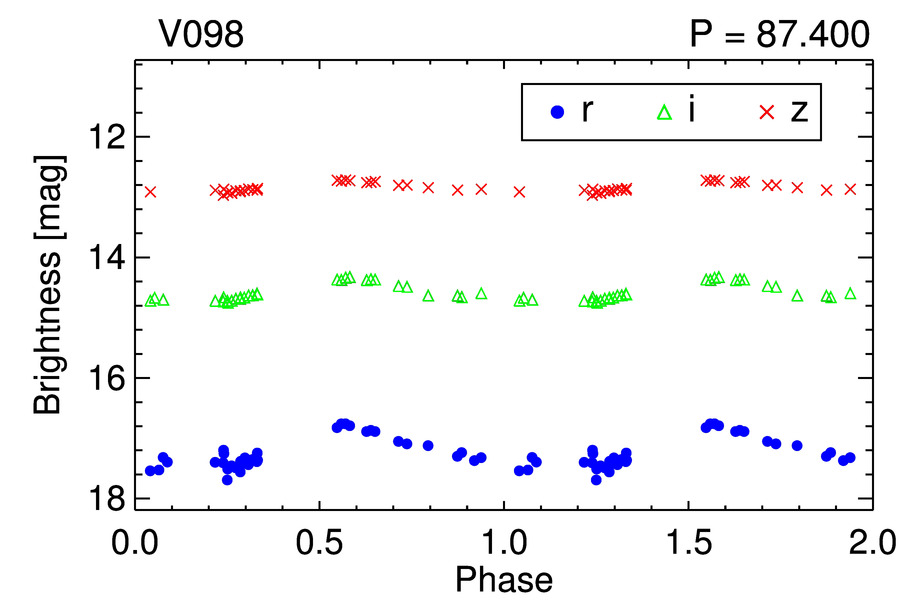}
\includegraphics[width=.24\textwidth]{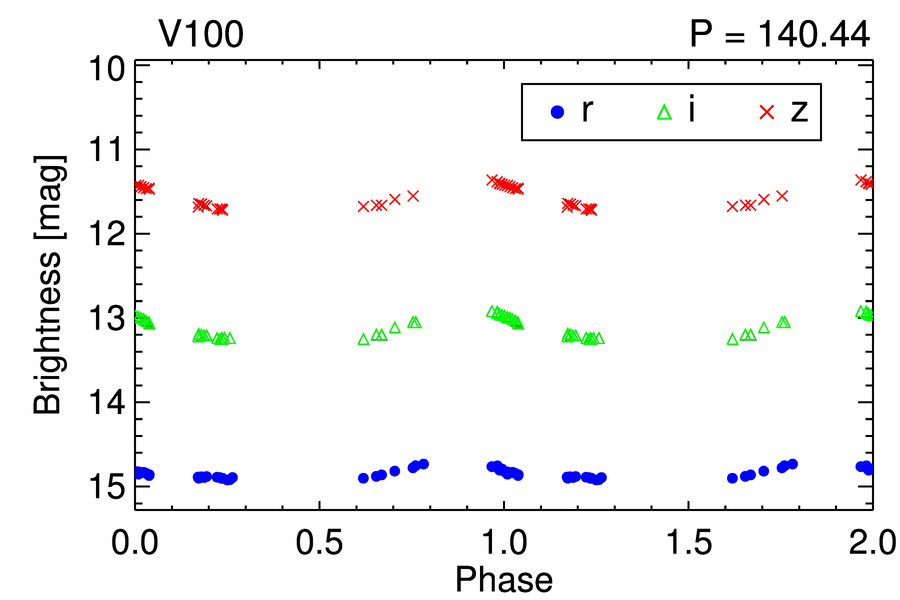}
\includegraphics[width=.24\textwidth]{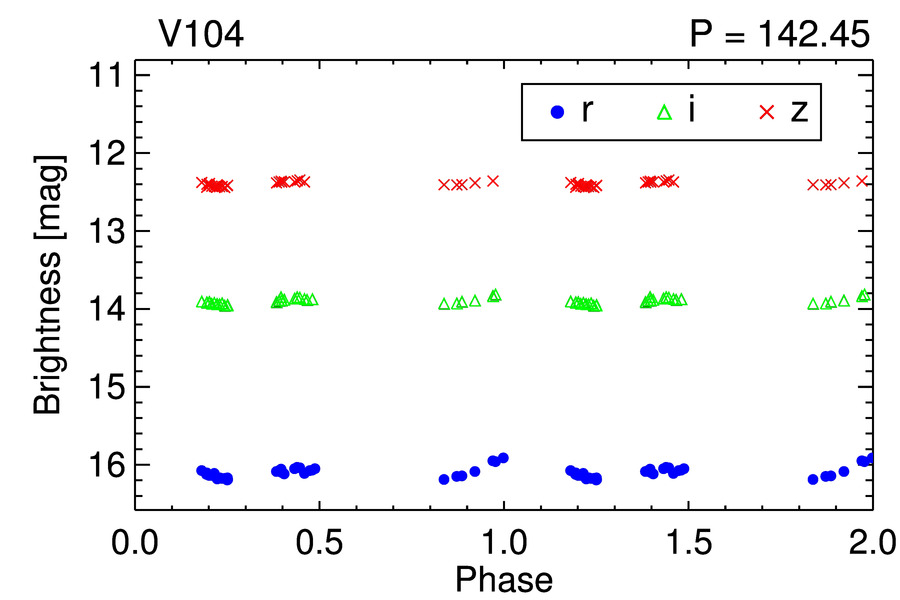}
\includegraphics[width=.24\textwidth]{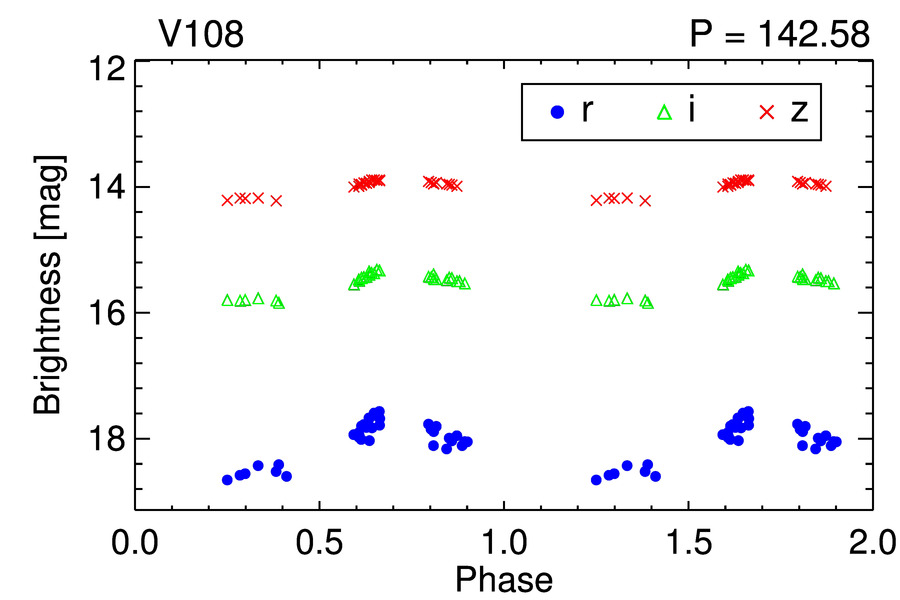}
\includegraphics[width=.24\textwidth]{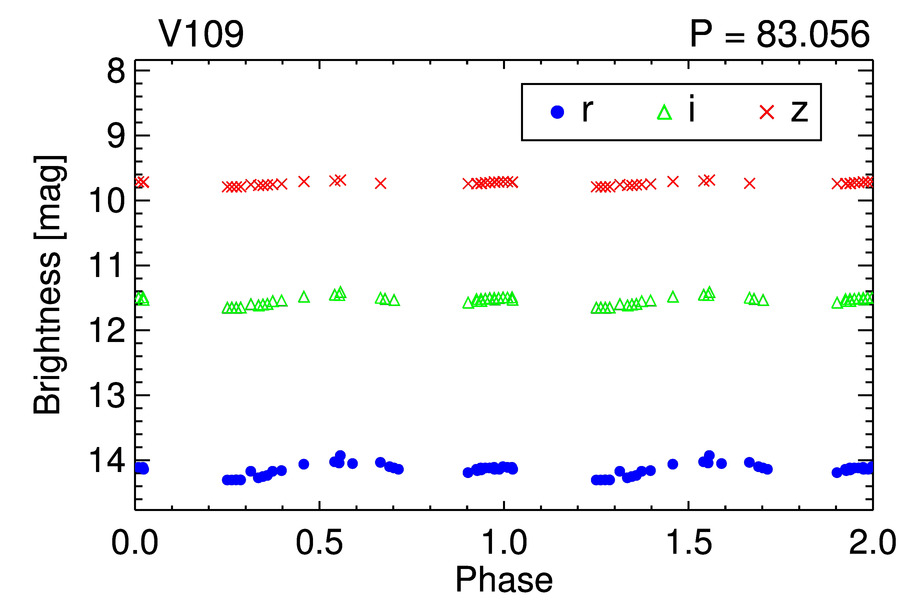}
      \caption{Light curves of the LP variables. The typical photometric error is less than 0.01\,mag.}\label{fig:lp_lc}
\end{figure}

\setcounter{figure}{1}
\renewcommand{\thefigure}{A\arabic{figure}}

\begin{figure}[!h]
 \centering
\includegraphics[width=.24\textwidth]{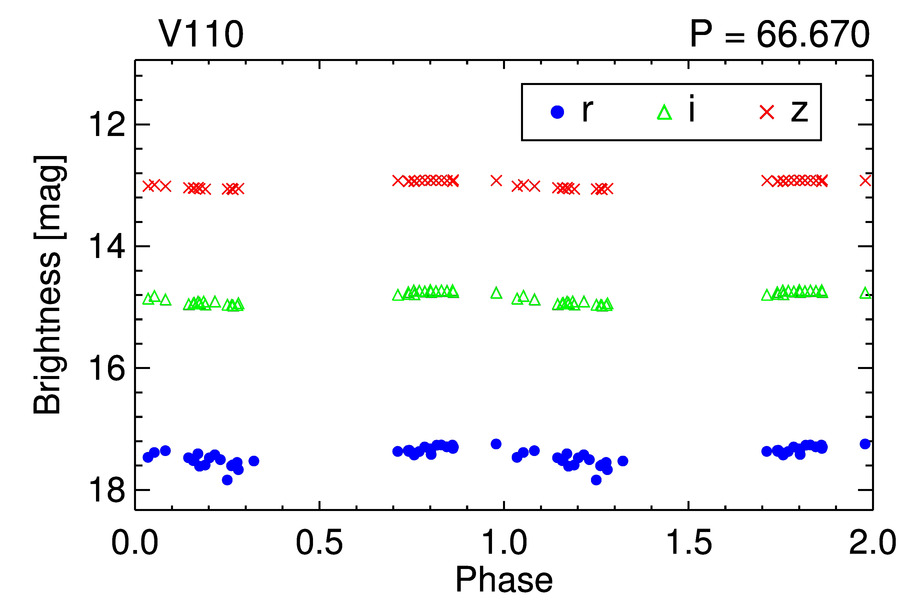}
\includegraphics[width=.24\textwidth]{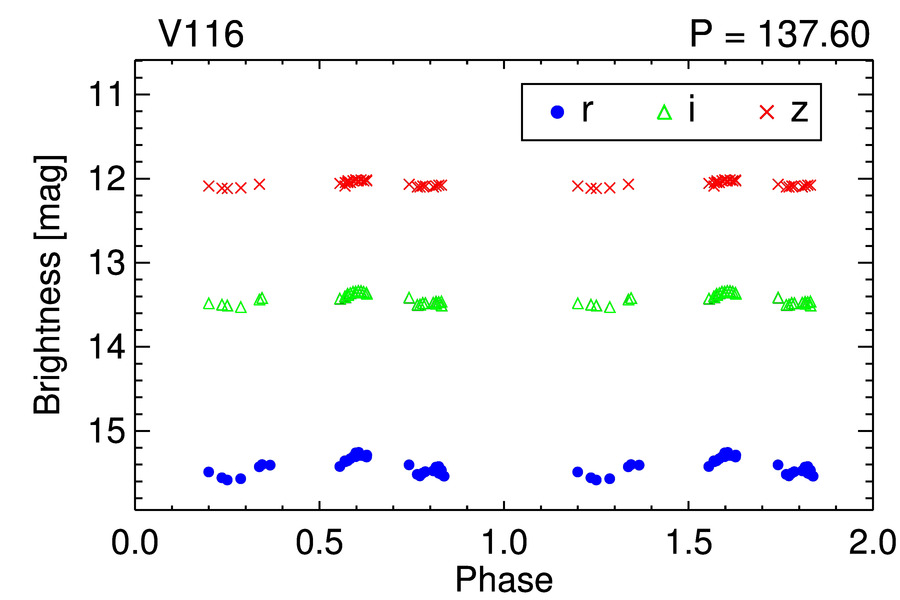}
\includegraphics[width=.24\textwidth]{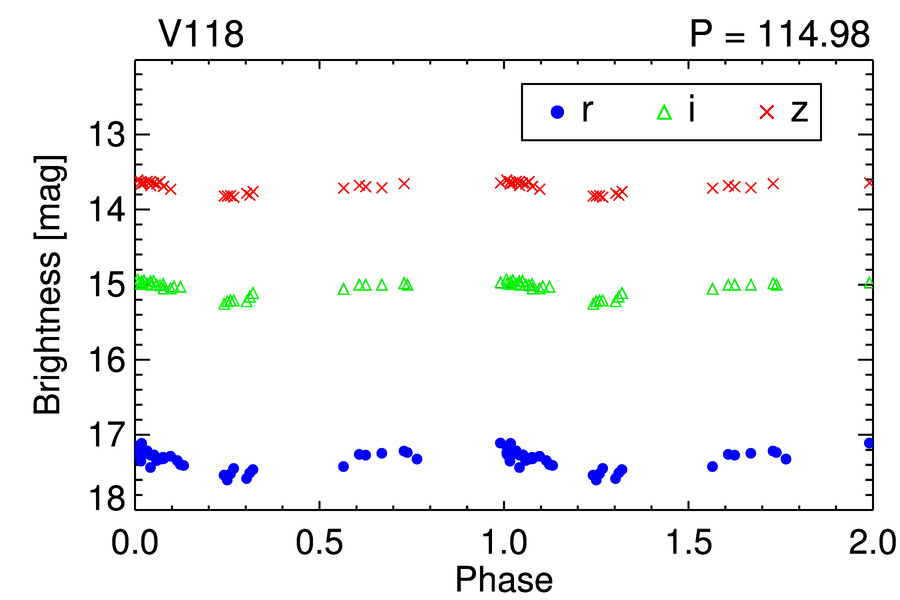}
\includegraphics[width=.24\textwidth]{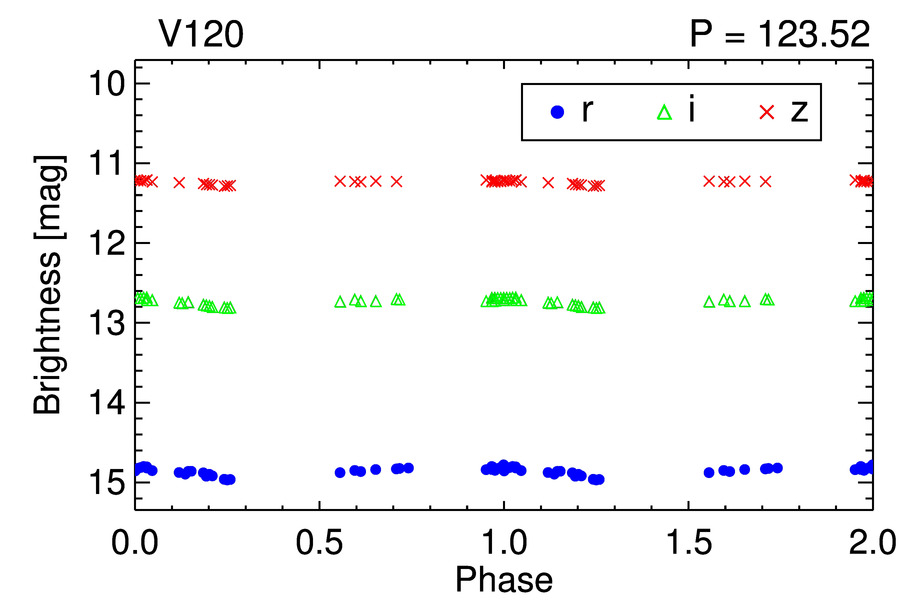}
\includegraphics[width=.24\textwidth]{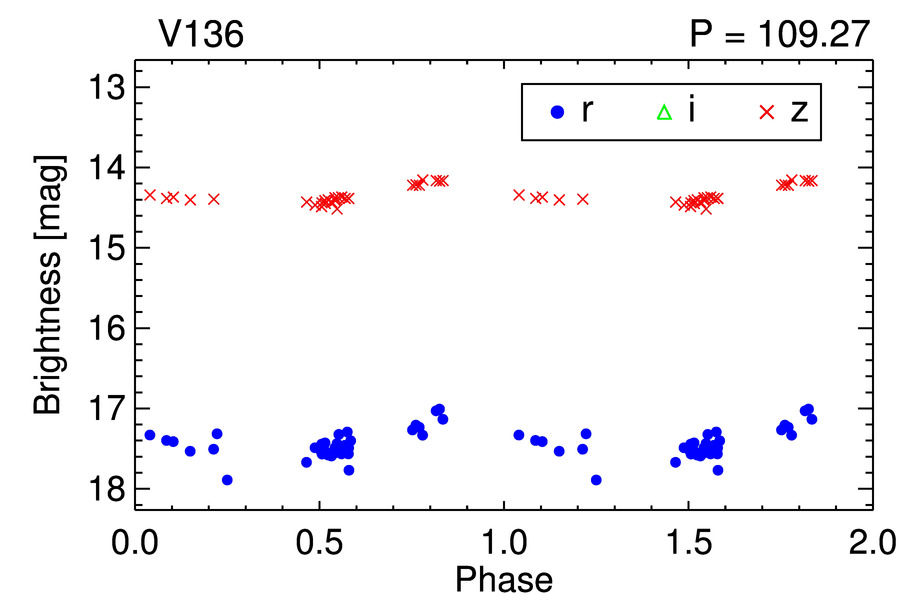}
\includegraphics[width=.24\textwidth]{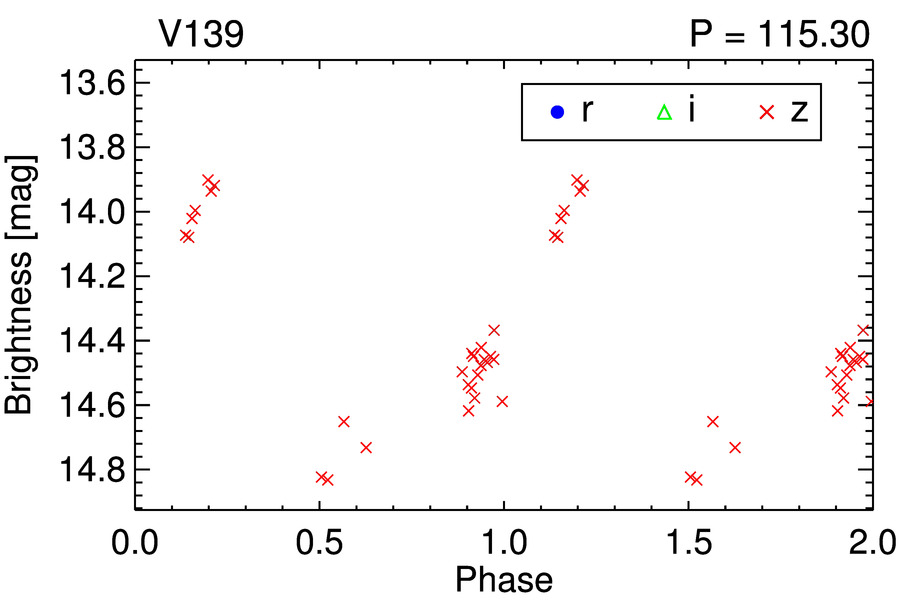}
\includegraphics[width=.24\textwidth]{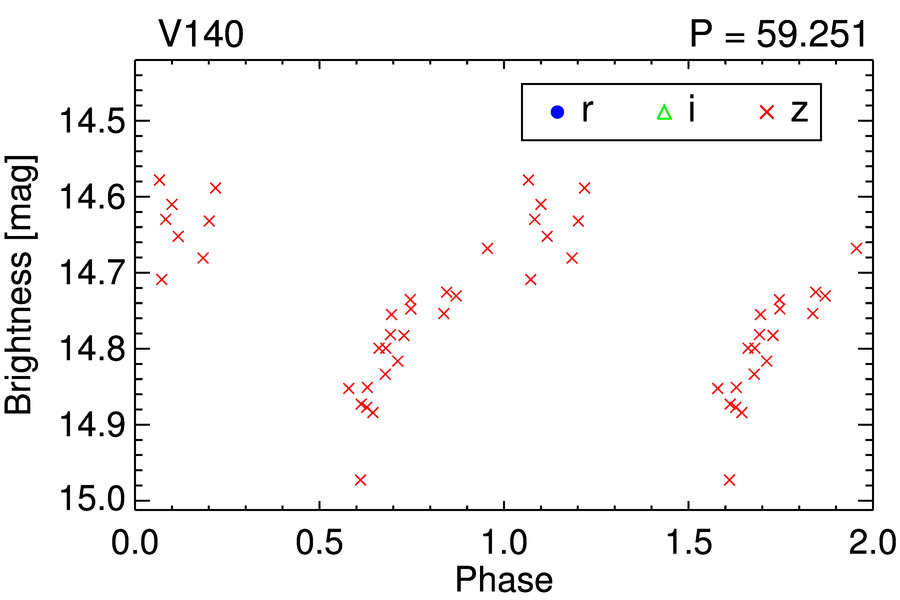}
      \caption{(Continued)}
\end{figure}

\setcounter{figure}{2}
\renewcommand{\thefigure}{A\arabic{figure}}

\begin{figure}[!h]
 \centering
     \includegraphics[width=.24\textwidth]{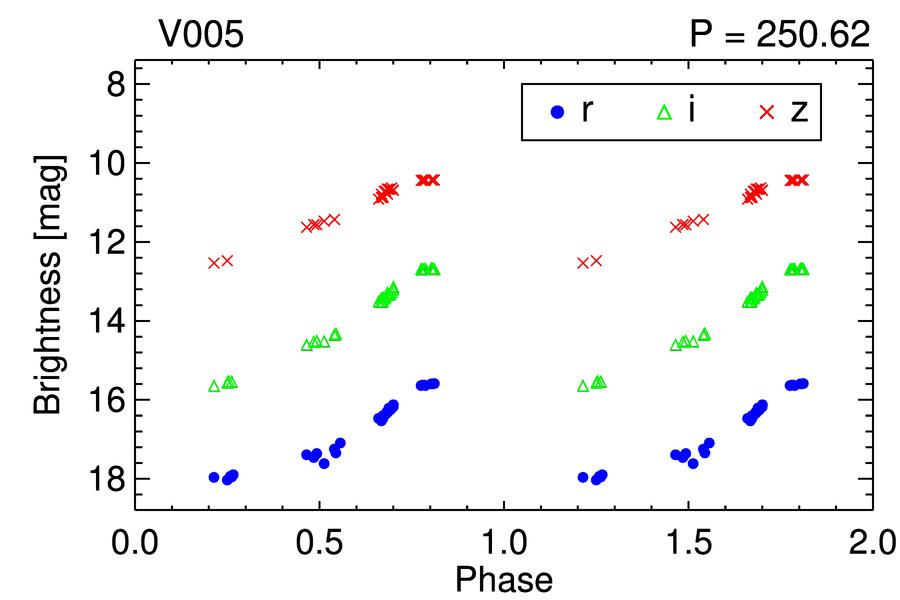}
\includegraphics[width=.24\textwidth]{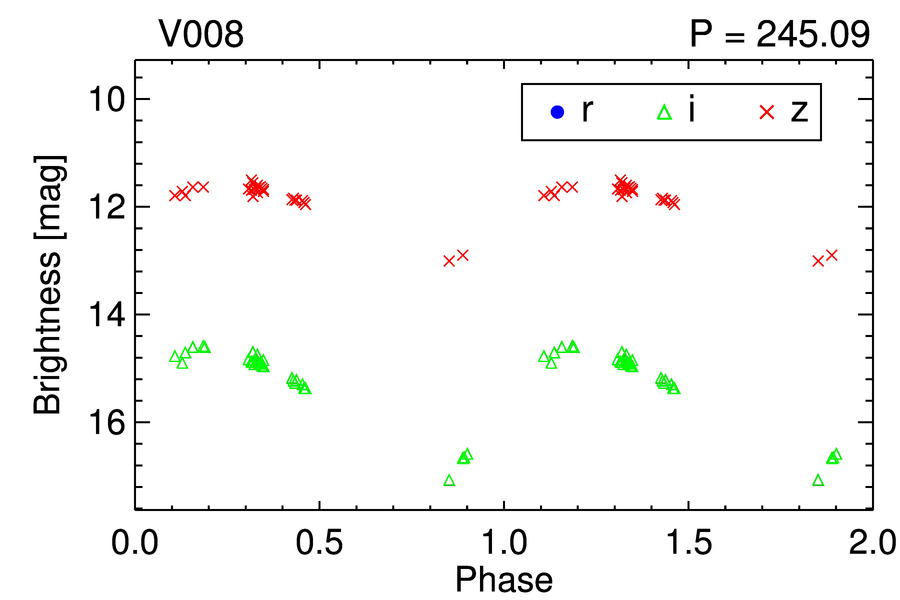}
\includegraphics[width=.24\textwidth]{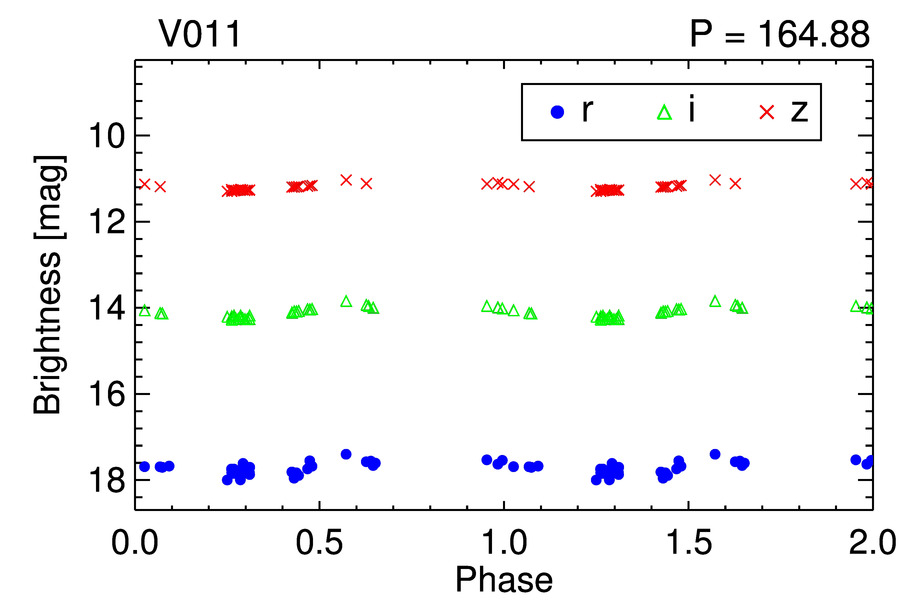}
\includegraphics[width=.24\textwidth]{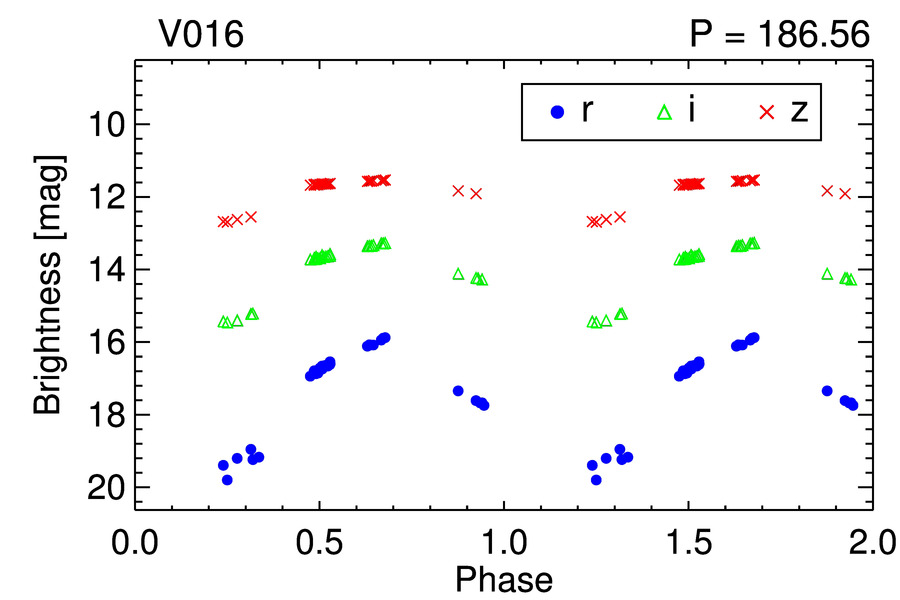}
\includegraphics[width=.24\textwidth]{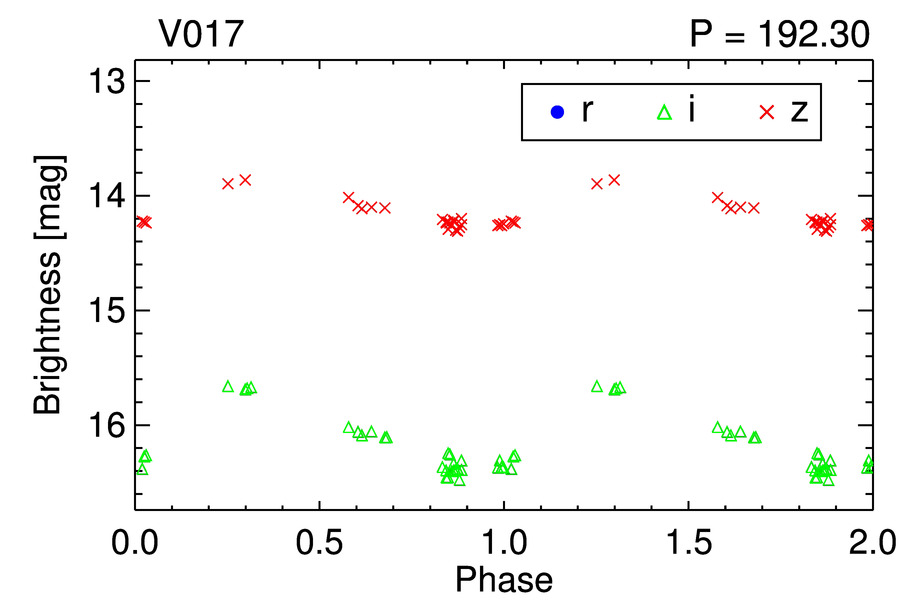}
\includegraphics[width=.24\textwidth]{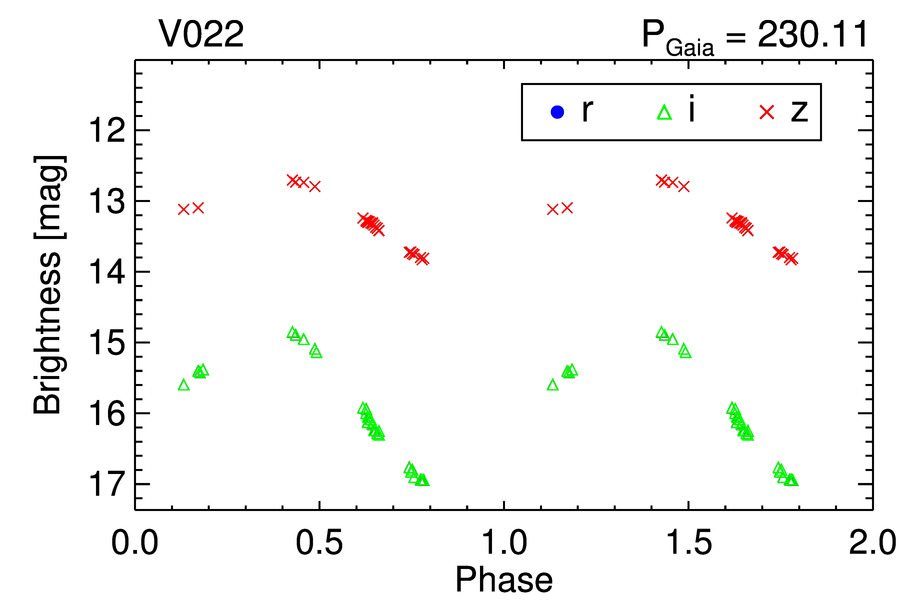}
\includegraphics[width=.24\textwidth]{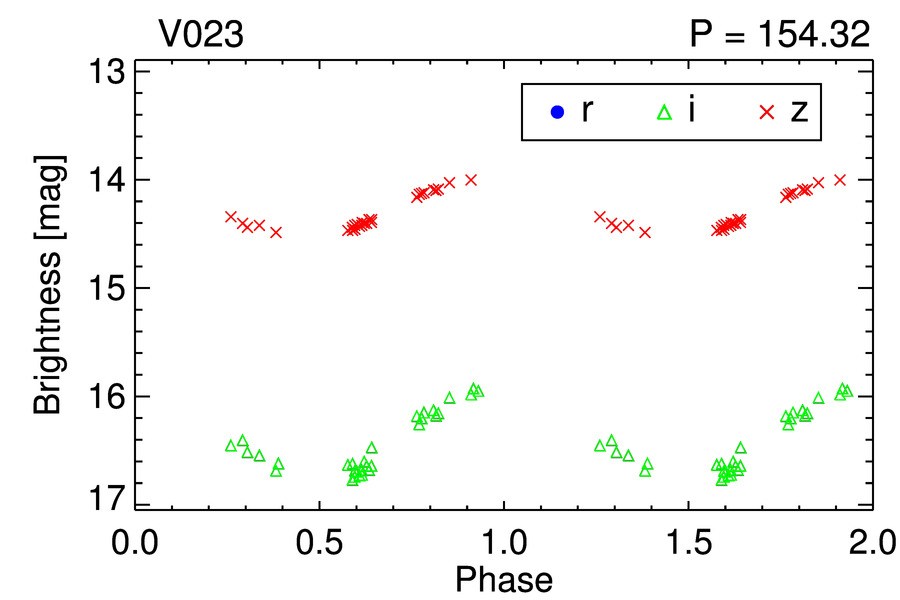}
\includegraphics[width=.24\textwidth]{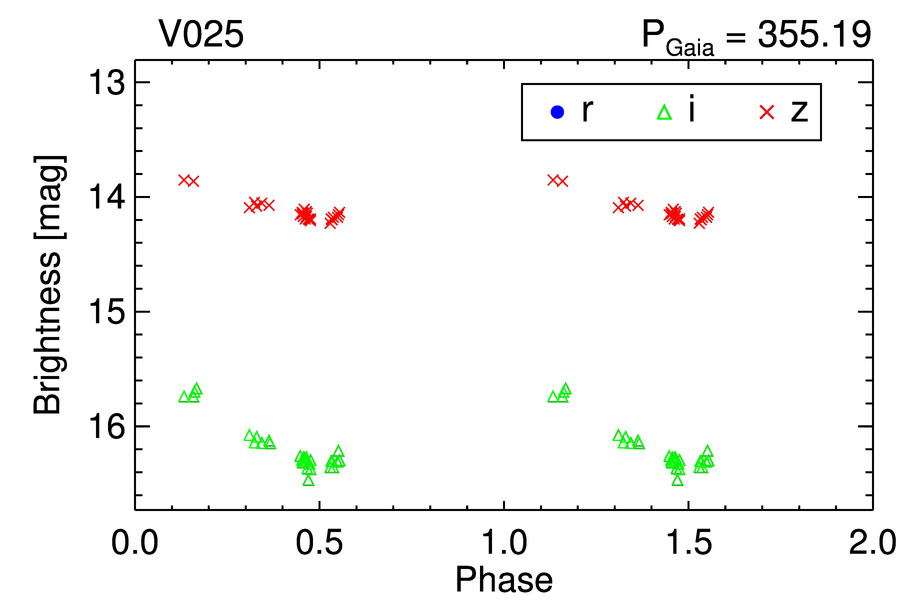}
\includegraphics[width=.24\textwidth]{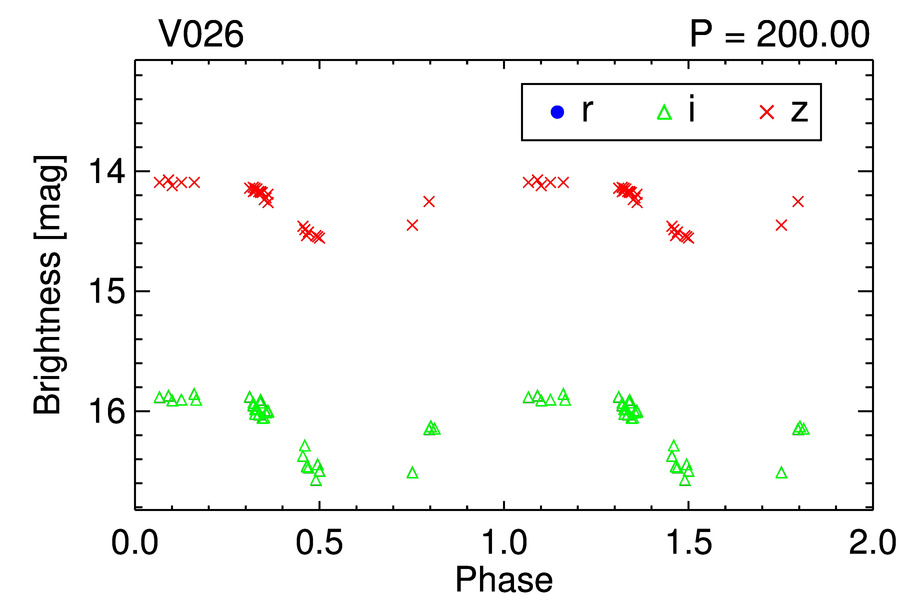}
\includegraphics[width=.24\textwidth]{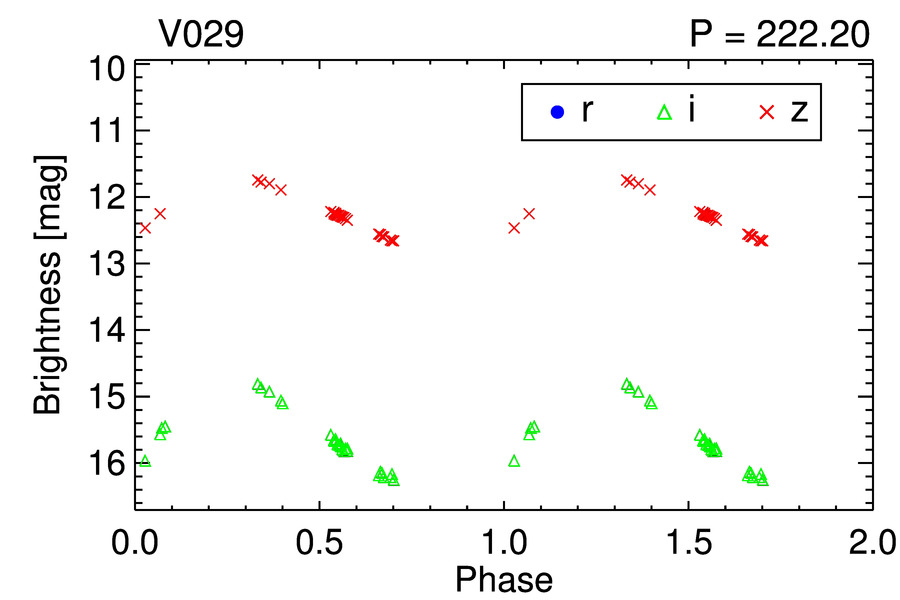}
\includegraphics[width=.24\textwidth]{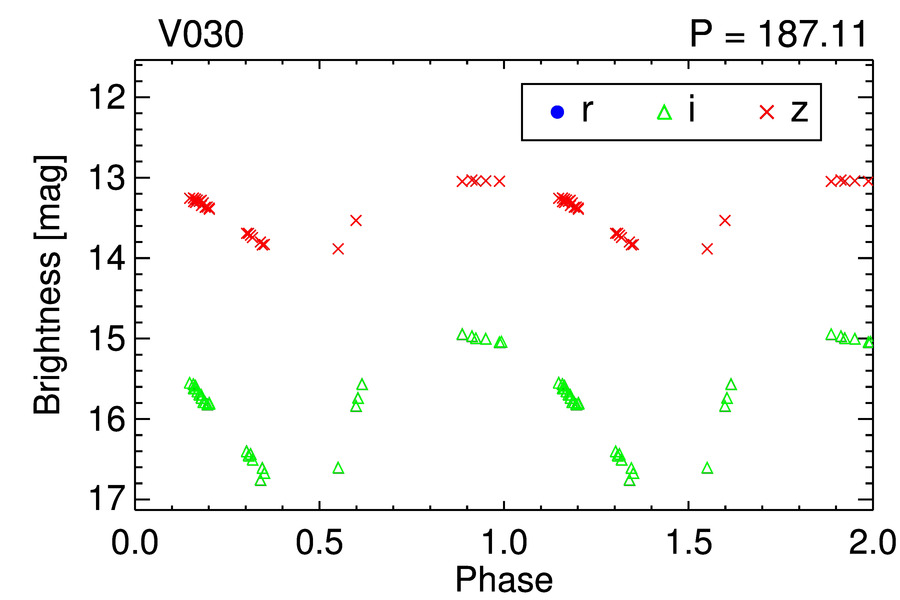}
\includegraphics[width=.24\textwidth]{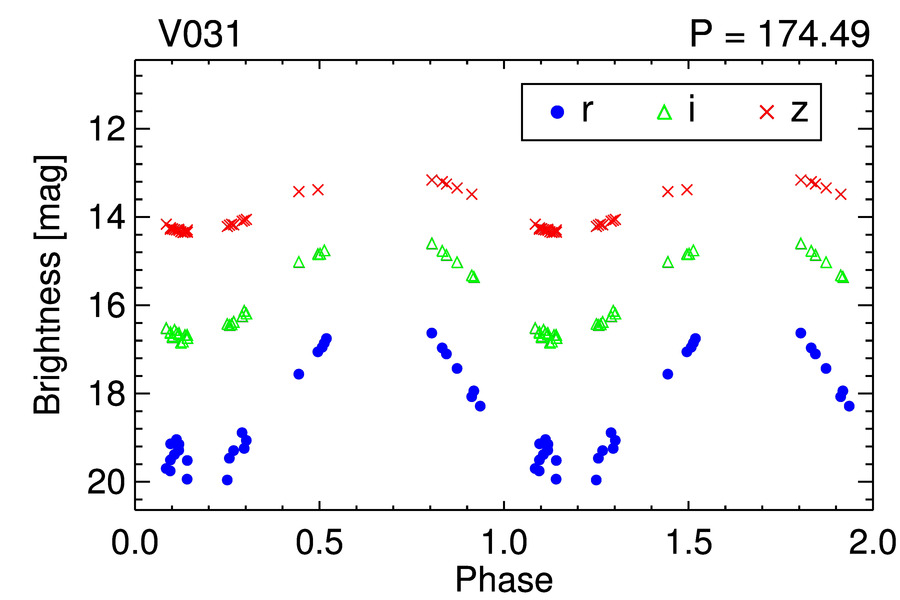}
\includegraphics[width=.24\textwidth]{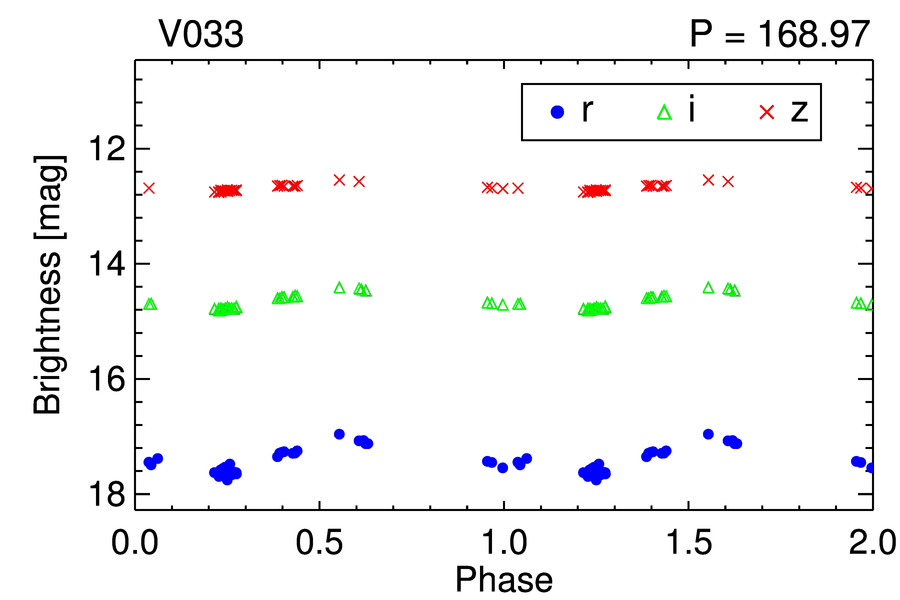}
\includegraphics[width=.24\textwidth]{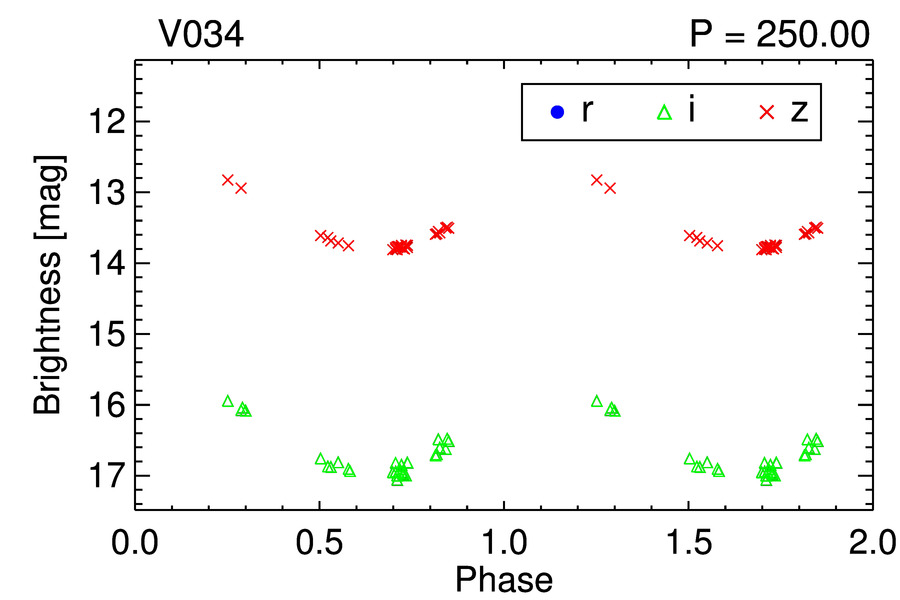}
\includegraphics[width=.24\textwidth]{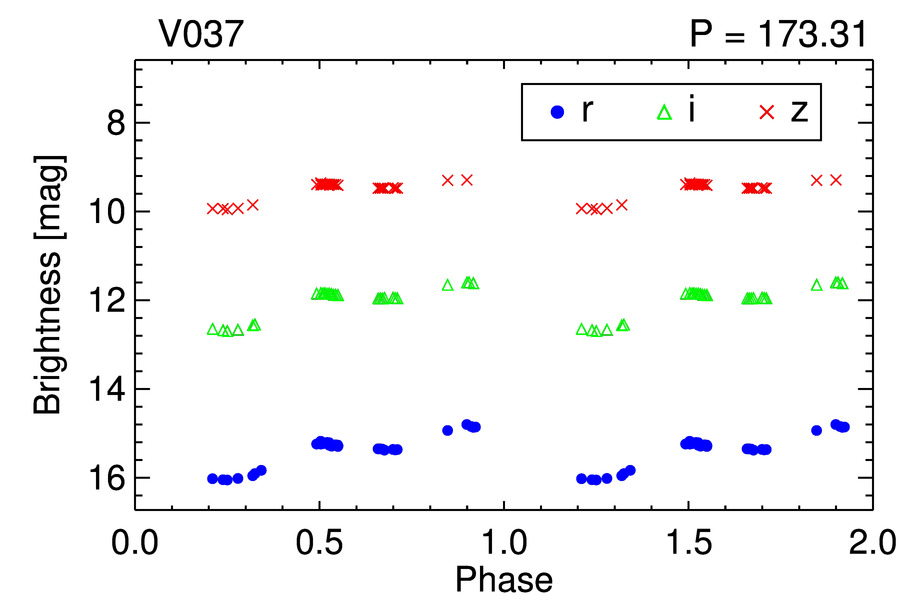}
\includegraphics[width=.24\textwidth]{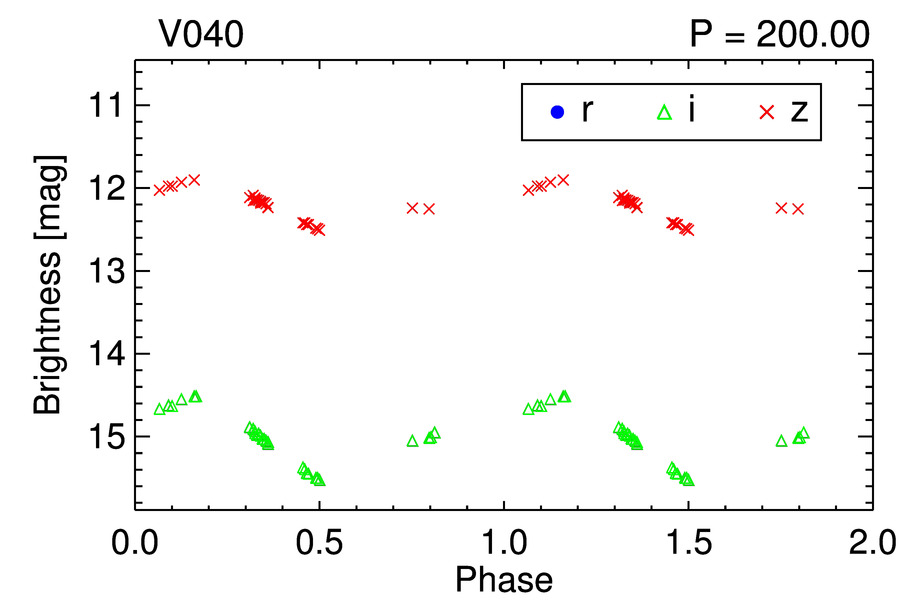}
\includegraphics[width=.24\textwidth]{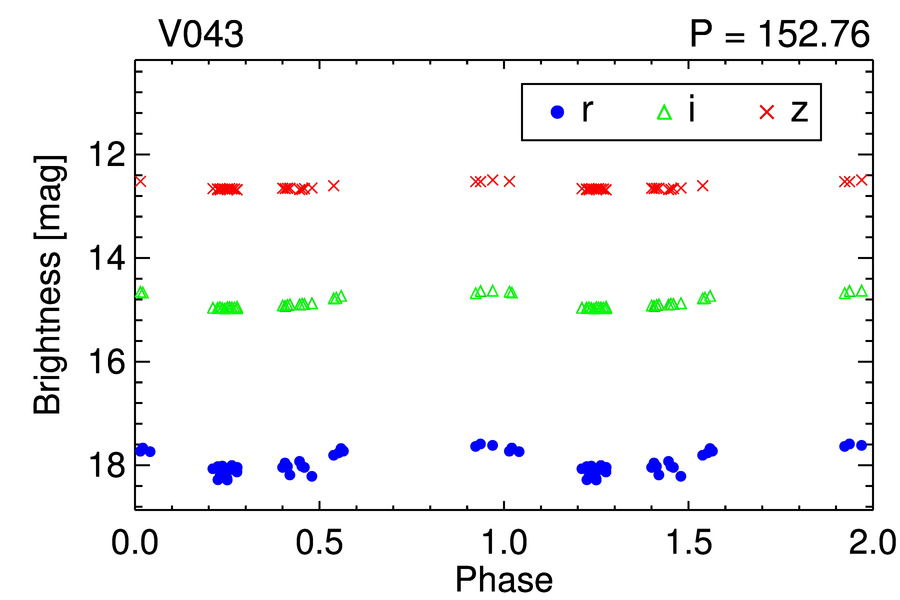}
\includegraphics[width=.24\textwidth]{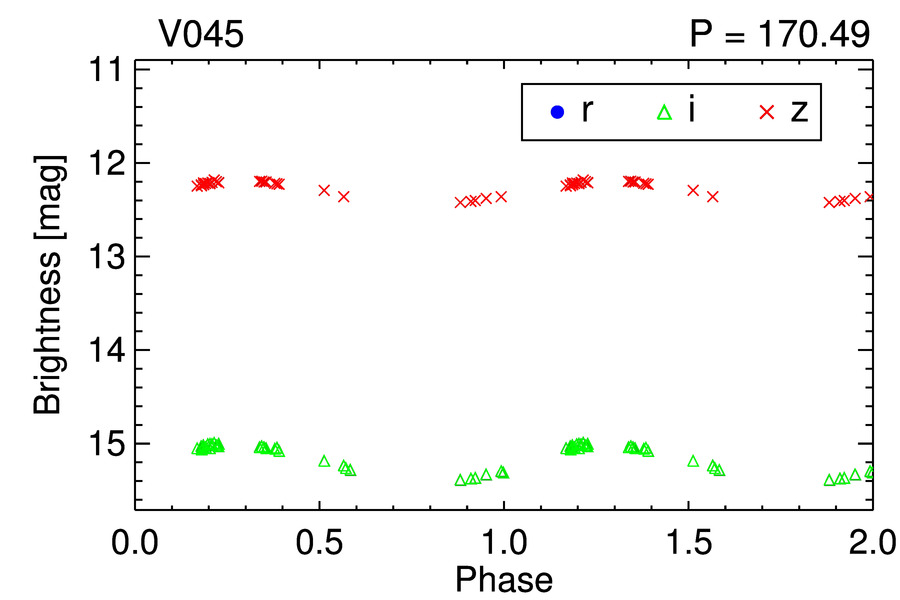}
\includegraphics[width=.24\textwidth]{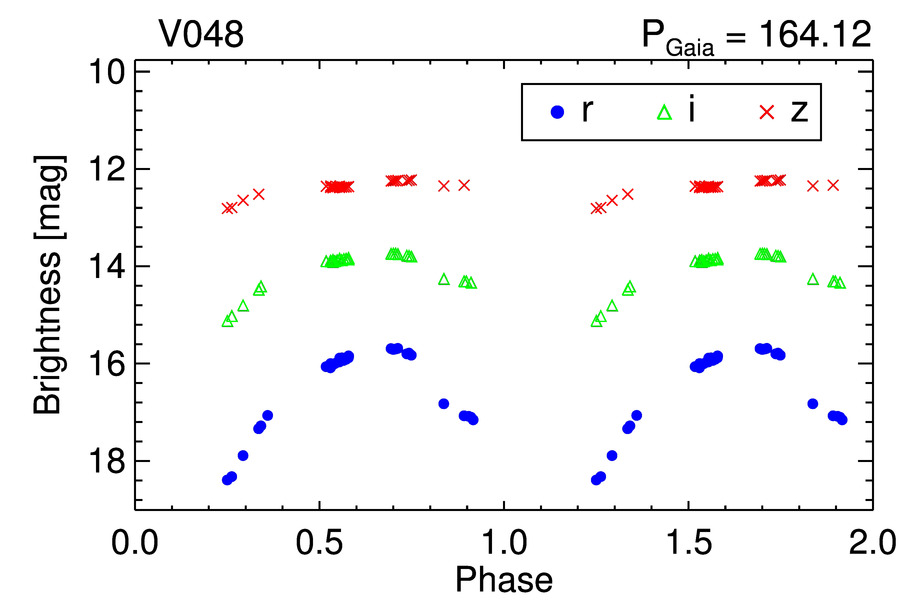}
\includegraphics[width=.24\textwidth]{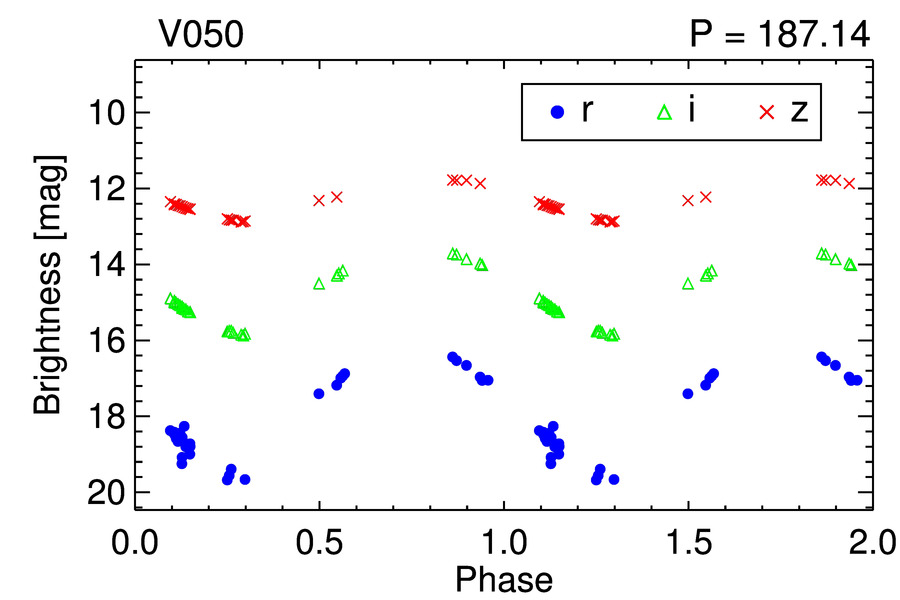}
      \caption{Light curves of the SR\_M variables. The typical photometric error is less than 0.01\,mag.}\label{fig:sr_lc}
\end{figure}

\setcounter{figure}{2}
\renewcommand{\thefigure}{A\arabic{figure}}

\begin{figure}[!h]
 \centering
\includegraphics[width=.24\textwidth]{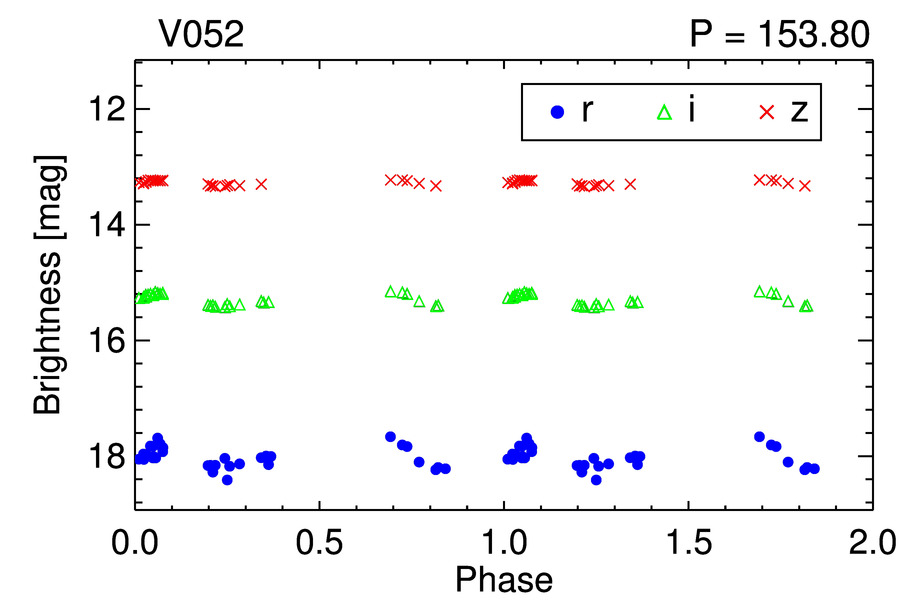}
\includegraphics[width=.24\textwidth]{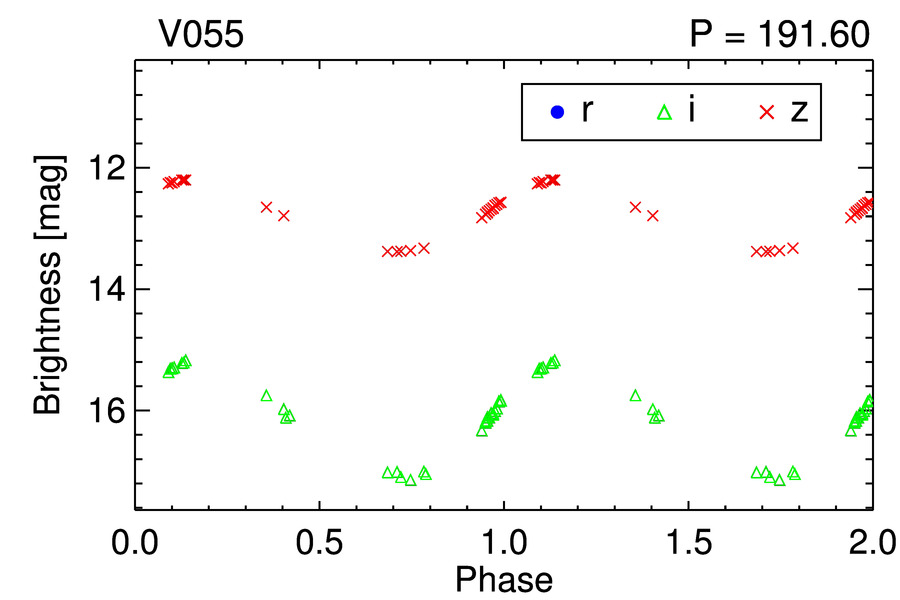}
\includegraphics[width=.24\textwidth]{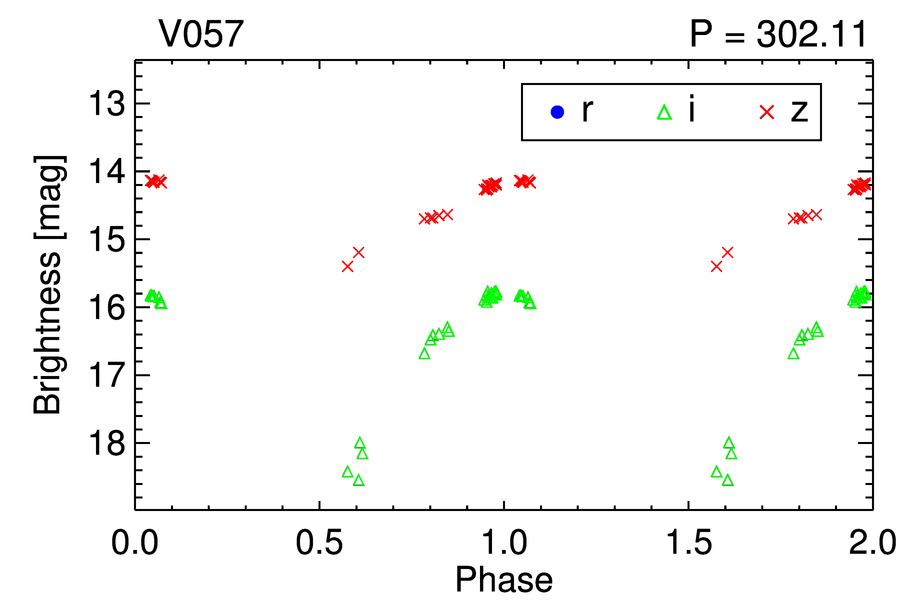}
\includegraphics[width=.24\textwidth]{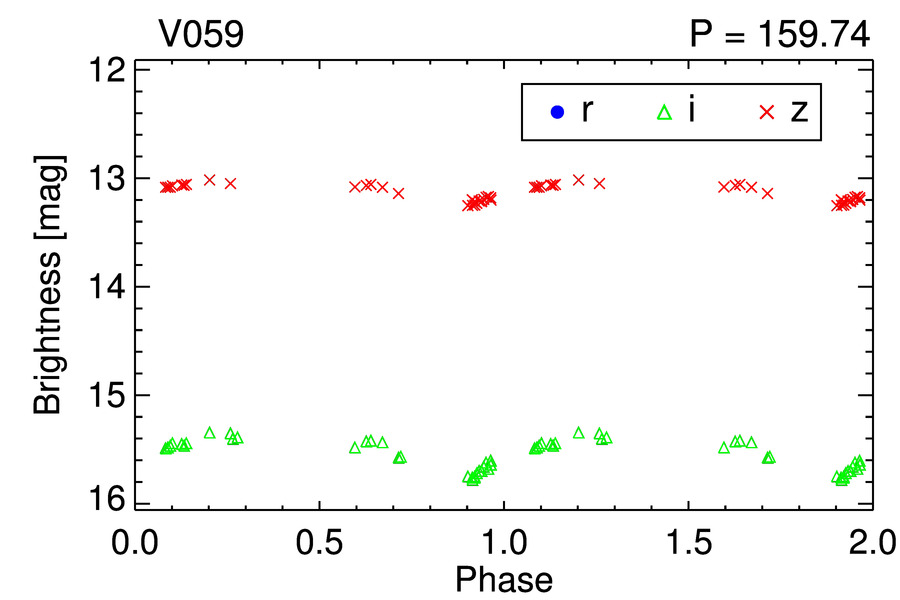}
\includegraphics[width=.24\textwidth]{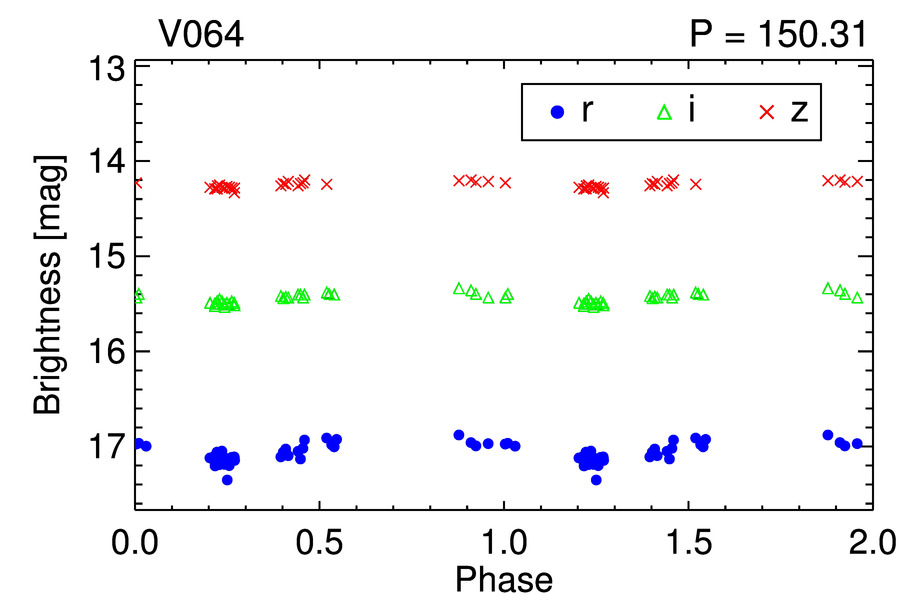}
\includegraphics[width=.24\textwidth]{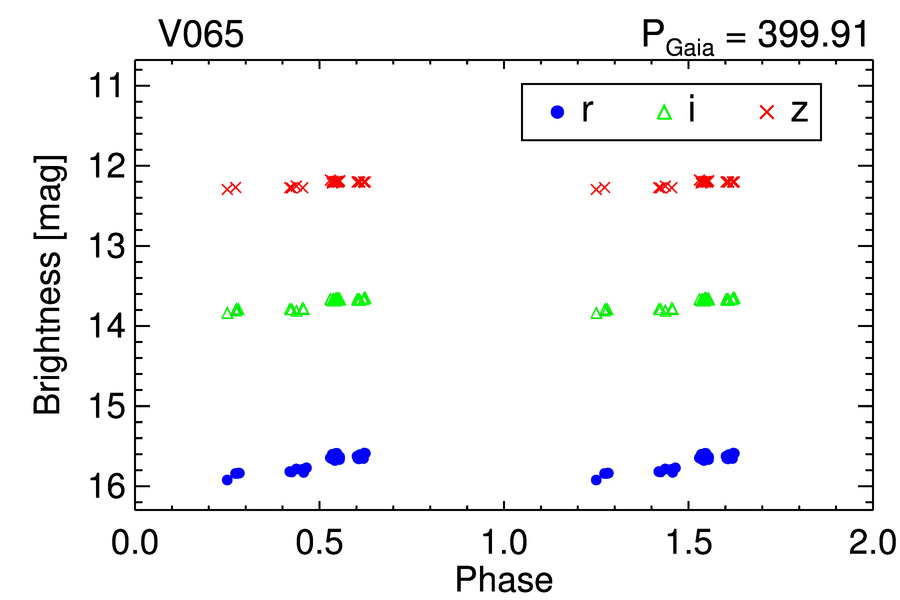}
\includegraphics[width=.24\textwidth]{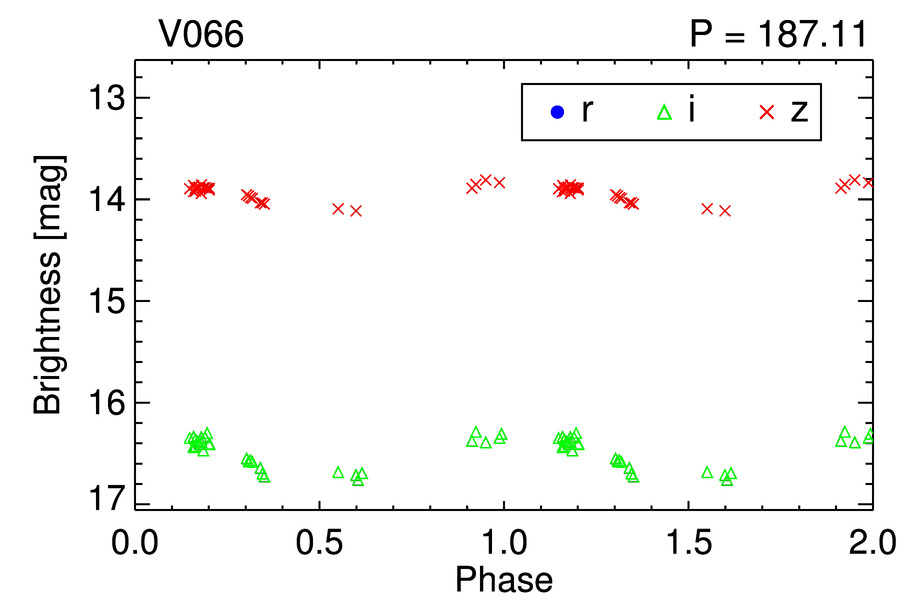}
\includegraphics[width=.24\textwidth]{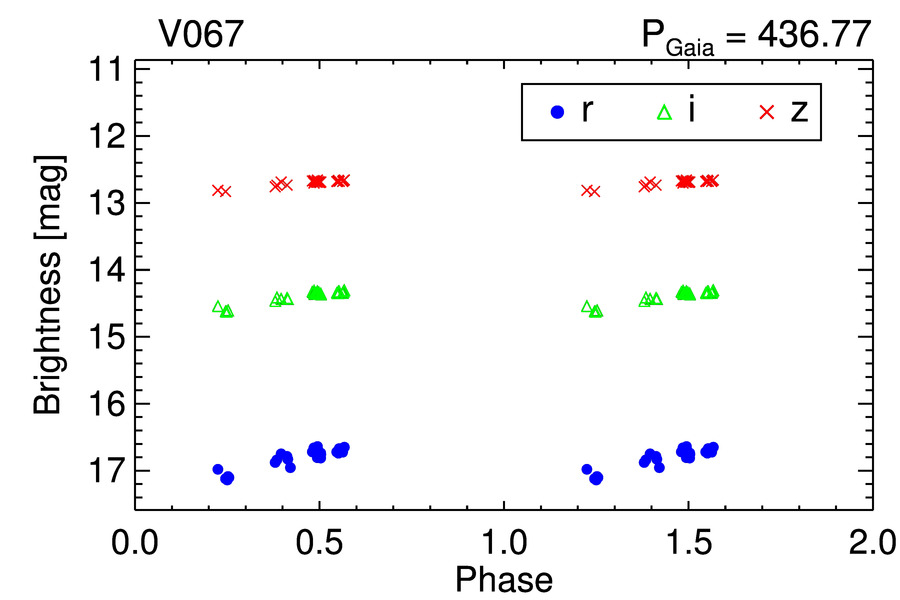}
\includegraphics[width=.24\textwidth]{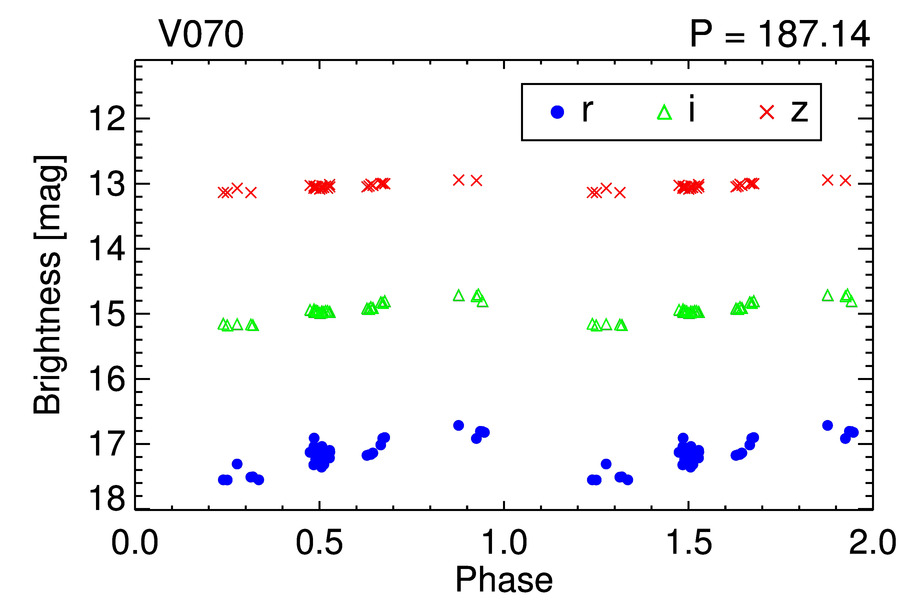}
\includegraphics[width=.24\textwidth]{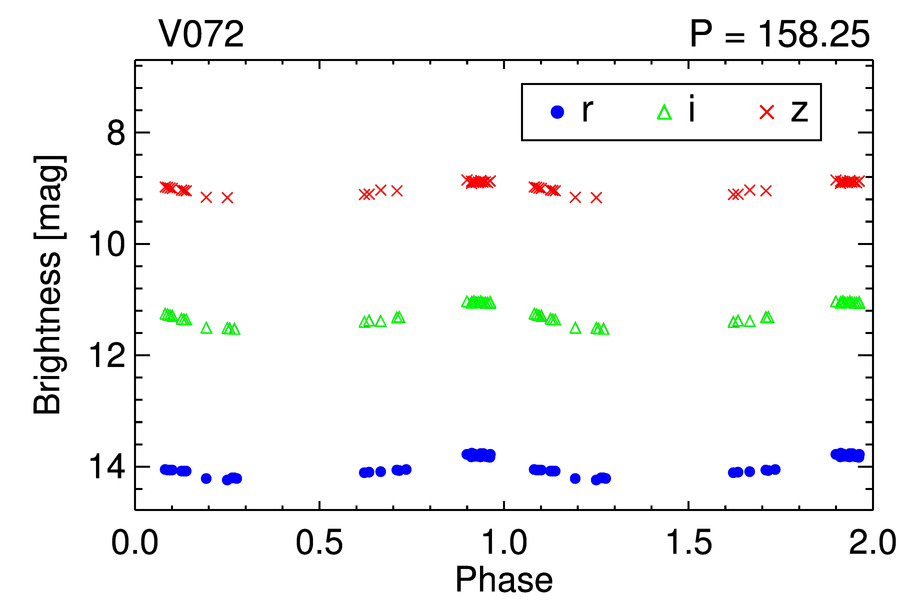}
\includegraphics[width=.24\textwidth]{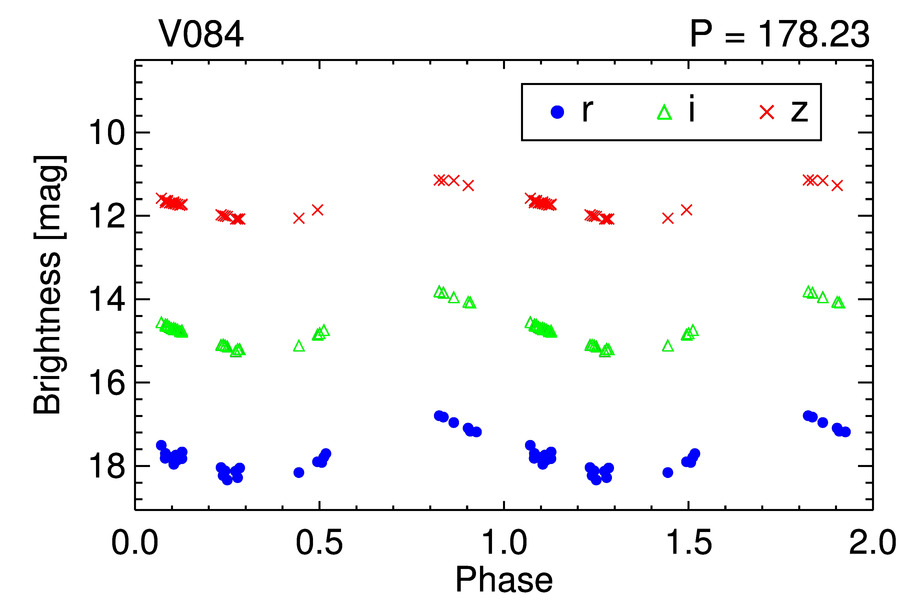}
\includegraphics[width=.24\textwidth]{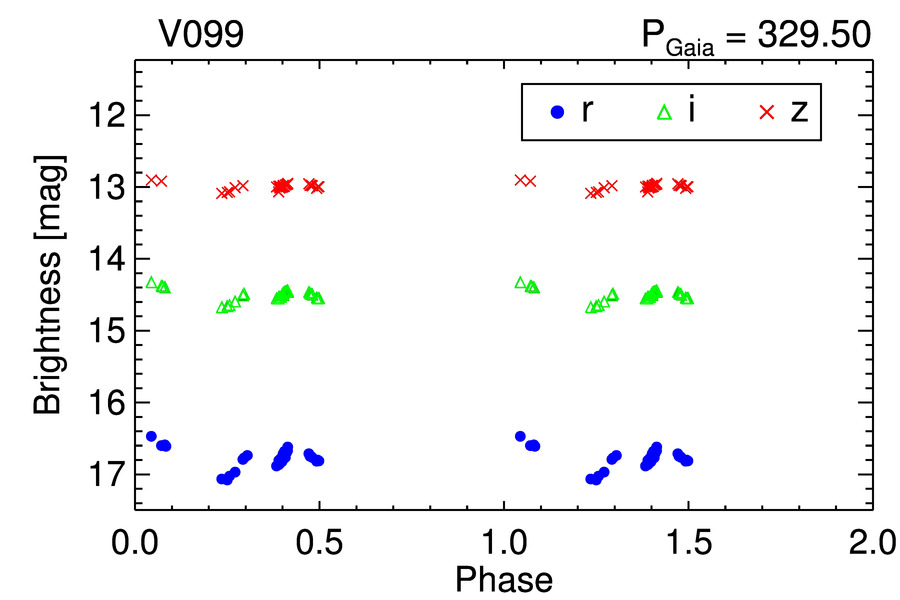}
\includegraphics[width=.24\textwidth]{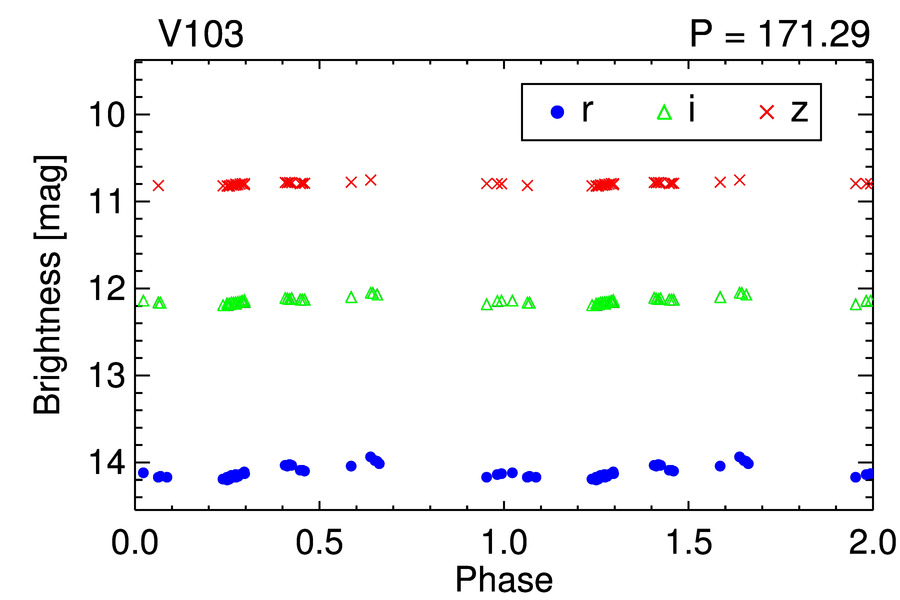}
\includegraphics[width=.24\textwidth]{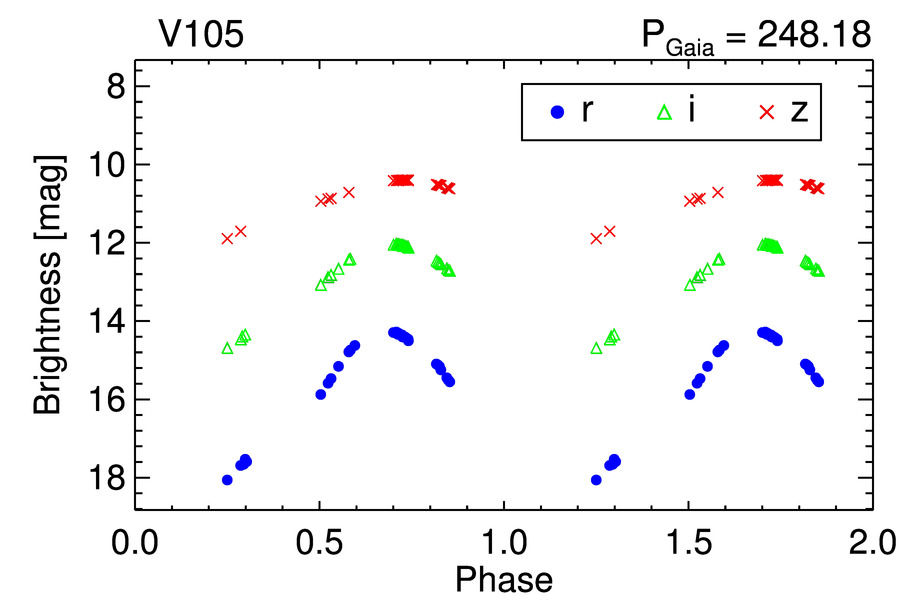}
\includegraphics[width=.24\textwidth]{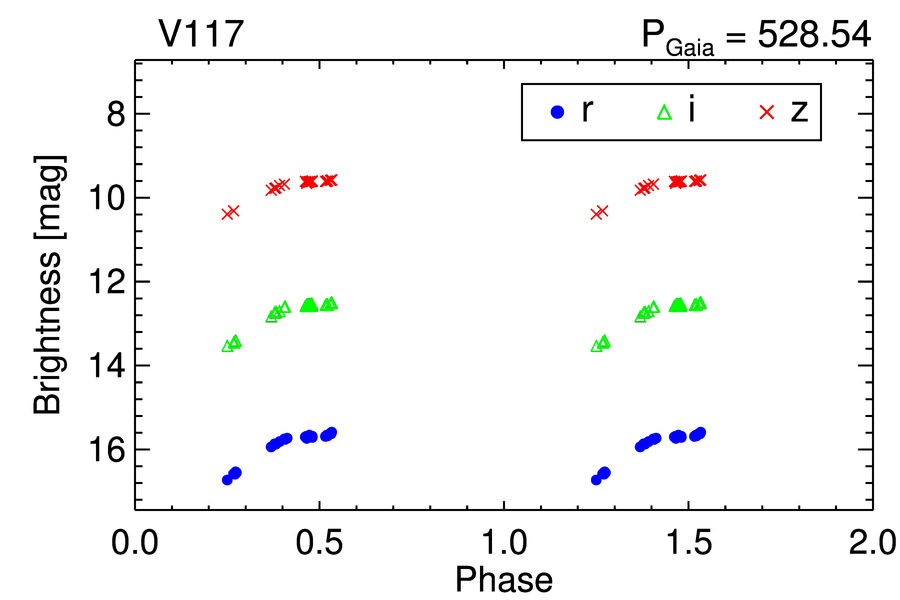}
\includegraphics[width=.24\textwidth]{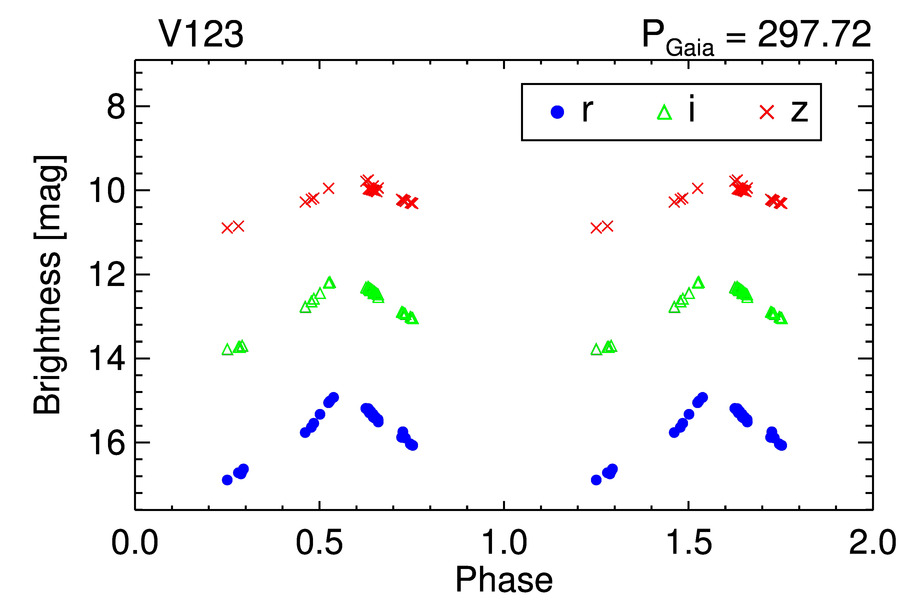}
\includegraphics[width=.24\textwidth]{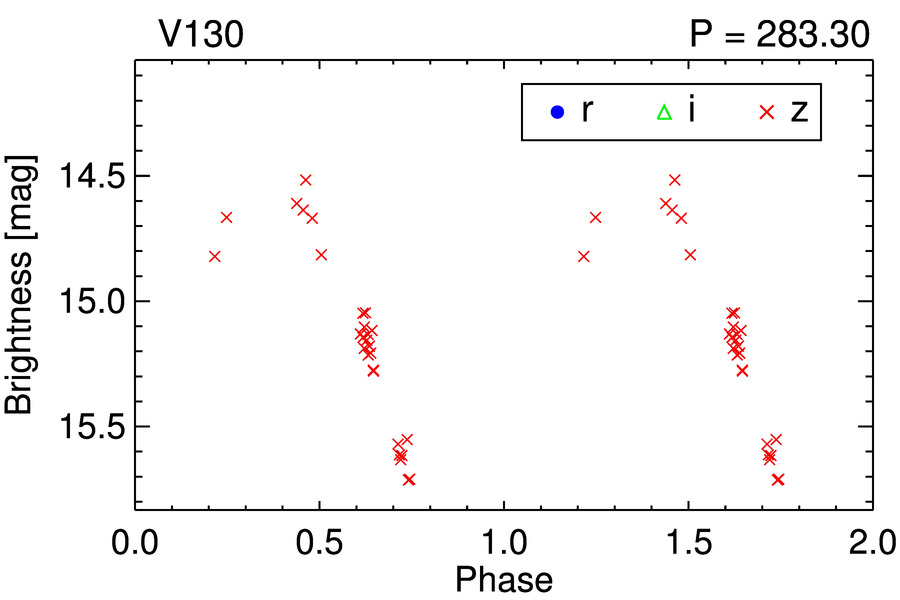}
\includegraphics[width=.24\textwidth]{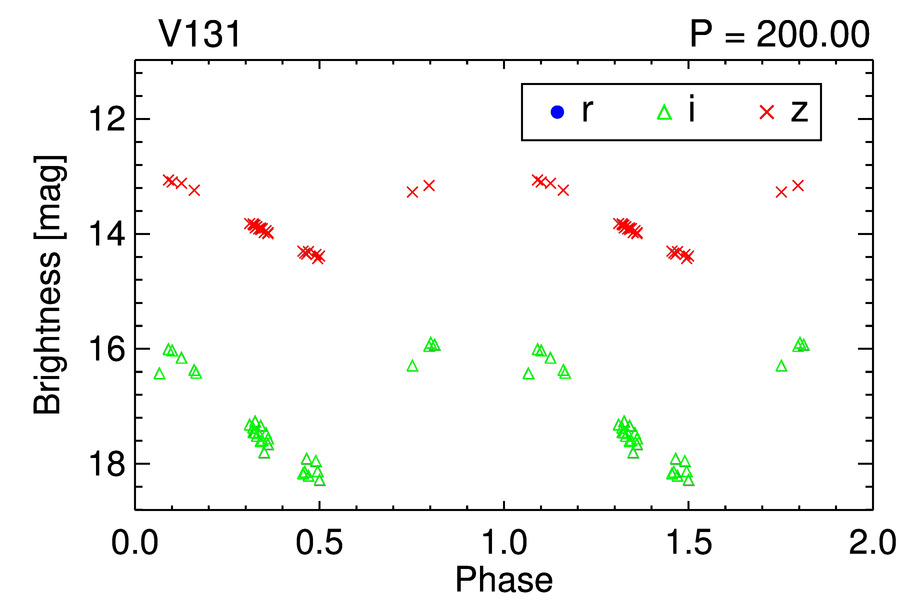}
\includegraphics[width=.24\textwidth]{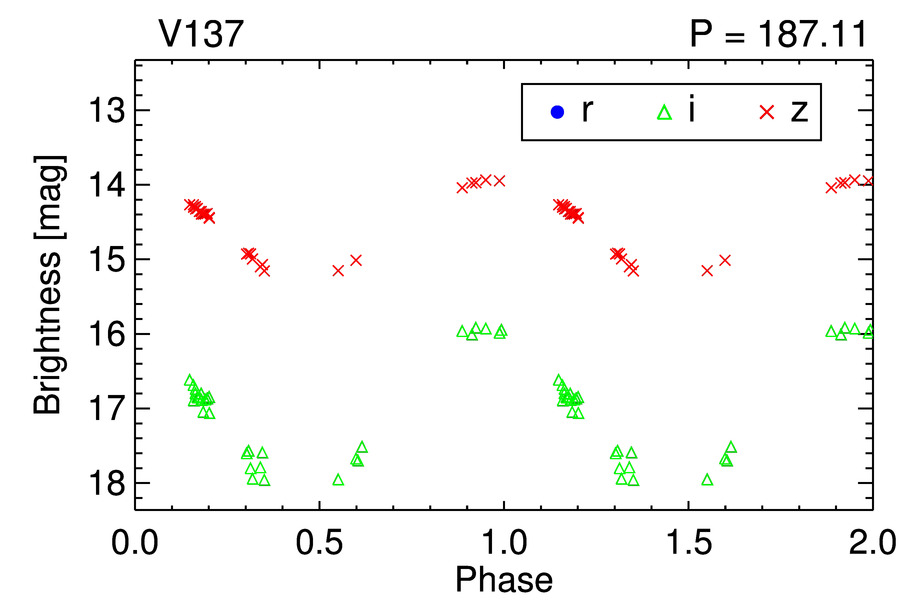}
\includegraphics[width=.24\textwidth]{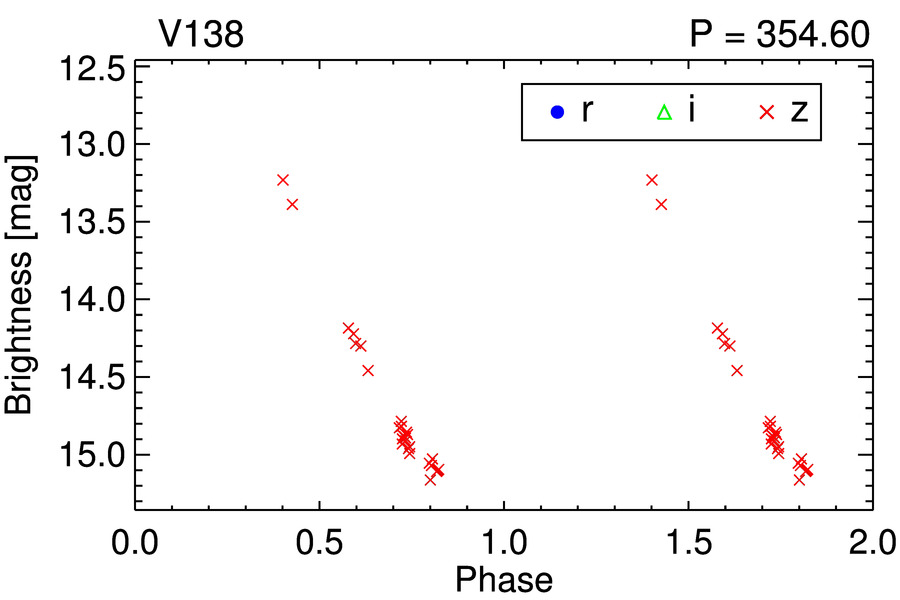}
\includegraphics[width=.24\textwidth]{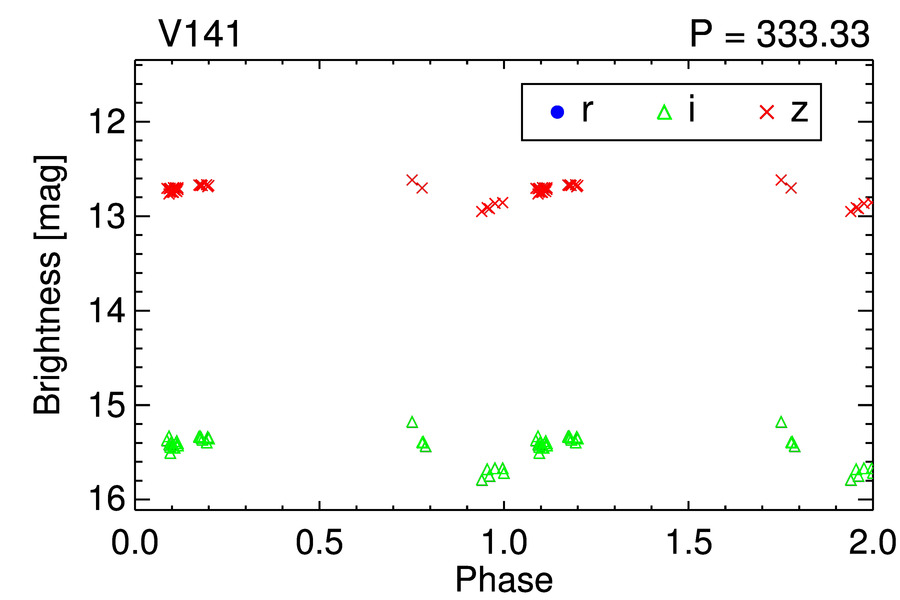}
\includegraphics[width=.24\textwidth]{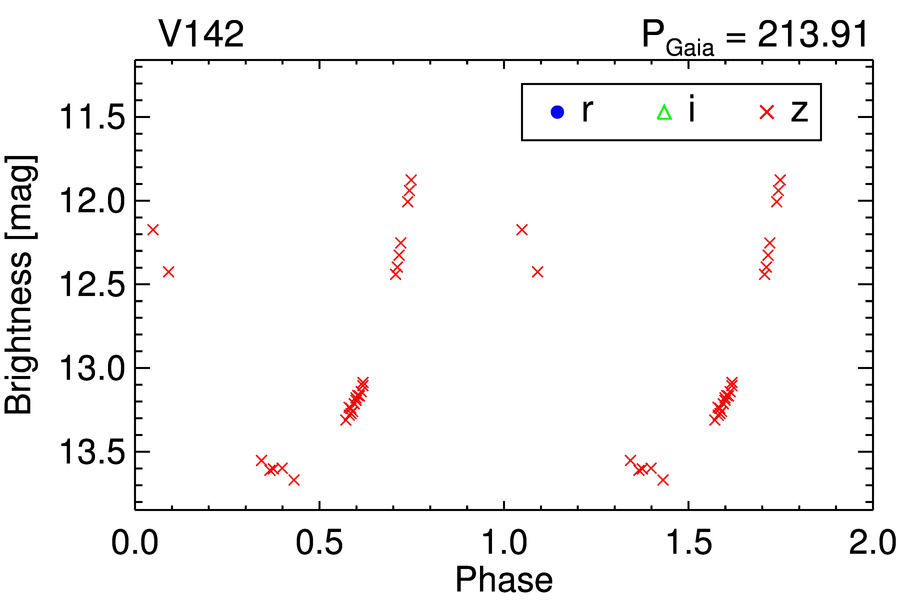}
      \caption{(Continued)}
\end{figure}

\clearpage
\section{Parameters of the Eclipsing Binaries}
\label{app:model}

The PHOEBE model, based on the widely used Wilson--Devinney code \citep{1971ApJ...166..605W} is often used to reproduce and fit the light curves, radial velocity curves, and spectral line profiles of eclipsing systems though modeling eclipse events with high precision using the triangulation method of meshing surfaces \citep{2016ApJS..227...29P}. We make use of PHOEBE~2.3 to derive the parameters of the eclipsing binaries found with our survey.

\begin{deluxetable*}{lcccccccc}
\tablecaption{Fitted parameters of the modeled binary systems\label{tab:phoebe}}
\tablewidth{700pt}
\tabletypesize{\tiny}
\tablehead{
\colhead{Obs\_ID}                       & \colhead{V013}         & \colhead{V038}         & \colhead{V079}     & \colhead{V080}         & \colhead{V086}         & \colhead{V106}         & \colhead{V113}         & \colhead{V129} \\
\colhead{Classification}                & \colhead{EW}                  & \colhead{EW}                  & \colhead{EW}              & \colhead{EA}                  & \colhead{EA}                  & \colhead{EW}                  & \colhead{EA}                  & \colhead{EW}
}
\startdata
\multicolumn{9}{c}{Measured Parameters}\\
\hline
Magnitude $Z_m$ (mag)
& 14.16                       & 12.48                        & 13.80                   & 14.13                       & 13.69                      & 14.29                       & 15.08                       & 14.99 \\
Amplitude $A_z$ (mag)
& 0.4                           & 0.51                          & 0.33                     & 0.44                          & 0.26                        & 0.52                         & 1.10                          & 0.45 \\
Period $P$ (day)
& 18.02                      & 0.2905                      & 7.730                   & 2.888                        & 2.286                      & 0.2955                     & 0.8701                     & 0.4266 \\
\hline
\multicolumn{9}{c}{Fitting Parameters}\\
\hline
Gaia Temperature& 3707.7 			& 4701.3 			& 3498.0 			& 4032.3 			& 3977.8 			& 3784.0 			& 3940.0 			& 3918.5 \\
$T1$ (K)			& 		 			&		 			& 		 			& 					& 		 			& 		 			& 		 			& 		 \\
SEDs Temperature& 3615 (fixed) 		& 4618 (fixed) 		& 3545 (fixed) 		& 5072 (fixed) 		& 4420 (fixed) 		& 3779 (fixed) 		& 5735 (fixed) 		& 4738 (fixed) \\
$T1$ (K)			& 		 			&		 			& 		 			& 					& 		 			& 		 			& 		 			& 		 \\
Epoch $t0$ (MJD-& 4.856 & 0.109 	& 5.025   & 1.851    & 1.509     & 0.184  & 0.087  & 0.0129 \\
57490)		& 	 (2.22E-8)	&	(5.82E-4)		& 	(3.95E-2)		& 	(1.19E-8)		& 	(3.01E-2)	& 	(7.54E-4)	& 	(9.85E-9)&  (1.43E-3) \\
Mass ratio $Q$& 0.960          & 0.830 (0.039)    & 0.960   & 0.84      & 1.00    & 0.79 (0.17)        & 0.95        & 0.87 (0.11) \\
			& 		(0.062)	&		 			& 		 (5.14E-9)	& 	 (4.88E-9) 		& 	 (1.10E-8)	& 		 			& 	 (9.56E-8) 	& 		 \\
Inclination $i$ (deg)& 63.00 (2.20)           & 72.51 (0.02)       & 58.00 (5.31)        & 85.44        & 73.50     & 64.28 (5.94)       & 85.33       & 65.83 (0.83) \\
			& 		 			&		 			& 		 			& 		(1.33E-7) & 		(2.23E-7)& 		 			& 	 (3.3E-6) & 		 \\
Temperature $T2$ (K)& 3531.94  & 4547.61     & 3508.63  & 4982.39        & 3871.13  & 3463.67  & 4577.24    & 4174.52 \\
			& 		(72.79)	&		 (74.80)	& 		 (147.50)	& 		(57.41)	& 		(10.54)	& 	 (175.50)  	& 	 (148.55)	&  (165.05) \\
Semi-major axis $a$ & 3.20       & 3.53 (0.03)             & 3.47      & 6.76          & 3.98      & 3.68      & 3.22       & 3.55 (0.05) \\
(R$_{\odot}$)			& 	(5.42E-9)	&		 			& 	 (2.28E-2)	& 	 (1.15E-7)	& 	 (7.07E-9)	& 	 (3.46E-8) 	&  (2.91E-7) 	& 		 \\
Equivalent Radius $R1$ & 1.52 (0.06)            & 1.51              & 1.53 (0.176)         & 1.18 (0.03)         & 0.906    & 1.62 (0.17)                & 0.91      & 1.53 (0.04) \\
(R$_{\odot}$)			& 		 			&	 (1.32E-8) 	& 		 			& 					& 	 (2.28E-8)	& 		 			& 	(1.56E-7) 	& 		 \\
Equivalent Radius $R2^a$ 
& 1.50                       & 1.39                          & 1.53                      & 1.0                            & 1.0                          & 1.47                           & 1.0                           & 1.44 \\
(R$_{\odot}$)			& 		 			&		 			& 		 			& 					& 		 			& 		 			& 		 			& 		 \\
Potentials$^a$
& 3.25                       & 3.31                          & 3.39                      & -                                & -                              & 3.21                           & -                                & 3.34 \\
Fill-out factor $f^a$
& 0.836                     & 0.331                       & 0.565                    & -                                 & -                             & 0.431                         & -                                & 0.404 \\
pbLum$_r^b$
& 0.91 (0.09)            & 3.85 (0.18)               & 1.12 (0.23)           & 2.50 (0.08)               & 2.51 (0.005)          & 2.77 (0.39)                & 2.32 (0.14)               & 2.26 (0.20) \\
pbLum$_i^b$
& 2.52 (0.15)            & 5.19 (0.21)               & 2.73 (0.31)           & 4.66 (0.14)               & 4.69 (0.03)            & 4.81 (0.54)                & 4.52 (0.22)               & 4.60 (0.35) \\
pbLum$_z^b$
& 5.97 (0.27)            & 6.33 (0.23)               & 5.47 (0.46)           & 7.22 (0.19)               & 7.22 (0.04)            & 7.23 (0.74)                & 6.95 (0.31)               & 7.23 (0.17) \\
${\chi}^2$
& 3.20E-2                 & 5.63E-2                    & 5.72E-2                & 1.43E-2                    & 1.18E-2                 & 0.178                        & 0.207                        & 8.13E-2 \\
\enddata
\tablenotetext{a}{The constraint parameters of the PHOEBE model. The value of primary equivalent radius constrains the potential and fillout factor of the envelope, as well as the equivalent radius of the secondary.}
\tablenotetext{b}{The best-fitting results of bandpass luminosity for the normalized fluxes in $r$, $i$, and $z$.}
\end{deluxetable*}

The parameters of binary systems from PHOEBE have default values, and there are constraints among some of these parameters; for example, the mass of a component is constrained by the semi-major axis, period, and mass ratio of the system. In this work, the default values of some parameters and all of the constraints were preserved, and the values of the other parameters were obtained from the observations, as shown in Table~\ref{tab:phoebe}, including the fluxes and period, $P$, of each the binary system and the effective temperature, $T1$, of the primary therein. The fluxes in the $r$, $i$, and $z$ bands were normalized by dividing by the maximum flux in the $z$ band, and the flux ratios of the three bands were invariant.

Generally, there are two minima in the light curves of eclipsing binaries. However, in most case, there is only one minimum, either because the primary eclipse and secondary eclipses are similar or because the secondary transit is not detected. Therefore, for the possible eclipsing systems with only one minimum (V013, V038, V079, V080, V086, and V129), the double period parameter was used for fitting.

The effective temperature of each primary star was obtained from fitting stellar model grids to the SEDs \citep{2017A&A...600A..11R}, where the SEDs were constructed using the multiband photometry from Gaia/PanSTARRS, 2MASS, and WISE\footnote{Multi-band photometric data: $G$, $G_{\rm BP}$, and $G_{\rm RP}$ bands from \emph{Gaia} DR2, \url{https://www.cosmos.esa.int/web/gaia/data-release-2}; $g$, $r$, $i$, $z$, and $y$ bands from PanSTARRS, \url{https://panstarrs.stsci.edu/}; $J$, $H$, and $K_{\rm s}$ bands from 2MASS, \url{https://old.ipac.caltech.edu/2mass/}; and 3.4\,$\mu$m, 4.6\,$\mu$m, 12\,$\mu$m, and 22\,$\mu$m bands from WISE, \url{http://wise2.ipac.caltech.edu/docs/release/allsky/}}. The effective temperatures of the eight eclipsing binaries from the Gaia catalog \citep{2018A&A...616A...8A} are listed in Table~\ref{tab:phoebe} as a reference. The SED temperatures of the primaries are all below 6000\,K. The bolometric gravity brightening, $g = 0.32$, and albedo, $A = 0.6$, values suggested by PHOEBE were used when modeling the light curves. The eccentricity of the contact binaries (EW) was set to a default value of zero, while for the detached binaries (EA), there were no obvious deviations of the secondary eclipses; thus, their eccentricities were set to zero ($e = 0$).

The least-squares minimization method was used to obtain the free parameters, such as the effective surface temperature, $T2$, of the secondary star; mass ratio, $Q$; inclination, $i$; semi-major axis, $a$; the equivalent radius of the primary, $R1$; and the epoch, $t0$ (i.e., the time of primary eclipse). In the meantime, the other important constrained parameters, including the equivalent radius, $R2$, of the secondary star; the potential of the envelope; and the filling factor, $f$, of the contact binary were derived from the model. The results are summarized in Table~\ref{tab:phoebe}.

For the case of two similar MS components, the temperature of the primary star ascertained from the combined SED may have a systematic error of $\sim$100K, suggested by \citet{2006Ap&SS.304..347P}. To estimate the uncertainty of the free parameters caused by the uncertainty of the primary star's temperature, the initial temperature of the primary star was increased by 200 or 300\,K (e.g., SED temperature deviations of about 3\% or 5\%). It was found that the corresponding uncertainties of the free parameters were between $\pm$3\% and $\pm$5\%.

Figure~\ref{fig:phoebe} shows the modeling results of the binary systems. Object V038 was the brightest EW-type binary system, and its observed light curves fit well with those from the spotless model. The heights of the maxima in the light curves of V038 showed some slight differences, such as the O'Connell effect, possibly caused by spots. The addition of spots to the spotless model did not yield a better-fitting model, probably due to the small cadence of our observations. The fitted curves of the other contact binary systems (V013, V079, V106, and V129) were basically consistent with the observed light curves. As shown in Table~\ref{tab:phoebe}, the best-fitting results ($T2$, $i$, and $R1$) of V013, V079, and V106 were similar, probably caused by their same variability type and similar initial parameters, implying they could be K-type stars with similar brightnesses, variability amplitudes, and distances. For the detached binaries (V080 and V086), the fitted light curves agreed well with the observed light curves. The light curves of V113 presented obvious random fluctuations near the secondary eclipse, which likely had an adverse effect on the fitted results.

\setcounter{figure}{0}
\renewcommand{\thefigure}{B\arabic{figure}}
\begin{figure}[!h]
 \centering
          \includegraphics[width=.45\textwidth]{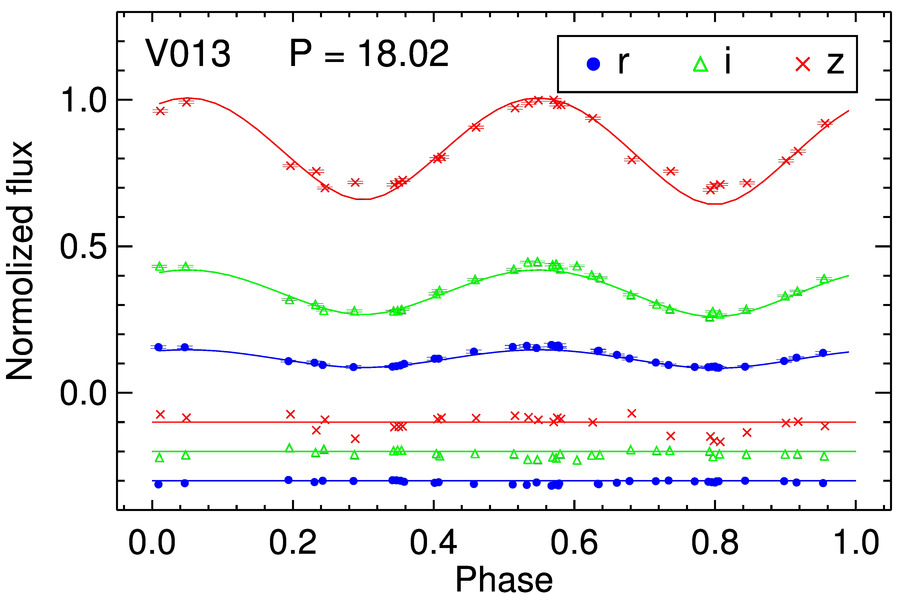}
          \includegraphics[width=.45\textwidth]{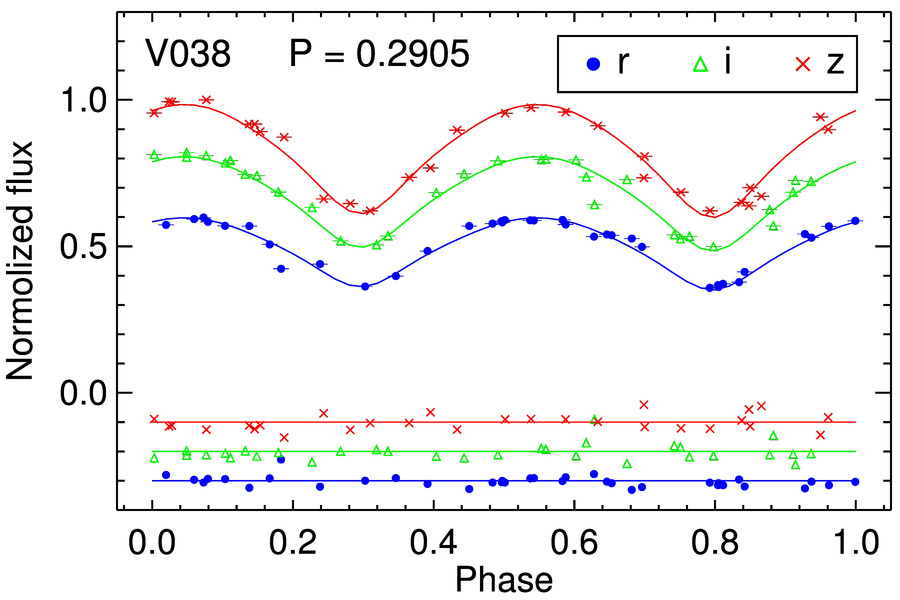}
          \includegraphics[width=.45\textwidth]{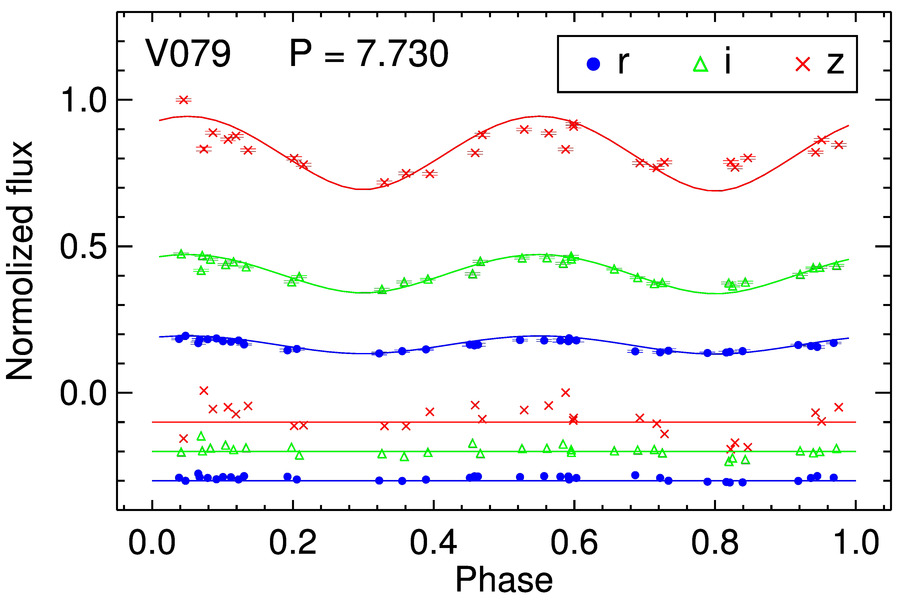}
          \includegraphics[width=.45\textwidth]{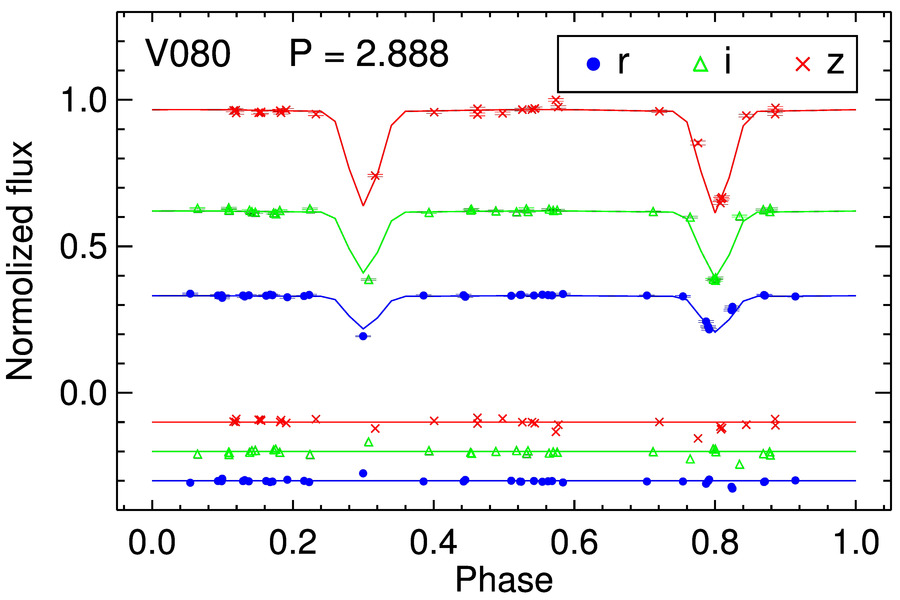}
          \includegraphics[width=.45\textwidth]{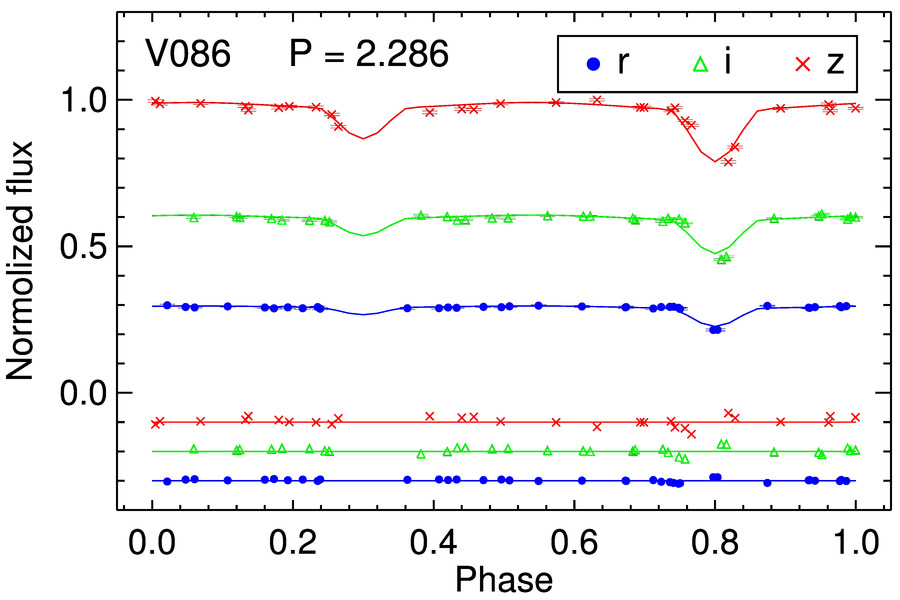}
          \includegraphics[width=.45\textwidth]{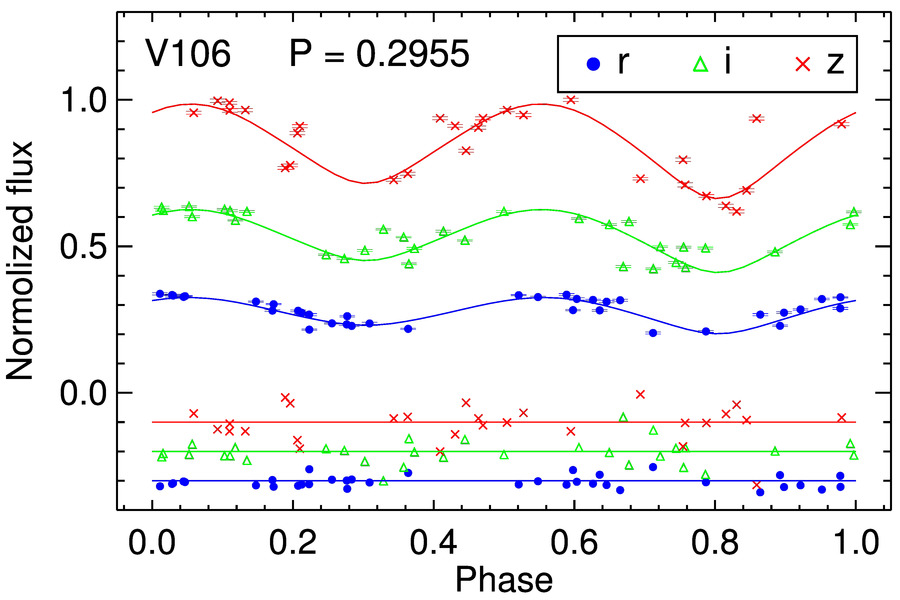}
          \includegraphics[width=.45\textwidth]{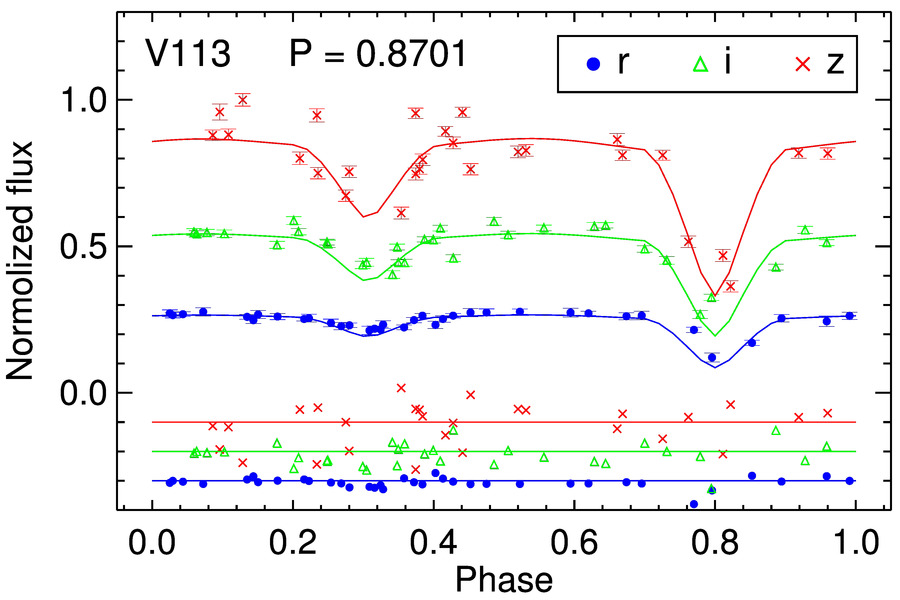}
          \includegraphics[width=.45\textwidth]{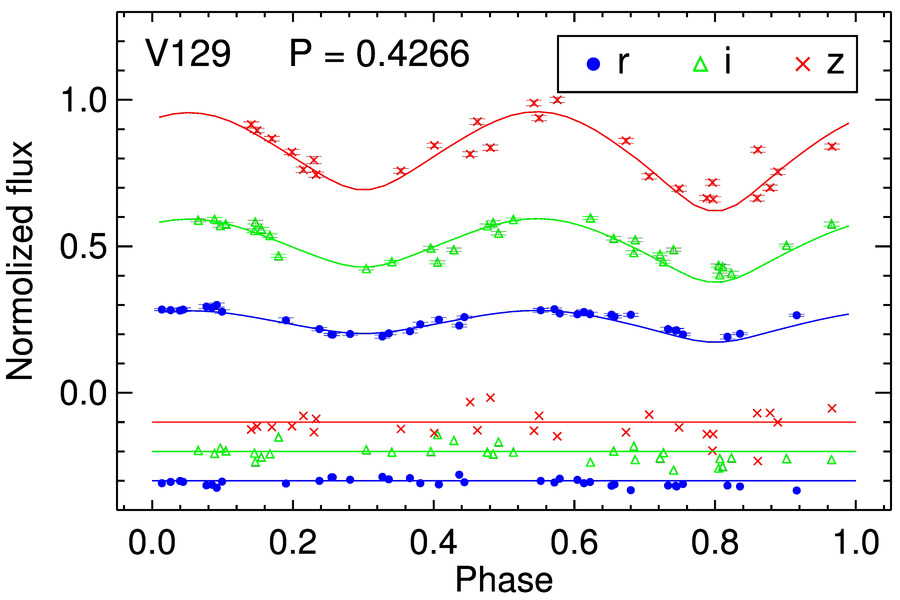}
      \caption{Light-curve modeling results of eight binary systems in the Serpens Main field. The symbols with different colors represent the observations in $r$ (blue circles), $i$ (green triangles), and $z$ (red crosses); the curves denote the corresponding fits. The residuals ($\delta r$, $\delta i$, and $\delta z$) are shown at the bottom of the figures; constants have been added for better visibility (e.g., $\delta r$-0.3, $\delta i$-0.2, and $\delta z$-0.1). The typical photometric error is less than 0.01\,mag. The previously known variable star is V038.}\label{fig:phoebe}
\end{figure}

\section{LP Variables}
\label{subsec:lps}

\setcounter{figure}{0}
\renewcommand{\thefigure}{C\arabic{figure}}
\begin{figure}[!h]
 \centering
          \includegraphics[width=.9\textwidth]{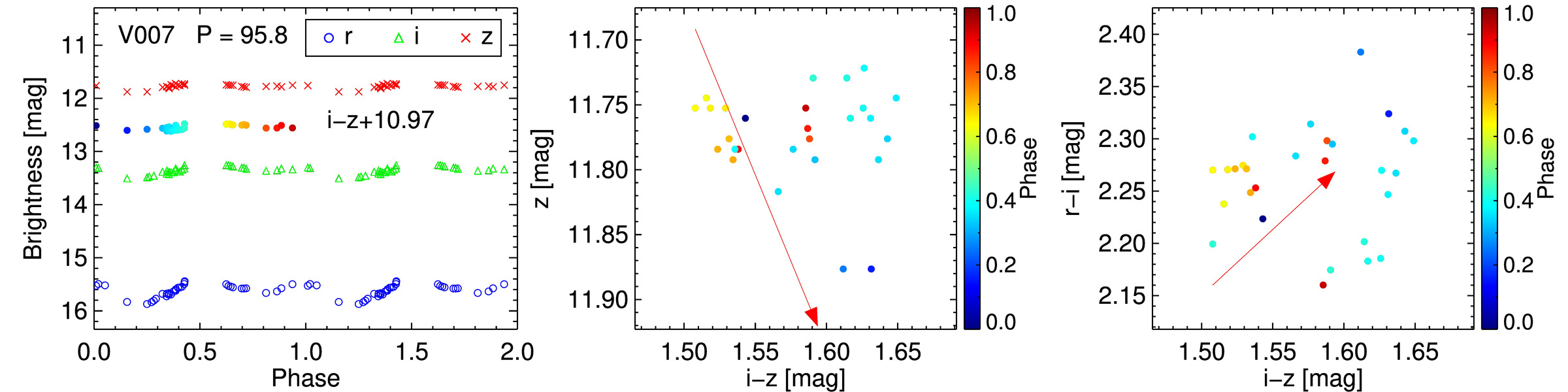}
          \includegraphics[width=.9\textwidth]{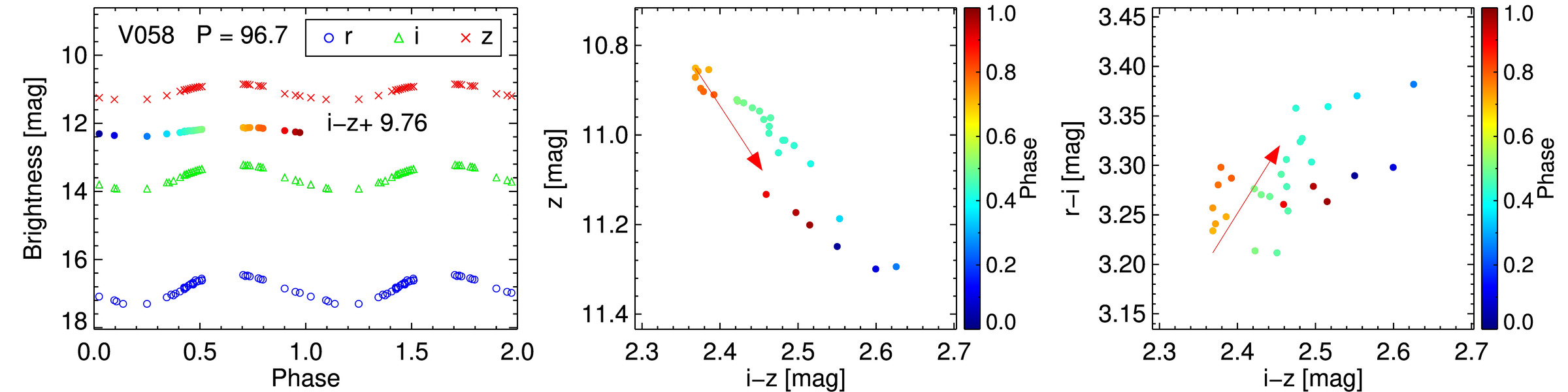}
          \includegraphics[width=.9\textwidth]{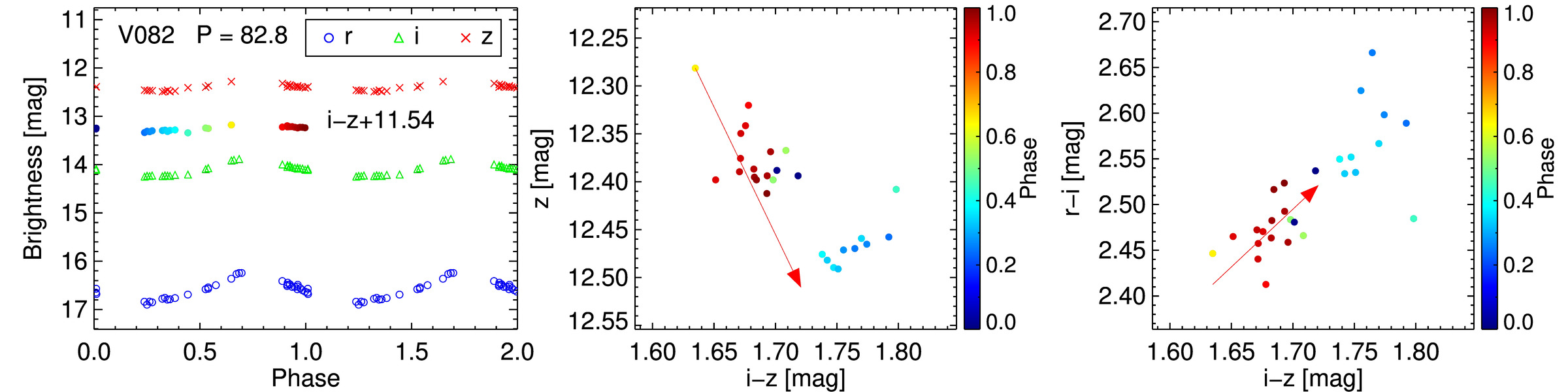}
          \includegraphics[width=.9\textwidth]{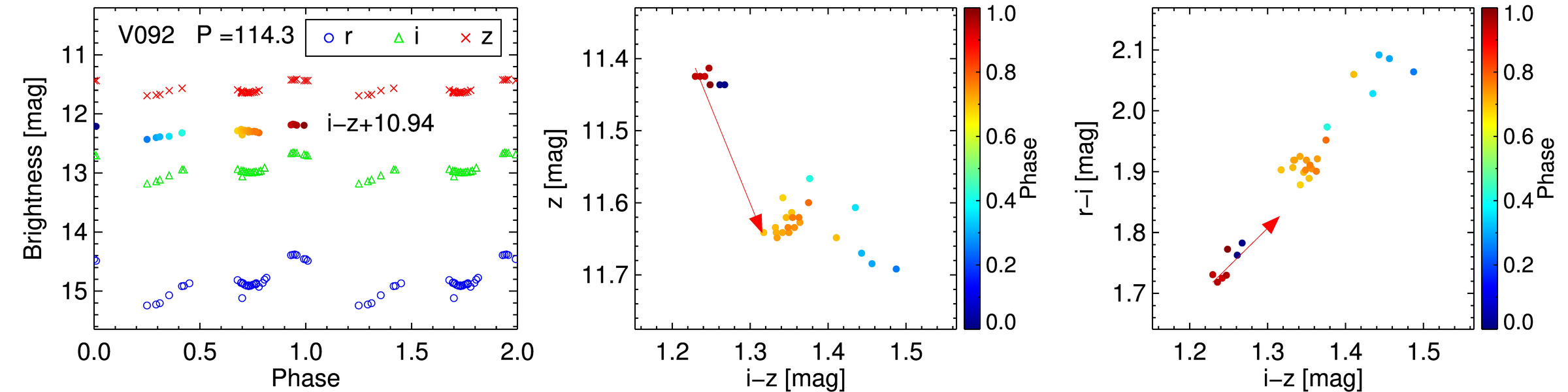}
      \caption{Folded light curves ($r$: blue; $i$: green; and $z$: red), CMDs, and C-CDs of the LPs (V007, V058, V082, and V092). The dots in the left column show the colors ($i-z$ + constant) varying with the phase. The typical photometric error is less than 0.01\,mag. The color bar indicates the phase. The phases of color ($i-z$) in the CMDs and C-CDs are indicated by color, and the arrows indicate an extinction vector of $A_V = 0.5$\,mag.}\label{fig:cmds}
\end{figure}

The LP variables could be pulsating stars, the brightness and radius of which vary due to intrinsic variations of temperature and pressure. The light curves appeared to be strictly periodic, such as seen for Cepheid variables, or merely cyclic, such as observed for RV\,Tauri and Mira variables \citep{2009ebs..book.....K}. Luminosity, pressure, and temperature vary with the energy released by nuclear reactions, and the energy is transferred by the contraction and expansion of the stellar layers. When the star is compressed, the temperature in the layers in which the nuclear reactions take place increases, and the temperature decreases during expansion \citep{2015pust.book.....C}. The driving/damping of pulsations in different layers and the turbulence of convective envelopes make the variations of light curves and their colors complicated to interpret. As \citet{2019A&A...623A.110G} mentioned, the variability-induce motions of pulsating stars in CMDs appear as loop-like shapes, and the directions of their motions are similar.

A simple Planck model was used to investigate the color variabilities of the pulsating stars, and we assumed that a star cools as its expands, from a high temperature, $T_{r}$, at the minimum radius, $r$, to a low temperature, $T_{R}$, at the maximum radius, $R$. The variation of photospheric emission can be expressed as $\Delta m(\lambda) = -2.5 {\rm log} \{R^2/r^2 \times [B_{\lambda}(T_{\rm R})/B_{\lambda}(T_{r})]\}$, where $B_{\lambda}(T)$ is the Planck function. The variability-induced motion of pulsating stars in CMDs can be described simply by a combination of two vectors: one is a vertical vector (-90$^{\circ}$) produced by radial variations of the star (i.e. $R/r$), and the length of the vector indicates the distance of the motion, which depends on $R/r$; the other is a vector with an angle of about 80$^{\circ}$ induced by variations of the ratio $T_{R}/T_{r}$, where the length of the vector depends on $T_{R}/T_{r}$.  For a pulsating star with a stellar temperature of 4000\,K, when $R/r=2.0$ and $T_{R}/T_{r}=0.7$, the vector will show an angle of 60.4$^{\circ}$, and a vector length of 0.43\,mag, as shown by the red arrow in Figure \ref{fig:ysos}.

For the LPs and SR\_Ms shown in Figure \ref{fig:cmds}, their motions show linear and loop-like shapes, and their magnitudes are highly correlated with their colors. The light curves of V058 are consistent with those of pulsating stars, showing a loop-like shape of motions in the CMD. The light curves of V007 and V092 are SR, with different paths of motions. The light curves of V082 are periodic, and the motions are not clear; indeed, the difference of the motions may be caused by a single pulsation or a combination of multiple pulsation mechanisms. In general, the directions of motion of most LPs are similar (about 50$^{\circ}$) in CMDs, as shown in Figure~\ref{fig:ysos}, and their color variabilities may be uncorrelated with extinction. According to the relationship of color--temperature \citep{2000AJ....120.1072S}, the temperature is related to color, metallicity, and surface gravity, and motion in the CMD induced by the color variability can be affected by metallicity and surface gravity; a more detailed analysis of the effects of these physical parameters which could be helpful for explaining the different motions induced by different mechanisms.

\section{Effect of Phase Difference}
\label{subsec:phase}
To analyze the effect of nonsimultaneous observations on the color variability, the PHOEBE model was used to simulate the light curves of eclipsing binaries; CMDs and C-CDs were obtained at the same time. As shown in Figure~\ref{fig:app}, the motion induced by color variability in the CMD ($i-z$ vs. $z$) changed when the phase difference was greater than 0.04$\pi$ (2\% of the period). The motion in the C-CD changed when the phase difference was greater than 0.08$\pi$ (4\% of period). In this study, the motions of variables with periods less than 1.5\,days were affected by the phase difference.

\setcounter{figure}{0}
\renewcommand{\thefigure}{D\arabic{figure}}
\begin{figure}[!h]
 \centering
          \includegraphics[width=.99\textwidth]{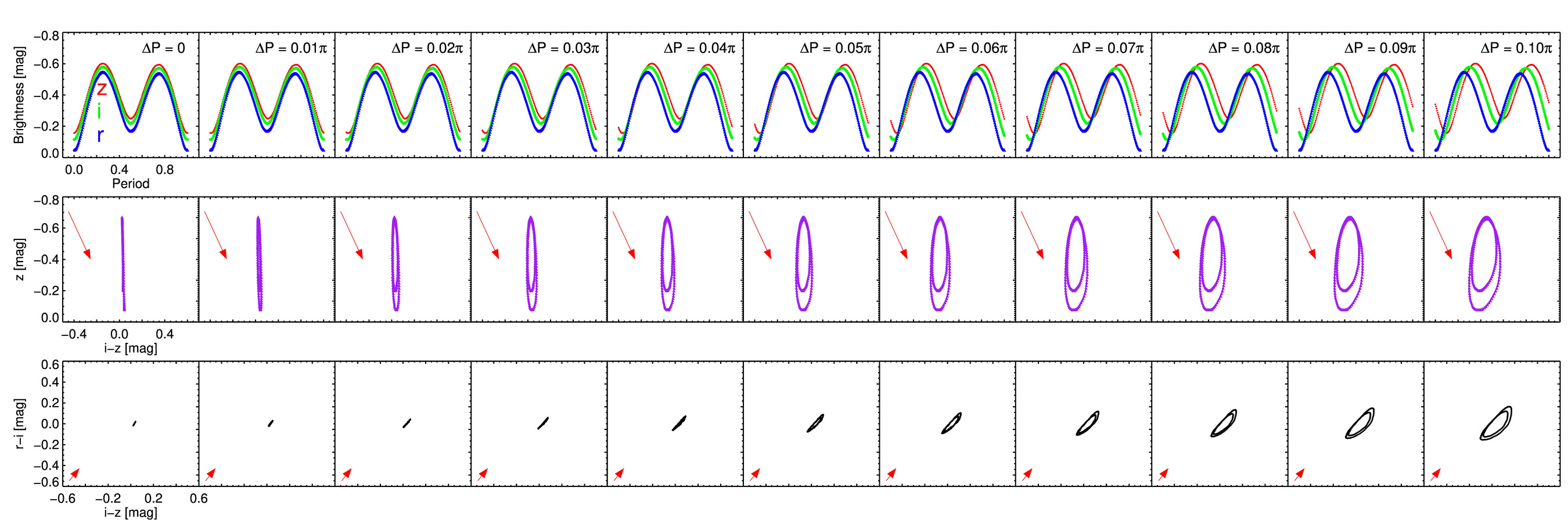}
          \includegraphics[width=.99\textwidth]{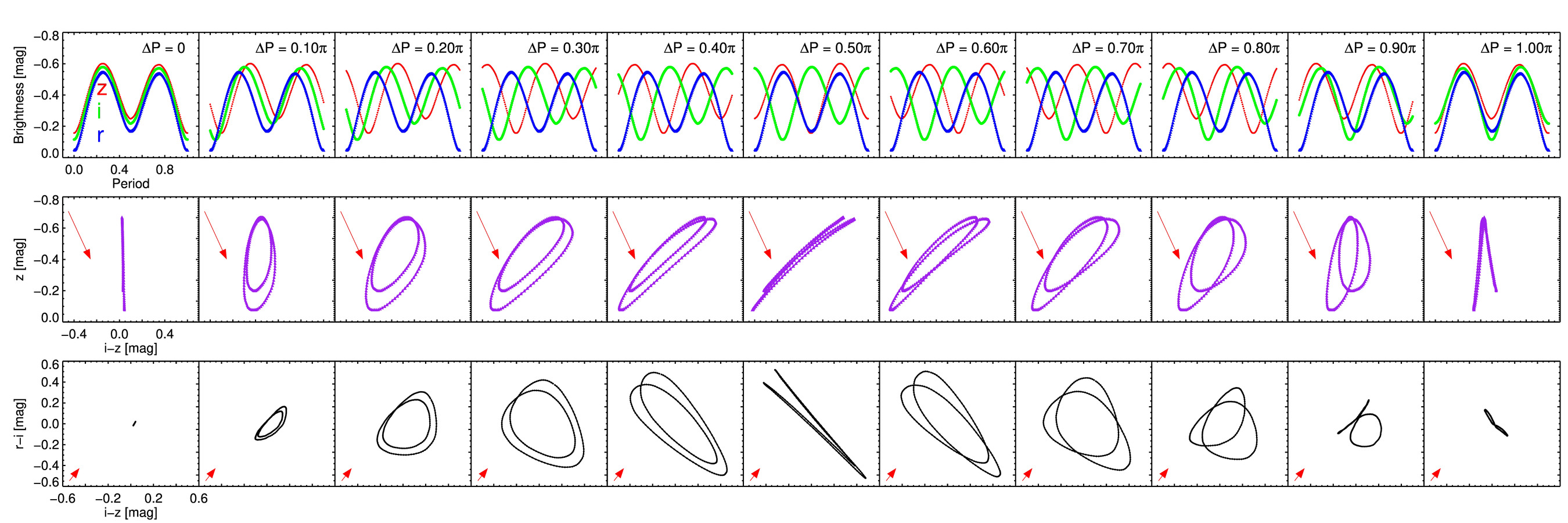}
      \caption{Simulated light curves, CMDs, and C-CDs of the eclipsing binary systems obtained from nonsimultaneous observations. The first three rows show light curves ($r$: blue; $i$: green; and $z$: red) with different phase differences and the corresponding CMDs and C-CDs, where the phase difference varies from 0 to 0.1$\pi$, with an interval of 0.01$\pi$. In the last three rows, the phase difference ranges from 0 to 1.0$\pi$, with an interval of 0.1$\pi$. The arrows indicate an extinction vector of $A_V = 0.5$\,mag.}\label{fig:app}
\end{figure}




\bibliographystyle{aasjournal}
\bibliography{serpens_vstar_accepted}



\end{document}